\documentclass[11pt,a4paper]{article}
\usepackage{amssymb,amsmath,amsfonts,amsthm,amstext,amscd,array}
\usepackage{mathrsfs}
\usepackage{hyperref}
\usepackage{pdfsync}
\usepackage{bbm}
\usepackage{bm}
\usepackage[arrow,matrix,curve]{xy}
\usepackage{bbding}
\usepackage{wasysym}

\newcommand{\mbf}[1]{{\boldsymbol {#1} }}
\def\ii{{\,{\rm i}\,}}
\def\dd{{\rm d}}

\def\A{{\sf A}}
\def\V{{\sf V}}

\newcommand{\unit}{\mathbbm{1}}   			

\def\mfs{{\mathfrak s}}
\newcommand{\fru}{\mathfrak{u}}				

\newcommand{\CA}{\mathcal{A}}

\newcommand{\CB}{\mathcal{B}}

\newcommand{\eq}{\begin{equation}}
\newcommand{\eqend}{\end{equation}}
\newcommand{\eqa}{\begin{eqnarray}}
\newcommand{\nonueqa}{\begin{eqnarray*}}
\newcommand{\eqaend}{\end{eqnarray}}
\newcommand{\nonueqaend}{\end{eqnarray*}}

\newcommand{\bma}[1]{\begin{array}{#1}}
\newcommand{\ema}{\end{array}}
\newcommand{\bc}{\begin{center}}
\newcommand{\ec}{\end{center}}

\renewcommand{\thefootnote}{\fnsymbol{footnote}}
\newcommand{\newsection}{\setcounter{equation}{0}\section}

\newcommand{\complex}{{\mathbb C}} 
\newcommand{\zed}{{\mathbb Z}} 
\newcommand{\real}{{\mathbb R}} 

\newcommand{\quat}{{\mathbb H}} 
\newcommand{\oct}{{\mathbb O}}
\def\alg{{\mathcal A}}

\newif\ifold             \oldtrue

\def\e{{\,\rm e}\,}

\def\be{\begin{equation}}
\def\ee{\end{equation}}
\def\bea{\begin{eqnarray}}
\def\eea{\end{eqnarray}}
\def\bd{\begin{displaymath}}
\def\ed{\end{displaymath}}

\newcommand{\beq}{\begin{eqnarray}}
\newcommand{\eeq}{\end{eqnarray}}

\makeatletter
\newdimen\normalarrayskip              
\newdimen\minarrayskip                 
\normalarrayskip\baselineskip
\minarrayskip\jot
\newif\ifold             \oldtrue            
\def\arraymode{\ifold\relax\else\displaystyle\fi} 
\def\@arrayskip{\ifold\baselineskip\z@\lineskip\z@
     \else
     \baselineskip\minarrayskip\lineskip2\minarrayskip\fi}
\def\@arrayclassz{\ifcase \@lastchclass \@acolampacol \or
\@ampacol \or \or \or \@addamp \or
   \@acolampacol \or \@firstampfalse \@acol \fi
\edef\@preamble{\@preamble
  \ifcase \@chnum
     \hfil$\relax\arraymode\@sharp$\hfil
     \or $\relax\arraymode\@sharp$\hfil
     \or \hfil$\relax\arraymode\@sharp$\fi}}
\def\@array[#1]#2{\setbox\@arstrutbox=\hbox{\vrule
     height\arraystretch \ht\strutbox
     depth\arraystretch \dp\strutbox
     width\z@}\@mkpream{#2}\edef\@preamble{\halign \noexpand\@halignto
\bgroup \tabskip\z@ \@arstrut \@preamble \tabskip\z@ \cr}%
\let\@startpbox\@@startpbox \let\@endpbox\@@endpbox
  \if #1t\vtop \else \if#1b\vbox \else \vcenter \fi\fi
  \bgroup \let\par\relax
  \let\@sharp##\let\protect\relax
  \@arrayskip\@preamble}
\makeatother

\def\be{\beta}

\theoremstyle{definition}

\usepackage{hyperref}
\hypersetup{colorlinks,%
citecolor=black,%
filecolor=black,%
linkcolor=black,%
urlcolor=black}

\begin{document}

\begin{titlepage}

\begin{flushright}

\baselineskip=12pt

EMPG--17--01

\end{flushright}

\begin{center}

\vspace{1cm}

\baselineskip=24pt

{\Large\bf $\mbf{G_2}$-structures and quantization \\ of non-geometric M-theory backgrounds}

\baselineskip=14pt

\vspace{1cm}

{\bf Vladislav G. Kupriyanov}${}^{1}$ \ and \ {\bf Richard
  J. Szabo}${}^{2}$
\\[5mm]
\noindent  ${}^1$ {\it Centro de Matem\'atica, Computa\c{c}\~{a}o e
Cogni\c{c}\~{a}o}\\{\it Universidade de Federal do ABC}\\
{\it Santo Andr\'e, SP, 
Brazil}\\ and {\it 
Tomsk State University, Tomsk, Russia}\\
Email: \ {\tt
    vladislav.kupriyanov@gmail.com}
\\[3mm]
\noindent  ${}^2$ {\it Department of Mathematics, Heriot-Watt University\\ Colin Maclaurin Building,
  Riccarton, Edinburgh EH14 4AS, U.K.}\\ and {\it Maxwell Institute for
Mathematical Sciences, Edinburgh, U.K.} \\ and {\it The Higgs Centre
for Theoretical Physics, Edinburgh, U.K.}\\
Email: \ {\tt R.J.Szabo@hw.ac.uk}
\\[30mm]

\end{center}

\begin{abstract}

\baselineskip=12pt

\noindent
We describe the quantization of a four-dimensional locally
non-geometric M-theory background dual to a twisted three-torus by deriving a
phase space
star product for deformation quantization of quasi-Poisson
brackets related to the nonassociative algebra of octonions. The
construction is based on a choice of $G_2$-structure which defines a
nonassociative deformation of the addition law on the
seven-dimensional vector space of Fourier momenta. We
demonstrate explicitly that this star product reduces to that of the three-dimensional
parabolic constant $R$-flux model in the contraction of M-theory to
string theory, and use it to derive quantum phase space uncertainty
relations as well as triproducts for the
nonassociative geometry of the four-dimensional configuration
space. By extending the $G_2$-structure to a $Spin(7)$-structure, we
propose a 3-algebra structure on the full
eight-dimensional M2-brane phase space which reduces to the
quasi-Poisson algebra after imposing a particular gauge constraint, and
whose deformation quantisation simultaneously encompasses both the phase space star
products and the configuration space triproducts. We demonstrate how
these structures naturally fit in with previous occurences of
3-algebras in M-theory.

\end{abstract}

\end{titlepage}
\setcounter{page}{2}

\newpage

{\baselineskip=12pt
\tableofcontents
}

\bigskip

\renewcommand{\thefootnote}{\arabic{footnote}}
\setcounter{footnote}{0}

\newsection{Introduction and summary\label{sec:intro}}

Non-geometric backgrounds of string theory are of interest not only
because of their potential phenomenological applications, but also because they make
explicit use of string duality symmetries which allows them to probe
stringy regimes beyond supergravity. Many of them can be obtained by
duality transformations of geometric backgrounds in flux
compactifications of ten-dimensional and eleven-dimensional
supergravity (see e.g.~\cite{Grana,DKrev,BKLSrev} for reviews). 
They have been studied extensively for the NS--NS sector
of ten-dimensional supergravity which involves non-geometric
$Q$-fluxes and $R$-fluxes. 

One of the most interesting recent assertions concerning non-geometric strings
is that they probe noncommutative and nonassociative deformations of
closed string background
geometries~\cite{Blumenhagen2010,Lust2010}. This has
been further confirmed through explicit string computations in left-right asymmetric
worldsheet conformal field
theory~\cite{Blumenhagen2011,Condeescu2012,Andriot2013,Blair2014,Bakas2015} and in double
field theory~\cite{Blumenhagen2013}, and also in a topological open
membrane sigma-model~\cite{MSS2} and in Matrix
theory~\cite{Chatzistavrakidis2012} which both suggest origins for
non-geometric fluxes in M-theory. Quantization of these
backgrounds through explicit constructions of phase space star
products were provided in~\cite{MSS2,BaLu,Mylonas2013,Kupriyanov2015}, and
subsequently applied to building nonassociative models of quantum
mechanics~\cite{Mylonas2013} and field theories~\cite{Mylonas2014a,Barnes2016}; the physical significance and
viability of these nonassociative structures in quantum mechanics is
clarified in~\cite{Bojowald2014,Bojowald2015}. A nonassociative theory
of gravity describing the low-energy effective dynamics of closed strings in
locally non-geometric backgrounds is currently under
construction~\cite{Mylonas2013,Barnes2014,Aschieri2015,Barnes2015,Blumenhagen2016}.

Until very recently, however,
there had been two pieces missing from this story: Firstly, the M-theory
version of this deformation of geometry and, secondly, the role played
by the octonions which are the archetypical example of a
nonassociative algebra. In~\cite{GLM}, these two ingredients are
treated simultaneously and shown to be related. Their approach is
based on lifting the non-geometric string theory $R$-flux to M-theory
within the context of $SL(5)$ exceptional field theory,
following~\cite{BlairMalek2014}, which extends the $SL(4)=Spin(3,3)$
double field theory of string theory with three-dimensional target
spaces and is relevant for
compactifications of eleven-dimensional supergravity to seven
dimensions. They argue that the phase space of the four-dimensional
locally non-geometric M-theory background, which is dual to a twisted
three-torus, lacks a momentum mode and consequently is
seven-dimensional. The corresponding classical quasi-Poisson brackets
can then be mapped precisely onto the Lie 2-algebra generated by the
imaginary octonions. In the contraction limit $g_s\to0$ which reduces
M-theory to IIA string theory, the quasi-Poisson brackets contract
to those of the non-geometric string theory $R$-flux background
obtained via T-duality from a geometric three-torus with $H$-flux. The
goal of the present paper is to quantize these phase space quasi-Poisson
brackets, and to use it to describe various physical and geometrical
features of the non-geometric M-theory background.

For this, we derive a phase space star product which lifts that of the
three-dimensional string theory $R$-flux background~\cite{MSS2}, in
the sense that it reduces exactly to it in the appropriate contraction
limit which shrinks the M-theory circle to a point; our derivation is based on extending and elucidating deformation
quantization of the coordinate algebra related to the imaginary
octonions that was recently considered in~\cite{Kup24}. The
contraction limit reduces the complicated combinations of
trigonometric functions appearing in the resulting
star product to the elementary algebraic functions of the string
theory case. Our constructions
exploit relevant facts from calibrated geometry, particularly the theory of
$G_2$-structures and $Spin(7)$-structures, simplified to the case of flat
space, that may in future developments enable an extension of these
considerations to more general compactifications of M-theory on
manifolds of $G_2$-holonomy. In contrast to the usual considerations
of calibrated geometry, however, for deformation quantization our
structure manifolds involve corresponding bivectors and trivectors, respectively, rather than the
more conventional three-forms and four-forms. All of the relevant
deformation quantities are underpinned by vector cross products, whose
theory we review in the following.

In fact, in this paper we emphasise a common underlying mathematical feature of the star products which quantise non-geometric string
theory and M-theory backgrounds: They all originate, via the
Baker-Campbell-Hausdorff formula, 
from the theory of cross products on real vector spaces; non-trivial cross products only
exist in dimensions three (where they are associative) and seven
(where they are nonassociative). In the three-dimensional case,
relevant for the quantisation of the string theory $R$-flux background,
the vector cross product determines a 3-cocycle among Fourier momenta
that appears as a phase factor in the associator for the star product, whereas in the seven-dimensional
case, relevant for the quantisation of the M-theory $R$-flux
background, the vector cross product determines a nonassociative
deformation of the sum of Fourier momenta. In the generalisation to
the full eight-dimensional M-theory phase space, wherein the physical
seven-dimensional $R$-flux background arises as a certain gauge
constraint, triple cross products determine an underlying 3-algebraic
structure akin to those previously found in studies of multiple
M2-branes (see e.g.~\cite{Bagger2012} for a review). Higher
associativity of the 3-bracket is governed by a 5-bracket, but it is
not related in any simple way to a 5-vector. This parallels the
situation with 
the lift of non-geometric string theory fluxes: Unlike the NS--NS
$R$-flux, the M-theory $R$-flux is not a multivector. This point of
view should prove helpful in understanding generalisations of these
considerations to both higher dimensions and to the treatment of
missing momentum modes in M-theory backgrounds dual to non-toroidal
string vacua.

Armed with the phase space star product, we can use it to describe
various physical and geometrical features of the membrane phase
space. In particular, we derive quantum uncertainty relations which
explicitly exhibit novel minimal area cells in the M-theory phase
space, as well as minimal volumes demonstrating a coarse-graining of
both configuration space and phase space itself, in contrast to the
string theory case~\cite{Mylonas2013}. We also derive configuration space triproducts, in
the spirit of~\cite{Aschieri2015}, which quantize the four-dimensional
3-Lie algebra
$A_4$ and suggest an interpretation of the quantum geometry of the
M-theory $R$-flux background as a foliation by fuzzy membrane
worldvolume three-spheres; in the contraction limit $g_s\to0$, these
triproducts consistently reduce to those of the string theory
configuration space which quantize the three-dimensional
Nambu-Heisenberg 3-Lie algebra~\cite{Blumenhagen2011,Aschieri2015}. In
contrast to the string theory case, this curving of the configuration
space by three-spheres also results in a novel associative but
noncommutative deformation of the geometry of momentum space
itself. The origin of these configuration space triproducts in the
present case is most naturally understood in terms of quantisation of
the 3-algebraic
structure of the eight-dimensional membrane phase space: The $G_2$-structure, which determines the star product quantising the
seven-dimensional phase space, extends to a $Spin(7)$-structure
determining phase space triproducts that restrict to those on the
four-dimensional configuration space.

The organisation of the remainder of this paper is as follows. In
Section~\ref{sec:NAdef} we briefly review relevant aspects of the parabolic non-geometric string theory $R$-flux
model on a three-torus with constant fluxes and its deformation quantization; in particular, we point
out that star product algebras of functions generally spoil
the classical Malcev-Poisson algebraic structure that sometimes appears in
discussions of nonassociativity in physics, see
e.g.~\cite{Gunaydin2013,Bojowald2014}. As preparation for the M-theory lift of this
model, in Section~\ref{sec:star} we review pertinent properties of the algebra
of octonions and the associated linear algebra of vector cross
products, and use them to derive a deformed summation operation on
Fourier momenta that defines the pertinent star products. This
technical formalism is then applied in Section~\ref{sec:Mtheory} to
derive a star product quantising the phase space quasi-Poisson brackets
proposed by~\cite{GLM}, whose derivation we also review; we demonstrate in detail that it reduces
appropriately to that of Section~\ref{sec:NAdef} in the contraction
limit that sends M-theory to IIA string theory, and further apply it
to derive quantum uncertainty relations, as well as the nonassociative
geometry of configuration space induced by the M-theory
$R$-flux and the radius of the M-theory circle. Finally, after briefly reviewing how 3-algebra structures
have arisen in other contexts in M-theory as motivation, in
Section~\ref{sec:Mtheory3alg} we extend the vector cross products to
triple cross products and use them to postulate a novel 3-algebraic
structure of the full eight-dimensional membrane phase space which
reproduces the quasi-Poisson brackets of~\cite{GLM} upon imposing a
suitable gauge fixing constraint; we describe a partial quantisation
of this 3-algebra and show how it naturally encompasses both the phase
space and the configuration
space nonassociative geometry from Section~\ref{sec:Mtheory}. 

\newsection{Quantization of string theory $R$-flux background\label{sec:NAdef}}

In this section we review and elaborate on features of the quantization of the parabolic phase space model for the constant string $R$-flux background in three dimensions.

\subsection{Quasi-Poisson algebra for non-geometric string theory fluxes\label{sec:Rbrackets}}

In our situations of interest, a non-geometric string theory
$R$-flux background originates as a double T-duality transformation of
a supergravity background $M$ of dimension $D$ with geometric flux; this sends closed string winding number
into momentum. The non-trivial windings, and hence the momentum modes
in the $R$-flux background, are classified by the first homology group
$H_1(M,\zed)$. The $R$-flux is represented by a trivector or locally
by a totally antisymmetric rank three tensor $R^{ijk}$ in the
framework of double field theory: It is given by taking suitable ``covariant''
derivatives $\hat\partial^i$ along the string winding directions of a globally well-defined
bivector $\beta^{jk}$, which is related to an $O(D)\times O(D)$
rotation of the generalised vielbein containing the background metric and two-form $B$-field after T-duality.

In the parabolic flux model in three dimensions, the background $M$ is
a twisted three-torus which is a circle bundle over the two-torus
$T^2$ whose degree $d\in\zed$ coincides with the cohomology class of the
three-form $H$-flux $H=\dd\,{\rm vol}_{T^3}$ in the original T-duality
frame (consisting of a three-torus $T^3$). Since the first homology
group is $H_1(M,\zed)=\zed\oplus\zed\oplus\zed_d$, there are
non-trivial windings along all three directions of $M$ and the
non-torsion winding numbers map to momentum modes of the $R$-flux
model. In this way, the position and momentum coordinates $\mbf
x=(x^i)$ and $\mbf p=(p_i)$ of closed strings propagating in the
background of a constant $R$-flux $R^{ijk}=R\, \varepsilon^{ijk}$
(with $R=d$) define a quasi-Poisson structure on phase space $T^*M$ with the classical brackets~\cite{Lust2010}
\begin{equation}\label{rb}
\{x^i, x^j\}=\mbox{$\frac{\ell_s^3}{\hbar^2}$} \, R^{ijk} \, p_k \ , \qquad \{x^i,p_j\}=\delta^i_j \qquad \mbox{and} \qquad \{p_i,p_j\}=0 \ ,
\end{equation}
where $\ell_s$ is the string length and $\varepsilon^{ijk}$, $i,j,k=1,2,3$, is the alternating symbol in
three dimensions normalised as $\varepsilon^{123}=+1$; unless otherwise explicitly stated, in the following repeated indices are always understood to be summed over. It is convenient to rewrite (\ref{rb}) in a more condensed form as
\begin{eqnarray}\label{rb1}
\{x^I, x^J\}=\Theta^{IJ}(x)=\begin{pmatrix}
  \frac{\ell_s^3}{\hbar^2}\, R^{ijk}\, p_k &  -\delta^i_j   \\
  \delta^i_j  & 0 \end{pmatrix}
\qquad \mbox{with} \quad x=(x^I) =({\mbf x},{\mbf p}) \ ,
\end{eqnarray}
which identifies the components of a bivector $\Theta=\frac12\, \Theta^{IJ}(x)\,
\frac\partial{\partial x^I}\wedge\frac\partial{\partial x^J}$. 
Strictly
speaking, here the coordinates $\mbf x$ live on a three-torus $T^3$,
but as we are only interested in local considerations we take the
decompactification limit and consider $\mbf x\in\real^3$ throughout
this paper. From the perspective of double field theory, in this frame
the dual phase space coordinates $\tilde{\mbf x},\tilde{\mbf p}$
have canonical Poisson brackets among themselves and vanishing brackets with $\mbf
x,\mbf p$, and the totality of brackets among the double phase space
coordinates $(\mbf x,\mbf p,\tilde{\mbf x},\tilde{\mbf p})$ can be
rotated to any other T-duality frame via an $O(3,3)$ transformation~\cite{Blumenhagen2011,Blumenhagen2013};
the same is true of the star product reviewed below~\cite{BaLu}. For ease of
notation, in this paper we restrict our attention to the $R$-flux
frame and suppress the dependence on the dual coordinates $(\tilde{\mbf x},\tilde{\mbf p})$.

For any three functions $f$, $g$ and $h$ on phase space, the classical Jacobiator is defined as
\begin{equation}
\label{cjac}
\{f,g,h\}:=\{f,\{g,h\}\}-\{\{f,g\},h\}-\{g,\{f,h\}\} \ .
\end{equation}
By construction it is antisymmetric in all arguments, trilinear and satisfies the Leibniz rule. For the brackets (\ref{rb}) one finds
\begin{equation}\label{m4}
\{f,g,h\}=\mbox{$\frac{3\, \ell_s^3}{\hbar^2}$}\, R^{ijk}\, \partial_i f\, \partial_j g\, \partial_k h \ ,
\end{equation}
where $\partial_i=\frac{\partial}{\partial x^i}$, which further obeys
the fundamental identity of a 3-Lie algebra; in fact, when restricted
to functions on configuration space, it defines the standard Nambu-Poisson bracket on $\real^3$. Hence the classical brackets of the constant $R$-flux background generate a nonassociative phase space algebra.

\subsection{Phase space star product\label{sec:Rstar}}

To describe the quantization of closed strings in the $R$-flux
background and their dynamics, as well as the ensuing nonassociative
geometry of the non-geometric background, we define a star product by associating a (formal) differential operator $\hat f$ to a function $f$ as
\begin{equation}
(f\star g)(x)=\hat f \triangleright g(x)~,
\label{d3}
\end{equation}%
where the symbol $\triangleright$ denotes the action of a differential operator on
a function. In particular one has
\begin{equation}\label{op1}
x^I\star f=\hat x^I\triangleright f(x)~.
\end{equation}%
The operators
\begin{equation}\label{op2}
\hat x^I=x^I+\mbox{$\frac{\ii\hbar}2$}\, \Theta^{IJ}(x)\, \partial_J\ ,
\end{equation}
with $\partial_i=\frac{\partial}{\partial x^i}$ and $\partial_{i+3}=\frac{\partial}{\partial p_i}$ for $i=1,2,3$, close to an associative algebra of differential operators; in particular
\bea
[\hat x^i,\hat x^j]=\mbox{$\frac{\ii\ell_s^3}\hbar$}\, R^{ijk}\,
\big(\hat p_k+\ii\hbar\, \partial_k \big) \ .
\label{eq:stringhatxcomm}\eea
Taking (\ref{op1}) and (\ref{op2}) as a definition of the star product, one may easily calculate the quantum brackets
\bea
[x^I,x^J]_\star:= x^I\star x^J-x^J\star x^I=\ii\hbar\, \Theta^{IJ}
\qquad \mbox{and} \qquad [x^i,x^j,x^k]_\star =-3\, \ell_s^3\,R^{ijk} \ ,
\label{eq:Rbrackets}\eea
which thereby provide a quantization of the classical brackets \eqref{rb}; in particular, the quantum 3-bracket represents a Nambu-Heisenberg algebra which quantizes the standard classical Nambu-Poisson bracket \eqref{m4} on $\real^3$.

To define the star product $f\star g$ between two arbitrary functions on phase space, we introduce the notion of \emph{Weyl star product} by requiring that, for any $f$, the differential operator $\hat f$ defined by (\ref{d3})
can be obtained by symmetric ordering of the operators $\hat x^I$. Let
$\tilde f(k)$ denote the Fourier transform of $f(x)$, with $k=(k_I) =({\mbf k},{\mbf l})$ and ${\mbf k}=(k_i),{\mbf l}=(l^i) \in \real^3$. Then
\begin{equation}
\hat f=W( f)  :=\int\, \frac{\dd^{6}k}{( 2\pi ) ^{6}} \ 
\tilde{f}(k) \, \e^{-\ii k_{I}\, \hat{x}^{I}} \ .  \label{2}
\end{equation}
For example, $W(x^I\, x^J)=\tfrac 12\, (\hat x^I\,\hat x^J + \hat x^J\, \hat x^I)$. 
Weyl star products satisfy
   \begin{equation}\label{weyl}
    \big(x^{I_1}\cdots x^{I_n}\big)\star f=\frac 1{n!} \, \sum_{\sigma\in S_n}\, x^{I_{\sigma(1)}}\star\big(x^{I_{\sigma(2)}}\star \cdots \star (x^{I_{\sigma(n)}}\star f)\cdots\big)\ ,
\end{equation}
where the sum runs over all permutations in the symmetric group $S_n$ of degree $n$. It should be stressed that the correspondence $f\mapsto \hat f$ is not an algebra representation: Since the star product that we
consider here is not necessarily associative, in general $\widehat{ f\star g} \ne \hat f \circ \hat g$.

To obtain an explicit form for the corresponding star product we first observe that since $[k_{I}\, {x}^{I}, k_{J}\, \Theta^{JL}\, \partial_L]=0$ one can write
\begin{equation*}
\e^{-\ii k_{I}\, \hat{x}^{I}} =\e^{-\ii k\,\mbf\cdot\, {x}}\,
\e^{\frac\hbar2\, k_{I}\, \Theta^{IJ}(x) \, \partial_J} \ ,
\end{equation*}
with $\,\mbf\cdot\,$ the standard Euclidean inner product of vectors. By the relation $k_{I}\, k_{L}\, \Theta^{IJ}\, \partial_J
\Theta^{LM}\, \partial_M=\frac{\ell_s^3}{\hbar^2}\, k_i\, k_l\, R^{lki}\, \partial_k=0$ it follows that
\begin{equation*}
\big(k_{I}\,\Theta^{IJ}\,\partial_J\big)^n=k_{I_1}\cdots k_{I_n}\, \Theta^{I_1J_1}\cdots\Theta^{I_nJ_n}\, \partial_{J_1}\cdots\partial_{J_n} \ .
\end{equation*}
One may also write
\begin{equation*}
\big(\overleftarrow{\partial}_{I}\, \Theta^{IJ}\, \overrightarrow{\partial}_J\big)^n=\overleftarrow{\partial}_{I_1}\cdots \overleftarrow{\partial}_{I_n}\, \Theta^{I_1J_1}\cdots\Theta^{I_nJ_n}\, \overrightarrow{\partial}_{J_1}\cdots\overrightarrow{\partial}_{J_n} \ ,
\end{equation*}
where $\overleftarrow{\partial}_I$ and $\overrightarrow{\partial}_I$
stand for the action of the derivative $\frac\partial{\partial x^I}$
on the left and on the right correspondingly. Thus the Weyl star
product representing quantization of the quasi-Poisson bracket
(\ref{rb}) can be written in terms of a bidifferential operator as
\begin{equation}
(f\star g)(x) =\int\, \frac{\dd^{6}k}{( 2\pi) ^{6}} \ 
\tilde{f}(k) \, \e^{-\ii k_{I}\, \hat{x}^{I}}\triangleright
g(x)=f(x)\, \e^{\frac{\ii\hbar}{2}\, \overleftarrow{\partial}_{I}\,
  \Theta^{IJ}(x) \, \overrightarrow{\partial}_J} \, g(x) \ . \label{starrb}
\end{equation}
It is easy to see that (\ref{starrb}) is Hermitean, $(f\star
g)^\ast=g^\ast\star f^\ast$, and unital, $f\star1=f=1\star f$; it is moreover 2-cyclic and 3-cyclic under integration in the sense of~\cite{Mylonas2013}. This star product first appeared in
\cite{MSS2} where it was derived using the Kontsevich formula for
deformation quantization of twisted Poisson structures. Its realisation through an associative algebra of differential operators was first pointed out in~\cite{Mylonas2013}. The significance and utility of this star product in understanding non-geometric string theory is exemplified in~\cite{MSS2,BaLu,Mylonas2013,Aschieri2015}.

For later use, let us rewrite the star product $f\star g$ in integral
form through the Fourier transforms $\tilde{f}$ and $\tilde{g}$
alone. The star product of plane waves is given by
\begin{equation}\label{expR}
\e^{\ii k\,\mbf\cdot\, x}\star \e^{\ii k'\,\mbf\cdot\, x}=\e^{\ii\mathcal{ B}( k,k'\, )\,\mbf\cdot\, x}\ ,
\end{equation}
where
\begin{equation}\label{BR}
\mathcal{ B}( k, k'\,)\,\mbf\cdot\, x :=(\mbf k+\mbf k'\, )\,\mbf\cdot\, \mbf x+ (\mbf
l+\mbf l'\,)\,\mbf\cdot\, \mbf p-\mbox{$\frac{\ell_s^3}{2\hbar}$}\, R\, {\mbf
  p}\,\mbf\cdot\,({\mbf k}\,\mbf\times_\varepsilon \,{\mbf
  k}'\, )+\mbox{$\frac{\hbar}{2}$}\,\big({\mbf l}\,\mbf\cdot\, {\mbf k}'
-{\mbf k}\,\mbf\cdot\, {\mbf l}'\, \big) \ ,
\end{equation}
with
$(\mbf k\,\mbf\times_\varepsilon \,\mbf k'\,)_i=\varepsilon_{ijl}\, k_j\, k'_l$ the usual cross product of three-dimensional vectors.
Then
\begin{equation}\label{starrb1}
(f\star g)(x) =\int\, \frac{\dd^{6}k}{( 2\pi
) ^{6}}\ \frac{\dd^{6}k'}{( 2\pi
) ^{6}}\ \tilde{f}(k)\, \tilde{g}( k'\, )\, \e^{\ii\mathcal{ B}(k,k'\,
)\,\mbf\cdot\, x} \ .
\end{equation}
In this form, the star product follows from application of the Baker-Campbell-Hausdorff formula to the brackets \eqref{eq:Rbrackets}~\cite{MSS2,BaLu}, whereas the nonassociativity of the star product is encoded in the additive associator
\bea\label{eq:addassRfluxstring}
\alg(k,k',k^{\prime\prime}\,):=\big(\mathcal{ B}(\mathcal{ B}( k,
k'\,), k^{\prime\prime} \,)) -\mathcal{ B}( k, \mathcal{ B}( k',
k^{\prime\prime}\,))\big) \,\mbf\cdot\, x = \mbox{$\frac{\ell_s^3}2$}
\, R\, \mbf k\,\mbf\cdot\,(\mbf k'\,\mbf\times_\varepsilon \,\mbf k^{\prime\prime}\, ) \ ,
\eea
which is antisymmetric in all arguments, and in fact defines a certain 3-cocycle~\cite{MSS2,BaLu,Mylonas2013}.

\subsection{Alternativity and Malcev-Poisson identity\label{subsec:Ralternative}}

A star product is \emph{alternative} if the star associator of three functions 
$$
A_\star(f,g,h):=f\star(g\star h)-(f\star g)\star h
$$
vanishes whenever any two of them are equal (equivalently
$A_\star(f,g,h)$ is completely antisymmetric in its arguments, or
`alternating'). For such products the Jacobiator is proportional to
the associator. Since the function \eqref{eq:addassRfluxstring} vanishes whenever any
two of its arguments are equal, it follows that the star product
(\ref{starrb1}) restricted to Schwartz functions is
alternative. However, for generic smooth functions on phase space this
property is violated; in fact, the simple example $A_\star\big(|\mbf x|^2,|\mbf
x|^2,|\mbf x|^2 \big)= 2\ii\frac{\ell_s^6}{\hbar^4}\, R^2\, \mbf p\,\mbf\cdot\, \mbf
x$ shows that alternativity is even violated on the phase space
coordinate algebra $\complex[\mbf x,\mbf p]$.\footnote{To be more
  precise, alternativity generally holds only for functions with function-valued (rather than distribution-valued) Fourier transform, such as Lebesgue integrable functions. For
  some accidental cases, such as linear coordinate functions or powers
of single coordinate generators, the star product is trivially
alternative. The star products considered in this paper
are best behaved on algebras of Schwartz functions, so we will
often make this restriction. It is interesting to understand more
precisely the class of functions on which the star product is
alternative, but this is not relevant for the present paper;
see~\cite{Bojowald2016} for a systematic and general analysis of the
non-alternativity of a class of star products containing ours.}

Another way of understanding this violation, which will be relevant in later sections, is via the observation of~\cite{Kup24} that a necessary condition for the star product $f\star g$ to be alternative is that the corresponding classical bracket $\{f,g\}$ satisfies the Malcev-Poisson identity~\cite{Ivan}.
For any three functions $f$, $g$ and $h$ the Malcev-Poisson identity can be written as
\begin{equation}\label{m3}
  \{f,g,\{f,h\}\}=\{\{f,g,h\},f\} \ .
\end{equation}
As a simple example, let us check both sides of (\ref{m3}) for the three functions
$f=x^1$, $g=x^3\, p_1$ and $h=x^2$. 
Since $\{f,h\}=\{x^1,x^2\}=\frac{\ell_s^3}{\hbar^2}\, R \, p_3$
does not depend on $\mbf x$, the left-hand side of (\ref{m3}) vanishes by (\ref{m4}). On the other hand, one has $\{f,g,h\}=-\frac{3\ell_s^3}{\hbar^2}\, R\, p_1 $
and consequently for the right-hand side of (\ref{m3}) one finds
$\{\{f,g,h\},f\}=-\frac{3\ell_s^3}{\hbar^2}\, R\,
\{p_1,x^1\}=\frac{3\ell_s^3}{\hbar^2}\, R$. It follows that the
classical string $R$-flux coordinate algebra $\complex[\mbf x,\mbf p]$
is \emph{not} a Malcev algebra (beyond linear order in the phase space coordinates), and consequently no star product representing a quantization of (\ref{rb}) can be alternative.\footnote{The counterexamples
  always involve momentum dependence, and so alternativity can be
  restored in a suitable sense by restricting to configuration space
  $\mbf p=\mbf 0$. We return to this point in Section~\ref{sec:TriNAG}.}

\newsection{$G_2$-structures and deformation quantization\label{sec:star}}

In this paper we are interested in the lift of the string theory
phase space model of Section~\ref{sec:NAdef} to M-theory. As the conjectural
quasi-Poisson structure from~\cite{GLM}, which we review in Section~\ref{sec:Mtheory}, is intimately related to the nonassociative
algebra of octonions, in this section we shall take a technical
detour, recalling some of the algebraic and geometric features of
octonions, together with their related linear algebra, in the form that we need in this paper. In particular
we will derive, following~\cite{OR13,Kup24}, a 
star product decribing quantization of the coordinate algebra based on the
imaginary octonions, elucidating various aspects which will be
important for later sections and which are interesting in their own right. The reader uninterested in these technical
details may temporarily skip ahead to
Section~\ref{sec:Mtheory}.

\subsection{Octonions}

The algebra $\oct$ of octonions is the best known example of a nonassociative but alternative algebra. Every octonion $X\in \mathbb{O}$ can be written in the form
\begin{equation}\label{oct}
X=k^0\, \unit+ k^A\,e_A
\end{equation}
where $k^0,k^A\in\real$, $A=1,\dots,7$, while $\unit$ is the identity element and the imaginary unit octonions $e_A$ satisfy the multiplication law
\begin{equation}\label{oct1}
e_A\, e_B=-\delta_{AB}\, \unit +\eta_{ABC}\, e_C \ .
\end{equation}
Here $\eta_{ABC}$ is a completely antisymmetric tensor of rank three
with nonvanishing values
\bea
\eta_{ABC}=+1 \qquad \mbox{for} \quad ABC = 123 , \ 435, \ 471, \ 516,
\ 572, \ 624, \ 673 \ .
\label{eq:etadef}\eea
Introducing $f_i:=e_{i+3}$ for $i=1,2,3$, the algebra (\ref{oct1}) can be rewritten as
\begin{eqnarray}\label{oct1a}
e_i\, e_j&=&-\delta_{ij}\, \unit +\varepsilon_{ijk}\, e_k\ ,\\[4pt]
e_i\, f_j&=&\delta_{ij}\, e_7-\varepsilon_{ijk}\, f_k\ ,\nonumber\\[4pt]
f_i\, f_j&=&\delta_{ij}\, \unit -\varepsilon_{ijk}\, e_k\ ,\nonumber\\[4pt]
e_7\, e_i&=&f_i \qquad \mbox{and} \qquad f_i\, e_7 \ = \ e_i\ , \nonumber
\end{eqnarray}
which emphasises a subalgebra $\quat$ of quaternions generated by
$e_i$; we will use this component form of the algebra $\oct$ frequently in what follows.

The algebra $\oct$ is neither commutative nor associative. The commutator algebra of the octonions is given by
\begin{equation}\label{oct2}
[e_A,e_B]:=e_A\, e_B-e_B\, e_A=2\, \eta_{ABC}\, e_C\ ,
\end{equation}
which can be written in components as
\begin{eqnarray}\label{oct2a}
[e_i,e_j]&=&2\, \varepsilon_{ijk}\, e_k \qquad \mbox{and} \qquad [e_7,e_i] \ = \ 2\,
             f_i\ ,\\[4pt]
[f_i,f_j]&=&-2\, \varepsilon_{ijk}\, e_k \qquad \mbox{and} \qquad [e_7,f_i] \ = \
             -2\, e_i\ ,\nonumber\\[4pt]
[e_i,f_j]&=&2\, (\delta_{ij}\, e_7- \varepsilon_{ijk}\, f_k) \ .\nonumber
\end{eqnarray}
The structure constants $\eta_{ABC}$ satisfy the contraction identity
\begin{equation}\label{epsilon7}
\eta_{ABC}\, \eta_{DEC}=\delta_{AD}\, \delta_{BE}-\delta_{AE}\,
\delta_{BD}+\eta_{ABDE} \ ,
\end{equation}
where $\eta_{ABCD}$ is a completely antisymmetric tensor of rank four
with nonvanishing values
$$
\eta_{ABCD}= +1 \qquad \mbox{for} \quad ABCD = 1267, \ 1346, \ 1425, \
1537, \ 3247, \ 3256, \ 4567 \ .
$$
One may also represent the rank four tensor
$\eta_{ABCD}$ as the
dual of the rank three tensor $\eta_{ABC}$ through 
\begin{equation}\label{epsilon8}
\eta_{ABCD}=\mbox{$\frac{1}{6}$}\, \varepsilon_{ABCDEFG}\, \eta_{EFG}
\ ,
\end{equation}
where $\varepsilon_{ABCDEFG}$ is the alternating symbol in seven
dimensions normalized as $\varepsilon_{1234567}=+1$. Together they satisfy the contraction identity
\bea
\eta_{AEF}\, \eta_{ABCD} &=& \delta_{EB}\, \eta_{FCD}-\delta_{FB}\, \eta_{ECD} +\delta_{EC}\, \eta_{BFD}-\delta_{FC}\, \eta_{BED} \nonumber \\ && +\, \delta_{ED}\, \eta_{BCF}-\delta_{FD}\, \eta_{BCE} \ .
\label{eq:eta34}\eea
Taking into account (\ref{epsilon7}), for the Jacobiator we get
\begin{equation}\label{oct3}
[e_A,e_B,e_C]:=[e_A,[e_B,e_C]]+[e_C,[e_A,e_B]]+[e_B,[e_C,e_A]]=-12\,
\eta_{ABCD}\, e_D \ ,
\end{equation}
and the alternative property of the algebra $\oct$ implies that the
Jacobiator is proportional to the associator, i.e.,
$[X,Y,Z]=6\,\big((X\,Y)\, Z-X\,(Y\, Z) \big)$ for any three octonions
$X,Y,Z\in\oct$.

\subsection{Cross products\label{sec:crossproducts}}

An important related linear algebraic entity in this paper will be the notion of a
\emph{cross product} on a real inner product
space~\cite{HL82} (see~\cite{Joyce,Salamon2010} for nice introductions), generalising the well known cross
product of vectors in three dimensions. They are intimately related
to the four normed algebras over the field of real numbers $\real$
(namely $\real$, $\complex$, $\quat$ and $\oct$),
and likewise cross products only exist for vector spaces of real
dimensions $0$, $1$, $3$ and $7$. In dimensions $0$ and $1$ the cross
product vanishes, in three dimensions it is the standard one
$\,\mbf\times_\varepsilon \,$ (up to sign) which has appeared already in our discussion of the star
product for the string theory $R$-flux background, while in seven
dimensions it can be defined (uniquely up to orthogonal
transformation) in a Cayley basis for vectors $\vec
k=(k^A),\vec k'=(k^{\prime\,A}) \in \real^7$ by
\bea
(\vec k\,\mbf\times_\eta\,\vec k'\,)^A := \eta^{ABC}\, k^B\, k^{\prime\,C}
\label{eq:7dvectorprod}\eea
with the structure constants $\eta^{ABC}$ introduced in
(\ref{eq:etadef}). To help describe and interpret the underlying
geometry of the seven-dimensional cross
product, it is useful to note that it can be expressed in terms of the
algebra of imaginary octonions by writing $X_{\vec k}:=k^A\, e_A$ and observing that
\bea\label{eq:Xveckcomm}
X_{\vec k\,\mbf\times_\eta\,\vec k'}=\mbox{$\frac12$}\, \big[X_{\vec k},X_{\vec k'} \big] \ .
\eea

This bilinear product satisfies
the defining properties of cross products~\cite{Salamon2010}:
\begin{description}
\item[(C1)]  $\vec k\,\mbf\times_\eta\, \vec k'=-\vec k'\,\mbf\times_\eta\, \vec k$\, ;
\item[(C2)] $\vec k\,\mbf\cdot\,(\vec k'\,\mbf\times_\eta\, \vec k^{\prime\prime}\,)=-\vec k'\,\mbf\cdot\,(\vec k\,\mbf\times_\eta\, \vec k^{\prime\prime}\,)$\, ;
\item[(C3)]  $|\vec k\,\mbf\times_\eta\,\vec k'\,|^2 = |\vec k|^2\, |\vec
  k' \,|^2 - (\vec k\,\mbf\cdot\, \vec k'\,)^2$, where $|\vec k|=\sqrt{\vec k\,\mbf\cdot\,\vec k}$ is the Euclidean vector norm.
 \end{description}
As usual property {\bf{(C1)}} is equivalent to the statement that the
cross product $\vec k\,\mbf\times_\eta\, \vec k'$ is non-zero if and only if
$\vec k$, $\vec k'$ are linearly independent vectors, property
{\bf{(C2)}} is equivalent to the statement that it is orthogonal to
both $\vec k$ and $\vec k'$, while property {\bf{(C3)}} states that
its norm calculates the area of the triangle spanned by $\vec k$
and $\vec k'$ in $\real^7$. However, unlike the three-dimensional cross product $\,\mbf\times_\varepsilon\,$, due to \eqref{oct3} it does not obey the Jacobi identity: Using \eqref{epsilon7} the Jacobiator is given by
\bea
\vec{J}_\eta(\vec k,\vec k ',\vec k^{\prime\prime}\,)&:=& (\vec k \,\mbf\times_\eta\, \vec k'\,)\,\mbf\times_\eta\,\vec k^{\prime\prime}+(\vec k' \,\mbf\times_\eta\, \vec k^{\prime\prime}\,)\,\mbf\times_\eta\,\vec k+(\vec k \,\mbf\times_\eta\, \vec k^{\prime\prime}\,)\,\mbf\times_\eta\,\vec k' \nonumber\\[4pt]
&=& 3\,\big(\,(\vec k \,\mbf\times_\eta\, \vec k'\,)\,\mbf\times_\eta\,\vec k^{\prime\prime}+(\vec k'\,\mbf\cdot\,\vec k^{\prime\prime}\,)\ \vec k -(\vec k\,\mbf\cdot\,\vec k^{\prime\prime}\,)\ \vec k'\, \big) \ ,
\label{eq:Jaceta}\eea
which can be represented through the associator on the octonion algebra
$\oct$ as
\bea\nonumber
X_{\vec{J}_\eta(\vec k,\vec k ',\vec k^{\prime\prime}\,)} =
\mbox{$\frac14$}\, \big[X_{\vec k},X_{\vec k'},X_{\vec k''}\big] =
\mbox{$\frac32$}\, \big((X_{\vec k}\,X_{\vec k'})\,X_{\vec k''}-
X_{\vec k}\,(X_{\vec k'}\,X_{\vec k''} ) \big) \ .
\eea
Hence properties {\bf{(C1)}} and {\bf{(C2)}} imply that the products $(\,\mbf\times_\eta\,\, ,\,\,\mbf\cdot\,\,)$ make the vector space
$V=\real^7$ into a pre-Courant algebra~\cite{Vaisman2005}. Only
rotations in the $14$-dimensional exceptional group $G_2\subset SO(7)$ preserve the
cross product $\,\mbf\times_\eta\,$, where the action of $G_2$ can be
described as the transitive action on the unit sphere $S^6\subset V$
identified with the homogeneous space $S^6\simeq G_2/
SU(3)$.\footnote{This means that the Lie group $G_2$ is the stabilizer
of a unit vector in $V$.} A
\emph{$G_2$-structure} on an oriented seven-dimensional vector space
$V$ is the choice of a cross product that can be written as
\eqref{eq:7dvectorprod} in a suitable oriented frame.

\subsection{Baker-Campbell-Hausdorff formula and vector star sums\label{sec:crossvector}}

Let us now work out the Baker-Campbell-Hausdorff formula for $\oct$,
which will be crucial for the derivations which follow. The alternative property
$$
(X\,Y)\,Y=X\,(Y\,Y) \qquad \mbox{and} \qquad X\,(X\,Y)=(X\,X)\, Y \ ,
$$
for any pair of octonions $X,Y\in\oct$, implies in particular that $(X\,X)\,X=
X\,(X\, X)$. Hence the quantity $X^n:=X\,(X\,
(\cdots(X\, X)\cdots))$ is well-defined independently of the ordering of
parantheses for all $n\geq0$ (with
$X^0:=\unit$). This implies that power series in octonions are
readily defined~\cite{Ludkovsky2007,PBMO2008}, and in particular one can introduce the octonion exponential
function $\e^X:=\sum_{n\geq0}\, \frac1{n!}\, X^n$. By setting
$X=X_{\vec k} =k^A\,e_A$ with $\vec
k=(k^A)\in\real^7$ and using the multiplication
law \eqref{oct1}, one can derive an octonionic version of de~Moivre's theorem~\cite{Ludkovsky2007,PBMO2008}
\bea
\e^{X_{\vec k}}=\cos|\vec k|\ \unit+ \frac{\sin|\vec
  k|}{|\vec k|}\ X_{\vec k} \ .
\label{eq:octexp}\eea
We can take \eqref{eq:octexp} to \emph{define} the octonion exponential
function $\e^{X_{\vec k}}\in\oct$.

One can now repeat the derivation of~\cite[Appendix~C]{OR13}
to obtain a closed form for the Baker-Campbell-Hausdorff
formula for $\oct$. Multiplying two octonion exponentials of
the form \eqref{eq:octexp} together using \eqref{oct1} we get
\bea\label{eq:octexpprod}
\e^{X_{\vec k}}\, \e^{X_{\vec k'}} &=&
\Big(\cos|\vec k| \cos|\vec k'|-\frac{\sin|\vec k| \sin|\vec
  k'|}{|\vec k|\, |\vec k'|}\, \vec k\,\mbf\cdot\, \vec k'\Big)\
\unit \\ && + \, \frac{\cos|\vec k'|\sin|\vec k|}{|\vec k|}\
X_{\vec k}+\frac{\cos|\vec k|\sin|\vec k'|}{|\vec k'|}\ X_{\vec k'}
- \frac{\sin|\vec k|\sin|\vec k'|}{|\vec k|\,|\vec k'|}\ X_{\vec
k\,\mbf\times_\eta\, \vec k'} \ , \nonumber
\eea
where $\vec
k\,\mbf\times_\eta\, \vec k'$ is the seven-dimensional vector cross product
\eqref{eq:7dvectorprod} on $V$. On the other hand, the Baker-Campbell-Hausdorff expansion is defined by
\bea
\e^{X_{\vec k}}\, \e^{X_{\vec k'}}=: \e^{
  X_{\vec\CB^{\,\prime}_\eta(\vec k,\vec k'\,)}}= \cos\big|\vec\CB^{\,\prime}_\eta(\vec k,\vec
k'\,)\big|\ \unit + \frac{\sin\big|\vec\CB^{\,\prime}_\eta(\vec k,\vec
  k'\,)\big|}{\big|\vec\CB^{\,\prime}_\eta(\vec k,\vec k'\,)\big|}\ X_{\vec\CB^{\,\prime}_\eta(\vec k,\vec
k'\,)} \ .
\label{eq:BCHdefO}\eea
By comparing \eqref{eq:octexpprod} and \eqref{eq:BCHdefO} we arrive at
\bea\label{eq:BCHclosed}
\vec\CB^{\,\prime}_\eta(\vec k,\vec k'\,) &=& \frac{\cos^{-1}\big(\cos|\vec k| \cos|\vec k'|-\frac{\sin|\vec k| \sin|\vec
  k'|}{|\vec k|\, |\vec k'|}\, \vec k\,\mbf\cdot\, \vec k'\, \big)}{\sin \cos^{-1}\big(\cos|\vec k| \cos|\vec k'|-\frac{\sin|\vec k| \sin|\vec
  k'|}{|\vec k|\, |\vec k'|}\, \vec k\,\mbf\cdot\, \vec k'\, \big)} \\
&& \times \ \Big(\, \frac{\cos|\vec k'|\sin|\vec k|}{|\vec k|}\
\vec k+\frac{\cos|\vec k|\sin|\vec k'|}{|\vec k'|}\ \vec k'
- \frac{\sin|\vec k|\sin|\vec k'|}{|\vec k|\,|\vec k'|}\ \vec
k\,\mbf\times_\eta\, \vec k'\, \Big) \ . \nonumber
\eea

To rewrite \eqref{eq:BCHclosed} in a more manageable form, we use the
$G_2$-structure on $V$ to define a binary operation $\circledast_\eta$
on the unit ball
$B^7\subset V$ consisting of vectors $\vec p$ with $|\vec p\,|\leq1$. 
To any pair of vectors $\vec p,\vec p\,^{\prime}\in B^7$, it assigns the vector
\begin{equation}\label{vstar}
\vec p\circledast_\eta\vec p\,^{\prime}= \epsilon_{\vec p,\vec
  p\,'}\,\big(\, \sqrt{1-| \vec p\,^{\prime}|^2}\,\,\vec p+ {\sqrt{1-|
    \vec p\,|^2} }\,\,\vec p\,^{\prime}-\vec p\,\mbf\times_\eta\, \vec
p\,^{\prime}\, \big) \ ,
\end{equation}
where $\epsilon_{\vec p_1,\vec p_2}=\pm\,1$ is the sign of $\sqrt{1-|\vec p_1|^2}\, \sqrt{1-|\vec p_2|^2} -\vec p_1\,\mbf\cdot\, \vec p_2$ satisfying
\bea\label{eq:sumsign}
\epsilon_{\vec p_1,\vec p_2}\, \epsilon_{\vec p_1\circledast_\eta\vec
  p_2,\vec p_3} = \epsilon_{\vec p_1,\vec p_2\circledast_\eta\vec
  p_3}\, \epsilon_{\vec p_2,\vec p_3} \ ,
\eea
which follows by properties {\bf{(C1)}} and {\bf{(C2)}} of the cross
product from Section~\ref{sec:crossproducts}; these sign factors have
a precise intrinsic origin that we shall describe in
Section~\ref{sec:trisums}. 
Using the properties {\bf{(C1)}}--{\bf{(C3)}} from
Section~\ref{sec:crossproducts} we find
    \begin{equation}\label{eq:vstarid}
1-|\vec p\circledast_\eta \vec p\,^{\prime}\,|^2=\big(\, \sqrt{1-|\vec  p\, |^2}\, \sqrt{1-|\vec p\,^{\prime}\, |^2}-\vec p\,\mbf\cdot\,\vec p\,^{\prime}\, \big)^2\geq0 \ ,
\end{equation}
and so the vector $\vec p\circledast_\eta\vec p\,^{\prime}$ indeed also belongs to
the unit ball $B^7\subset V$. We call the binary operation
\eqref{vstar} on $B^7$ the \emph{vector star sum} of $\vec p$ and $\vec
p\,^\prime$. It admits an identity element given by the zero vector in $V$,
\begin{equation}\label{eq:vecstarsumid}
\vec p \circledast_\eta\vec 0= \vec p = \vec 0 \circledast_\eta \vec p
\ ,
\end{equation}
and the inverse of $\vec p\in B^7$ is $-\vec p\in B^7$,
\bea\nonumber
\vec p \circledast_\eta(-\vec p\,)=\vec 0= (-\vec p\,)\circledast_\eta
\vec p \ .
\eea
It is noncommutative with commutator given by the $G_2$-structure as
\begin{equation}\label{eq:vstarcomm}
\vec p\circledast_\eta\vec p\,^{\prime}- \vec p\,^{\prime}
\circledast_\eta\vec p = -2\, \vec p \,\mbf\times_\eta\, \vec
p\,^{\prime} \ ,
\end{equation}
and using \eqref{eq:sumsign} we find that the corresponding associator
is related to the Jacobiator \eqref{eq:Jaceta} for the
cross product \eqref{eq:7dvectorprod} through
\begin{eqnarray}\label{assvstar}
\vec A_\eta(\vec p,\vec p\,^{\prime},\vec p\,^{\prime\prime}\,):= (\vec p\circledast_\eta\vec p\,^{\prime}\,)\circledast_\eta\vec p\,^{\prime\prime}-\vec p\circledast_\eta(\vec p\,^{\prime}\circledast_\eta\vec p\,^{\prime\prime}\,) = \mbox{$\frac23$}\, \vec J_\eta(\vec p,\vec p\,^{\prime},\vec p\,^{\prime\prime}\,) \ .
\end{eqnarray}
It follows that the components of the
associator (\ref{assvstar}) take the form
\begin{equation*}
A_\eta(\vec p,\vec p\,^{\prime},\vec p\,^{\prime\prime}\, )^A= \eta^{ABCD}\, p^B\, p^{\prime\, C}\, p^{\prime\prime\, D} \ .
\end{equation*}
It is non-vanishing but totally antisymmetric, and hence the
seven-dimensional vector star sum (\ref{vstar}) is nonassociative but
alternative, making the ball $B^7\subset V$ into a 2-group.

To extend the 2-group structure (\ref{vstar}) over the entire vector space $V$ we introduce the map
\begin{equation}\label{eq:pkmap}
 \vec p=\frac{\sin(\hbar\, |\vec k|)}{|\vec k|}\ \vec k \qquad \mbox{with} \quad k^A\in\mathbb{R} \ .
\end{equation}
The inverse map is given by
\begin{equation}\nonumber
 \vec k=\frac{\sin^{-1}|\vec p\,|}{\hbar\, |\vec p\,|}\ \vec p \ .
\end{equation}
Then for each pair of vectors $\vec k,\vec k'\in V$, following~\cite[Appendix~C]{OR13} we can use the trigonometric identities
\bea
\sin \cos^{-1} s= \cos \sin^{-1} s = \sqrt{1-s^2} 
\label{eq:trigid}\eea
for $-1\leq s\leq1$ to find that the deformed vector sum \eqref{eq:BCHclosed} can be written in terms of the vector star sum as
\begin{equation}
\label{Bk}
\vec{\mathcal{ B}}_\eta(\vec k,\vec k'\,):= \mbox{$\frac1\hbar$}\, \vec{\mathcal{ B}}^{\,\prime}_\eta(\hbar\, \vec k,\hbar\, \vec k'\,)= \left.\frac{\sin^{-1}|\vec p\circledast_\eta\vec p\,'\,|}{\hbar\,|\vec p\circledast_\eta\vec p\,'\,|}\ \vec p\circledast_\eta\vec p\,'\, \right|_{ \vec p=\vec k\sin(\hbar\, |\vec k|)/|\vec k|} \ .
\end{equation}
From (\ref{Bk}) one immediately infers the following properties:
\begin{description}
\item[(B1)]  $\vec{\mathcal{ B}}_\eta(\vec k,\vec k'\,)=-\vec{\mathcal{ B}}_\eta(-\vec k',-\vec k\,)$\, ;
\item[(B2)]  $\vec{\mathcal{ B}}_\eta(\vec k,\vec 0\,) =\vec k=\vec{\mathcal{ B}}_\eta(\vec 0,\vec k\,)$\, ;
\item[(B3)]  Perturbative expansion: \ $\vec{\mathcal{ B}}_\eta(\vec k,\vec k'\,)=\vec k+\vec k'-{2\, \hbar}\, \vec k\,\mbf\times_\eta\,\vec k'+O(\hbar^2)$\, ;
\item[(B4)] The associator $$\vec{\mathcal{A}}_\eta(\vec k,\vec k',\vec k^{\prime\prime}\,):=\vec{\mathcal{ B}}_\eta\big(\vec{\mathcal{ B}}_\eta(\vec k,\vec k'\,)\,,\,\vec k^{\prime\prime}\,\big)-\vec{\mathcal{ B}}_\eta\big(\vec k\,,\,\vec{\mathcal{ B}}_\eta(\vec k',\vec k^{\prime\prime}\,)\big)$$ is antisymmetric in all arguments. \end{description}
One can explicitly compute the products $\big(\e^{X_{\vec k}}\, \e^{X_{\vec k'}}\,\big)\, \e^{X_{\vec k^{\prime\prime}}}$ and  $\e^{X_{\vec k}}\,\big( \e^{X_{\vec k'}}\, \e^{X_{\vec k^{\prime\prime}}}\, \big)$ of octonion exponentials using \eqref{eq:octexp} and \eqref{eq:octexpprod}, and after a little calculation using the identities \eqref{eq:vstarid} and \eqref{eq:trigid} one finds for the associator
\bea
\vec{\mathcal{A}}_\eta(\vec k,\vec k',\vec k^{\prime\prime}\,) =
\left. \frac{\sin^{-1}\big|(\vec p\circledast_\eta \vec p\,'\,)
    \circledast_\eta\vec p\,^{\prime\prime}\, \big|}{\hbar \, \big|(
    \vec p\circledast_\eta \vec p\,'\, ) \circledast_\eta\vec p\,^{\prime\prime}\,\big|} \ \vec A_\eta(\vec p,\vec p\,',\vec p\,^{\prime\prime}\,)\right|_{ \vec p=\vec k\sin(\hbar\, |\vec k|)/|\vec k|} \ .
\label{eq:assBk}\eea

\subsection{Quasi-Poisson algebra and Malcev-Poisson identity\label{sec:Malceveta}}

Consider the algebra of classical brackets on the coordinate algebra
$\complex[\vec\xi \ ]$ which is isomorphic to the algebra (\ref{oct2}),
\begin{equation}\label{oct4}
\{\xi_A,\xi_B\}_\eta=2\, \eta_{ABC}\, \xi_C \ ,
\end{equation}
where $\vec\xi=(\xi_A)$ with $\xi_A\in\real$, $A=1,\dots,7$. This bracket is bilinear,
antisymmetric and satisfies the Leibniz rule by definition. Introducing $\sigma^i:=\xi_{i+3}$ for
$i=1,2,3$ and $\sigma^4:= \xi_7$, one may rewrite (\ref{oct4}) in components as
\begin{eqnarray}\label{mp1}
\{\xi_i,\xi_j\}_\eta&=&2\, \varepsilon_{ijk}\, \xi_k \qquad \mbox{and} \qquad
                   \{\sigma^4,\xi_i\}_\eta \ = \ 2\, \sigma^i \ , \\[4pt]
\{\sigma^i,\sigma^j\}_\eta&=&-2\, \varepsilon^{ijk}\, \xi_k \qquad \mbox{and} \qquad
                     \{\sigma^4,\sigma^i\}_\eta \ = \ -2\, \xi_i \ , \nonumber\\[4pt]
\{\sigma^i,\xi_j\}_\eta&=&-2\,( \delta_j^i\, \sigma^4- \varepsilon^{i}{}_{jk}\,
                         \sigma^k) \ . \nonumber
\end{eqnarray}
Using \eqref{oct3} the non-vanishing Jacobiators can be written as
\begin{eqnarray}\label{mp2}
\{\xi_i,\xi_j,\sigma^k\}_\eta&=&-12\, (\varepsilon_{ij}{}^{k}\,
                                \sigma^4+ \delta^{k}_{j}\,
                          \sigma_i- \delta^{k}_{i}\, \sigma_j ) \ ,\\[4pt]
\{\xi_i,\sigma^j,\sigma^k\}_\eta&=&12\, ( \delta_{i}^{j}\, \xi_k-\delta_{i}^{k}\, \xi_j)
                           \ ,\nonumber\\[4pt]
\{\sigma^i,\sigma^j,\sigma^k\}_\eta&=&12\, \varepsilon^{ijk}\, \sigma^4 \ ,\nonumber\\[4pt]
\{\xi_i,\xi_j,\sigma^4\}_\eta&=&12\, \varepsilon_{ijk}\, \sigma^k \ ,\nonumber\\[4pt]
\{\xi_i,\sigma^j,\sigma^4\}_\eta&=&12\, \varepsilon_{i}{}^{jk}\, \xi_k \ ,\nonumber\\[4pt]
\{\sigma^i,\sigma^j,\sigma^4\}_\eta&=&-12\, \varepsilon^{ijk}\, \sigma^k \ . \nonumber
\end{eqnarray}

The Malcev-Poisson identity (\ref{m3}) is satisfied for
monomials. However, we can show in an analogous way as in
Section~\ref{subsec:Ralternative} that it is violated in general on
$\complex[\vec\xi \ ]$. For this, consider $f=\xi_1$, $g= \xi_3\, \sigma^1$ and
$h=\xi_2$. Using (\ref{mp1}) and (\ref{mp2}) one finds that the
left-hand side of (\ref{m3}) is given by $
\{f,g,\{f,h\}_\eta\}_\eta=-24\, \xi_3\,
\sigma^3,$ while for the right-hand side one has
$\{\{f,g,h\}_\eta,f\}_\eta=24\, (\xi_2\, \sigma^2-\xi_3\, \sigma^3).$ We conclude that the Malcev-Poisson identity for the classical brackets (\ref{oct4}) is violated. This is in contrast to the well-known fact that the pre-Lie algebra (\ref{oct2}) of imaginary octonions
defines a Malcev algebra, due to the identity \eqref{eq:eta34} and the
multiplication law (\ref{oct1}) on the finite-dimensional algebra
$\oct$. Hence the Malcev identity \eqref{m3} holds for octonions,
while it is violated in general for the quasi-Poisson structure
(\ref{oct4}) on the infinite-dimensional polynomial algebra
$\complex[\vec\xi \ ]$; this is also implied by the general results of~\cite{Bojowald2016}.

\subsection{$G_2$-symmetric star product\label{sec:G2star}}

Let us now work out the quantization of the classical brackets
\eqref{oct4}. Consider the quasi-Poisson bivector
\bea\label{eq:Thetaeta}
\Theta_\eta:=\eta_{ABC}\,\xi_C\,\psi^A\wedge\psi^B \qquad \mbox{with}
\quad \psi^A=\partial^A=\mbox{$\frac\partial{\partial\xi_A}$}
\eea
defining the
brackets \eqref{oct4}. It can be regarded as a pre-homological potential on $T^*\Pi V$ with coordinates $(\psi^A,\xi_B)$ and canonical Poisson bracket; then the corresponding derived brackets are $[[\xi_A,\xi_B]]_{\Theta_\eta}=\{\xi_A,\xi_B\}_\eta$ giving $T^*\Pi V$ the structure of a symplectic nearly Lie 2-algebra~\cite{Roytenberg2002}. We can extend this structure to the entire algebra of functions by defining a star product through
\begin{equation}\label{w1}
(f\star_\eta g)( \vec \xi\ ) =\int \, \frac{\dd^{7}\vec k}{( 2\pi
) ^{7}} \ \frac{\dd^{7}\vec k'}{( 2\pi
) ^{7}} \ \tilde{f}( \vec k\,)\, \tilde{g}(\vec k'\,)\, \e^{\ii\vec{\mathcal{ B}}_\eta(\vec k,\vec k'\,)\,\mbf\cdot\,{\vec \xi}} \ ,
\end{equation}
where again $\tilde{f}$ stands for the Fourier transform of the
function $f$ and $\vec{\mathcal{ B}}_\eta(\vec k,\vec k'\,)$ is the deformed vector sum \eqref{Bk}. By definition it is the Weyl star product.

Due to the properties {\bf{(B1)}} and {\bf{(B2)}} from Section~\ref{sec:crossvector} of the deformed vector addition $\vec{\mathcal{ B}}_\eta(\vec k,\vec k'\,)$, this star product is Hermitean, $(f\star_\eta g)^\ast=g^\ast \star_\eta f^\ast$,
and unital, $f\star_\eta 1=f=1\star_\eta f$. It can be regarded as
a quantization of the dual of the pre-Lie algebra \eqref{oct2}
underlying the octonion algebra $\oct$; in particular, by property
{\bf{(B3)}} it provides a quantization of the quasi-Poisson bracket (\ref{oct4}): Defining $[f,g]_{\star_\eta}=f\star_\eta g-g \star_\eta f$, we have
\begin{equation}\label{41}
       \lim_{\hbar\to0}\, \frac{[f,g]_{\star_\eta}}{\ii\hbar}=2\, \xi_A\, \eta_{ABC}\, \partial^Bf\, \partial^Cg = \{f,g\}_\eta \ .
\end{equation}
Property {\bf{(B4)}} implies that the star product (\ref{w1}) is alternative on monomials and Schwartz functions, but not generally because of the violation of the Malcev-Poisson identity discussed in Section~\ref{sec:Malceveta}.

Let us calculate $\xi_A\star_\eta f$ explicitly using (\ref{w1}). We have
\begin{eqnarray}
 \xi_A\star_\eta f =-\int\, \frac{\dd^{7}\vec k'}{( 2\pi
) ^{7}} \ \xi_D\, \frac{ \partial  \mathcal{ B}_\eta(\vec k,\vec k'\, )^D}{\partial k^A}\bigg|_{\vec k=\vec 0}\ \tilde{f}(\vec k'\,)\, \e^{\ii\vec{\mathcal{ B}}_\eta(\vec 0,\vec k'\, )\,\mbf\cdot\,{\vec  \xi}}\nonumber
\end{eqnarray}
and after some algebra one finds
\begin{eqnarray*}
\frac{ \partial \mathcal{ B}_\eta(\vec k,\vec k'\, )^D}{\partial k^A}\bigg|_{\vec k=\vec 0}= -\hbar \, \eta_{ADE}\, k^{\prime\, E}+\delta_{AD}\, \hbar\, |\vec k'\,|\cot(\hbar\, |\vec k'\,|)
+\frac{k^{\prime}_{A}\, k'_D}{|\vec k'\, |^2}\, \Big(\hbar\, |\vec k'\, |\cot(\hbar\, |\vec k'\,|)-1\Big) \ .\nonumber
\end{eqnarray*}
Taking into account property {\bf{(B2)}} from Section~\ref{sec:crossvector} and integrating over $\vec k'$, we arrive at
\begin{eqnarray}\label{poly}
 \xi_A\star_{\eta} f &=& \Big(\xi_A+\ii\hbar\, \eta_{ABC} \,
                         \xi_C\, \partial^B \\ && \qquad +\, 
  \big( \xi_A\, {\mbf\triangle_{\vec\xi}}- (\vec\xi\,\mbf\cdot\,
                                                  \nabla_{\vec\xi}\, ) \, \partial_A
                                                  \big)\,
                                                  \mbf\triangle_{\vec\xi}^{-1}\,
                                                  \big( \hbar\,
                                                  \mbf\triangle_{\vec\xi}^{1/2}
                                                  \, \coth(\hbar\,
                                                  \mbf\triangle_{\vec\xi}^{1/2}
                                                  \,)-1\big)\Big)\triangleright
                                                  f(\vec\xi\ ) \nonumber
\end{eqnarray}
where ${\mbf\triangle_{\vec\xi}}=\nabla_{\vec\xi}^2= \partial_A\, \partial^A$ is the flat space Laplacian in seven dimensions. In particular for the Jacobiator one finds
\begin{equation}\label{oct5}
[ \xi_A, \xi_B, \xi_C]_{\star_\eta} =12\, \hbar^2\, \eta_{ABCD} \, \xi_D \ ,
\end{equation}
which thereby provides a quantization of the classical 3-brackets
\eqref{mp2}.

\subsection{$SL(3)$-symmetric star product\label{sec:quaternion}}

Setting $e_A=0$ for $A=4,5,6,7$ (equivalently $f_i=e_7=0$) reduces the
nonassociative algebra of octonions $\oct$ to the associative algebra of quaternions $\quat$,
whose imaginary units $e_i$ generate the $\mfs\fru(2)$ Lie algebra
$[e_i,e_j]=2\,\varepsilon_{ijk}\, e_k$. To see this reduction at the
level of our vector products, consider the splitting of the
seven-dimensional vector space $V$ according to the components of
Section~\ref{sec:Malceveta} with $\vec k=(\mbf l, \mbf k,k_4)$, where
$\mbf l=(l^i),\mbf k=(k_i) \in\real^3$. With respect to this decomposition, by using (\ref{oct2a}) the seven-dimensional cross product can be written in terms of the three-dimensional cross product as
\begin{eqnarray}\label{oct12}
\vec k\,\mbf\times_\eta\,\vec k'=\big(\mbf l\,\mbf\times_\varepsilon\,\mbf l'-\mbf k\,\mbf\times_\varepsilon\,\mbf k' +
  k_4'\, \mbf k-k_4\,\mbf k' \,,\, \mbf k\,\mbf\times_\varepsilon\,\mbf l'-\mbf l\,\mbf\times_\varepsilon\,\mbf
  k'+k_4\, \mbf l'-k_4'\, \mbf l \,,\, \mbf l\,\mbf\cdot\,\mbf k'-\mbf k\,\mbf\cdot\,
  \mbf l'\,\big) \ .
\end{eqnarray}
The symmetry group $G_2$ preserving $\,\mbf\times_\eta\,$ contains a closed
$SL(3)$ subgroup acting on these components as $\mbf l\mapsto g\,\mbf
l$, $\mbf k\mapsto (g^{-1})^\top \, \mbf k$ and $k_4\mapsto
k_4$ for $g\in SL(3)$.
In particular, reduction to the three-dimensional subspace spanned by
$e_i$ gives
\begin{equation}\label{oct12b}
(\mbf l,\mbf 0,0)\,\mbf\times_\eta\,(\mbf l',\mbf 0,0) = (\mbf l\,\mbf\times_\varepsilon\, \mbf
l',\mbf 0,0) \ ,
\end{equation}
and so yields the expected three-dimensional cross product. This reduction is implemented in all of our previous formulas by simply replacing $\,\mbf\times_\eta\,$ with $\,\mbf\times_\varepsilon\,$ throughout. In particular, the corresponding Jacobiator $\mbf
J_\varepsilon$ from \eqref{eq:Jaceta} now vanishes by a well-known identity for
the cross product in three dimensions; as a consequence, the pair
$(\,\mbf\times_\varepsilon\,,\,\,\mbf\cdot\,\,)$ defines a Courant algebra structure on
the vector space $\real^3$~\cite{Vaisman2005}. The cross product $\,\mbf\times_\varepsilon\,$ is preserved by
the full rotation group $SO(3)\subset G_2$ in this case, acting transitively on the unit sphere $S^2\simeq SO(3)/SO(2)$ in $\real^3$.

Similarly, the reduction of the seven-dimensional vector star sum
(\ref{vstar}) on this three-dimensional subspace reproduces the
three-dimensional vector star sum $\circledast_\varepsilon$ from~\cite{KV15},
\begin{equation}\label{vstar3}
({\mbf q},{\mbf 0},0)\circledast_\eta ({\mbf q}',{\mbf 0},0)=({\mbf
  q}\circledast_\varepsilon {\mbf q}',{\mbf 0},0) \ ,
\end{equation}
which by \eqref{assvstar} is now associative; as a consequence, it
makes the unit ball $B^3\subset\real^3$ into a non-abelian group. The reduction of the deformed vector sum \eqref{Bk} reproduces the
three-dimensional vector sum ${\mbf{\mathcal{ B}}}_\varepsilon({\mbf l},{\mbf l}'\, )$ from~\cite{OR13,KV15},
\begin{equation}
\label{Bk3}
\vec{\mathcal{ B}}_\eta\big( ({\mbf l},{\mbf 0},0)\,,\,({\mbf l}',{\mbf
  0},0) \big)=\big({\mbf{\mathcal{ B}}}_\varepsilon({\mbf l},{\mbf l}'\, ),{\mbf
  0},0 \big) \ ,
\end{equation}
with vanishing associator \eqref{eq:assBk}. From (\ref{Bk3}) it follows that, for functions $f,g$ on this three-dimensional
subspace, the
corresponding star product $(f\star_\varepsilon g)(\mbf\xi,\mbf0,0)$
from (\ref{w1}) reproduces the associative star product of~\cite{OR13,KV15} for the quantization
of the dual of the Lie algebra $\mfs\fru(2)$. In the general case \eqref{oct12}, the
evident similarity with the terms in the vector sum \eqref{BR} will be
crucial for what follows. 

\newsection{Quantization of M-theory $R$-flux background\label{sec:Mtheory}}

In this section we use the constructions of Section~\ref{sec:star} to
derive a suitable star product which quantizes the four-dimensional
locally non-geometric M-theory background which is dual to a twisted
torus~\cite{GLM}. We demonstrate explicitly that it is the lift of the
star product which quantizes the string theory $R$-flux background of
Section~\ref{sec:NAdef}, by showing that it reduces to the star
product of Section~\ref{sec:Rstar} in the weak string coupling limit
which reduces M-theory to IIA string theory. We apply this
construction to the description of the quantum mechanics of M2-branes in the non-geometric
background, as well as of the noncommutative and nonassociative geometry these membranes
probe.

\subsection{Quasi-Poisson algebra for non-geometric M-theory fluxes\label{sec:QPAMtheory}}

Let us start by reviewing the derivation of the classical
quasi-Poisson algebra for the four-dimensional non-geometric M-theory
background from~\cite{GLM}, beginning again with some general considerations. 
String theory on a background $M$ is dual to M-theory on the total space of an oriented circle bundle
$$
\xymatrix{
S^1 \ \ar@{^{(}->}[r] & \ \widetilde{M} \ar[d]^{\, \pi} \\
 & \ M
}
$$
over $M$, where the radius $\lambda\in\real$ of the circle fibre translates into the string coupling constant $g_s$. T-duality transformations become U-duality transformations sending membrane wrapping numbers to momentum modes, which are classified by the second homology group $H_2(\widetilde{M},\zed)$. The homology groups of $\widetilde{M}$ are generally related to those of $M$ through the Gysin exact sequence
\bea
\cdots \ \longrightarrow \ H_k(\widetilde{M},\zed) \ \xrightarrow{ \
  \pi_*  \ } \ H_k(M,\zed) \ \xrightarrow{ \ \cap\, e \ } \ H_{k-2}(M,\zed) \ \xrightarrow{ \ \pi^! \ } \ H_{k-1}(\widetilde{M},\zed) \ \longrightarrow \ \cdots
\label{eq:Gysin}\eea
where $\pi_*$ and $\pi^!$ are the usual pushforward and Gysin pullback on homology, and $\cap\, e$ is the cap product with the Euler class $e\in H^2(M,\zed)$ of the fibration. For instance, in the case of a trivial fibration $\widetilde{M}=M\times S^1$, wherein $e=0$, the Gysin sequence collapses to a collection of short exact sequences, and in particular by the K\"unneth theorem there is a splitting $H_2(\widetilde{M},\zed)\simeq H_1(M,\zed)\oplus H_2(M,\zed)$. More generally, for 
an automorphism $g$ of $M$ we can define a twisted lift to an $M$-bundle $\widetilde{M}_g$ over $S^1$ whose total space is the quotient of $M\times\real$ by the $\zed$-action\footnote{The precise sort of automorphism should be specified by the intended application; for example, $g$ could be an automorphism preserving some background form field.}
$$
(\mbf x,t)\longmapsto \big(g^n(\mbf x),t+2\pi\,n\,\lambda \big) \ ,
$$
where $\mbf x\in M$, $t\in\real$ and $n\in\zed$.
The Gysin sequence \eqref{eq:Gysin} shows that $H_1(M,\zed)$ generally
classifies ``vertical'' wrapping modes around the $S^1$-fibre which are dual to momenta along $M$, whereas $H_2(M,\zed)$ classifies ``horizontal'' wrapping modes dual to momenta along the $S^1$-fibre. 

In the situations we are interested in from
Section~\ref{sec:Rbrackets}, the lift of the non-geometric string
theory $R$-flux can be described locally within the framework of
$SL(5)$ exceptional field theory as a quantity
$R^{\mu,\nu\rho\alpha\beta}$: It is derived by taking suitable
``covariant'' derivatives $\hat\partial^{\mu\nu}$ along the membrane
wrapping directions of a trivector
$\Omega^{\rho\alpha\beta}$, which is related to an $SO(5)$ rotation of
the generalised vielbein containing the background metric and
three-form $C$-field after U-duality. Its first index is a vector index while the remaining indices define a completely antisymmetric rank four tensor. Because of the possibly non-trivial Euler class $e\in H^2(M,\zed)$, it is proposed in~\cite{GLM} that, generally, the phase space $T^*\widetilde{M}$ of the locally non-geometric background in M-theory is constrained to a codimension one subspace defined by the momentum slice
\bea\label{eq:Rpconstraint}
R^{\mu,\nu\rho\alpha\beta} \, p_\mu=0 \ ,
\eea
reflecting the absence of momentum modes in the dual $R$-flux background.

This proposal was checked explicitly in~\cite{GLM} for the parabolic toroidal flux model in three dimensions from Section~\ref{sec:Rbrackets}, wherein $M$ is a twisted three-torus. In the M-theory lift to the four-manifold $\widetilde{M}=M\times S^1$, with local coordinates $(x^\mu)=(\mbf x,x^4)$ where $\mbf x\in M$ and $x^4\in S^1$, by Poincar\'e duality the second homology group is $H_2(M,\zed)= H^1(M,\zed)=\zed\oplus\zed$, which does not contain the requisite non-trivial two-cycle that would allow for non-trivial wrapping modes dual to momenta along the $x^4$-direction, i.e., $p_4=0$. As a consequence the phase space of the M-theory lift of the $R$-flux background is only seven-dimensional and lacks a momentum space direction. The only non-vanishing components of the M-theory $R$-flux in this case are $R^{4,\mu\nu\alpha\beta}=R\, \varepsilon^{\mu\nu\alpha\beta}$, $\mu,\nu,\dots=1,2,3,4$, where $\varepsilon^{\mu\nu\alpha\beta}$ is the alternating symbol in
four dimensions normalised as $\varepsilon^{1234}=+1$. Note that this reduction only occurs in the presence of non-trivial flux: When $d=0$ there is no torsion in the homology or cohomology of the torus $M=T^3$ and Poincar\'e-Hodge duality implies $H_2(M,\zed)= H_1(M,\zed)=H^1(M,\zed)= \zed\oplus \zed\oplus\zed$, so that all two-cycles are homologically non-trivial.

The main conjecture of~\cite{GLM} is that the classical brackets of
this seven-dimensional phase space are given by the quasi-Poisson
brackets of Section~\ref{sec:Malceveta} after a suitable choice of
affine structure on the vector space $\real^7$, i.e., a choice of
linear functions. For this, let us introduce the $7\times7$ matrix 
\begin{equation}\label{oct6}
{\mit\Lambda} = \big({\mit\Lambda}^{AB}\big)= \frac1{2\hbar} \,
\begin{pmatrix}
0 & {\sqrt{\lambda\, \ell_s^3\, R}}\ \unit_3 & 0 \\
0 & 0 & \sqrt{\lambda^3\, \ell_s^3\, R} \\
 -\lambda\, \hbar\ \unit_3 & 0 & 0
\end{pmatrix}
\end{equation}
with $\unit_3$ the $3\times3$ identity matrix. The matrix $\mit\Lambda$
is non-degenerate as long as all parameters are non-zero, but it is
not orthogonal. Using it we define new coordinates
\begin{equation}\label{oct8}
\vec x = \big(x^A\big) = \big(\mbf x,x^4,\mbf p\big) := {\mit\Lambda}\,
\vec\xi= \mbox{$\frac1{2\hbar}$}\, \big(\sqrt{\lambda\,
      \ell_s^{3}\, R}\ \mbf\sigma\,,\,
\sqrt{\lambda^3\, \ell_s^{3}\, R}\ \sigma^4\,,\, -\lambda\,\hbar\ \mbf\xi \big) \ .
\end{equation}
From the classical brackets (\ref{oct4}) one obtains the quasi-Poisson algebra
\begin{equation}\label{oct4a}
\{x^A,x^B\}_\lambda=2\,\lambda^{ABC} \, x^C \qquad \mbox{with} \quad \lambda^{ABC}:= {\mit\Lambda}^{AA^\prime}\,{\mit\Lambda}^{BB^\prime}\,\eta_{A^\prime B^\prime C^\prime}\, {\mit\Lambda}^{-1}_{C^\prime C} \ ,
\end{equation}
which can be written in components as
\begin{eqnarray}\label{oct9}
\{x^i,x^j\}_\lambda &=&\mbox{$\frac{\ell_s^3}{\hbar^2}$}\,
                        R^{4,ijk4}\, p_k \qquad \mbox{and} \qquad
                        \{x^4,x^i\}_\lambda \ = \ \mbox{$\frac{\lambda\, \ell_s^3}{\hbar^2}$}\, R^{4,1234}\, p^i \ , \\[4pt]
\{x^i,p_j\}_\lambda &=&\delta^i_j\,x^4+\lambda\,
                        \varepsilon^i{}_{jk}\, x^k \qquad \mbox{and}
                        \qquad \{x^4,p_i\}_\lambda \ = \ \lambda^2\,x_i \ , \nonumber\\[4pt]
\{p_i,p_j\}_\lambda &=&-\lambda\, \varepsilon_{ijk}\, p^k\ , \nonumber
\end{eqnarray}
where we recall that the M-theory radius $\lambda$ incorporates the string coupling constant $g_s$. The corresponding Jacobiators are
\bea\nonumber
\{x^A,x^B,x^C\}_\lambda = -12\, \lambda^{ABCD} \, x^D \qquad \mbox{with} \quad \lambda^{ABCD}:= {\mit\Lambda}^{AA^\prime}\,{\mit\Lambda}^{BB^\prime}\,{\mit\Lambda}^{CC^\prime}\,\eta_{A^\prime B^\prime C^\prime D^\prime}\, {\mit\Lambda}^{-1}_{D^\prime D} \ ,
\eea
with the components
\bea\label{oct9a}
\{x^i,x^j,x^k\}_\lambda &=& \mbox{$\frac{3\,\ell_s^3}{\hbar^2}$}\, R^{4,ijk4} \, x^4 \ , \\[4pt]
\{x^i,x^j,x^4\}_\lambda &=& -\mbox{$\frac{3\,\lambda^2\, \ell_s^3}{\hbar^2}$}\, R^{4,ijk4} \, x_k \ , \nonumber \\[4pt]
\{p_i,x^j,x^k\}_\lambda &=& \mbox{$\frac{3\,\lambda\,\ell_s^3}{\hbar^2}$} \, R^{4,1234}\, \big(\delta^j_i\, p^k-\delta^k_i\, p^j \big) \ , \nonumber \\[4pt]
\{p_i,x^j,x^4\}_\lambda &=& \mbox{$\frac{3\,\lambda^2\, \ell_s^3}{\hbar^2}$}\, R^{4,ijk4}\, p_k \ , \nonumber \\[4pt]
\{p_i,p_j,x^k\}_\lambda &=& -3\, \lambda^2\, \varepsilon_{ij}{}^{k}\, x^4-3\,\lambda\, \big(\delta_j^k\, x_i-\delta_i^k\, x_j \big) \ , \nonumber \\[4pt]
\{p_i,p_j,x^4\}_\lambda &=& 3\,\lambda^3\, \varepsilon_{ijk}\, x^k \ , \nonumber \\[4pt]
\{p_i,p_j,p_k\}_\lambda &=& 0 \ . \nonumber
\eea

The crucial observation of~\cite{GLM} is that in the contraction limit $\lambda=0$ which shrinks the M-theory circle to a point, i.e., the weak string coupling limit $g_s\to0$ which reduces M-theory to IIA string theory, the classical brackets \eqref{oct9} and \eqref{oct9a} of the M-theory $R$-flux background reduce to the quasi-Poisson structure \eqref{rb} and \eqref{m4} of the string theory $R$-flux background; in this limit the circle fibre coordinate $x^4$ is central in the algebra defined by \eqref{oct9} and so may be set to any non-zero constant value, which we conveniently take to be $x^4=1$. In the following we will extend this observation to the quantum level. As in Section~\ref{sec:Malceveta} the classical coordinate algebra here is not a Malcev algebra, which is another way of understanding the violation of the Malcev-Poisson identity from Section~\ref{subsec:Ralternative} in the contraction limit $\lambda=0$.

\subsection{Phase space star product\label{sec:Mtheorystar}}

We will now quantize the brackets \eqref{oct9}. For this, we use the
$G_2$-symmetric star product (\ref{w1}) to define a star product of functions
on the seven-dimensional M-theory phase space by the prescription
\bea\label{oct10}
(f\star_\lambda g)(\vec x\, )=(f_{\mit\Lambda}\star_\eta g_{\mit\Lambda})( \vec\xi\ )\big|_{\vec\xi={\mit\Lambda}^{-1}\,\vec x}
\eea
where $f_{\mit\Lambda}(\vec\xi\ ):= f({\mit\Lambda}\,\vec\xi\ )$. Using the deformed vector addition $\vec{\mathcal{ B}}_\eta(\vec k, \vec k'\,)$ from (\ref{Bk}), we can write \eqref{oct10} as
\begin{equation}\label{oct11}
(f\star_\lambda g)(\vec x\, )= \int\, \frac{\dd^{7}\vec k}{( 2\pi
) ^{7}} \ \frac{\dd^{7}\vec k'}{( 2\pi
) ^{7}} \ \tilde{f}( \vec k\, )\, \tilde{g}( \vec k'\, )\, \e^{\ii\vec{\mathcal{ B}}_\eta({\mit\Lambda}\,\vec k,{\mit\Lambda}\, \vec k'\, )\,\mbf\cdot\, {\mit\Lambda}^{-1}\,\vec x} \ .
\end{equation}
The star product \eqref{oct11} may also be written in terms of a
(formal) bidifferential operator as
\bea\nonumber
(f\star_\lambda g)(\vec x\,) = f(\vec x\,) \, \e^{\ii\vec
  x\,\mbf\cdot\,({\mit\Lambda}^{-1}\,\vec{\mathcal{B}}_\eta(-\ii
  {\mit\Lambda}\,\overleftarrow{\partial},-\ii{\mit\Lambda}\,
  \overrightarrow{\partial}\,)
  +\ii\overleftarrow{\partial}+\ii\overrightarrow{\partial})} \,
g(\vec x\, ) \ ,
\eea
which identifies it as a cochain twist deformation~\cite{Mylonas2013}.
For the same reasons as (\ref{w1}) the star product (\ref{oct11}) is
unital, Hermitean and Weyl, and it is alternative on monomials and
Schwartz functions.

To show that (\ref{oct11}) provides a quantization of the brackets
(\ref{oct9}), we calculate $x^A\star_\lambda f$ by making the change
of affine structure \eqref{oct8} in (\ref{poly}) to get
\begin{eqnarray}\nonumber
 x^A\star_{\lambda} f = \hat x^A\triangleright f
\label{i2}\end{eqnarray}
where
\begin{eqnarray}\label{poly1}
\hat x^A= x^A+\ii\hbar\, \lambda^{ABC}\, x^C\, \partial_B
+\hbar^2\, \big( x^A\, {\tilde{\mbf\triangle}_{\vec x}}- (\vec x \,\mbf\cdot\, \tilde\nabla_{\vec x}\, ) \, \tilde\partial^A\big)\, \chi\big(\hbar^2\,\tilde{\mbf\triangle}_{\vec x}\big) \ ,
\label{i3}\end{eqnarray}
with
\bea\label{der}
\tilde\nabla_{\vec x}=\big(\tilde\partial^A\big):= \big(\, \Lambda^{BA}\,
\mbox{$\frac\partial{\partial x^B}$}\,\big) =
\mbox{$\frac1{2\hbar}$}\, \big(\sqrt{\lambda\,
\ell_s^3\, R}\ \nabla_{\mbf x}\,,\, \sqrt{\lambda^3\,\ell_s^3\, R}\
\mbox{$\frac\partial{\partial x^4}$}\,,\, -\lambda\,\hbar\
\nabla_{\mbf p}\big)
\eea
and
\bea\label{lap}
\tilde{\mbf\triangle}_{\vec x}=\tilde\nabla_{\vec x}^2=
\mbox{$\frac{\lambda}{4\hbar^2}$}\, \big(\ell_s^3\, R\, \mbf\triangle_{\mbf
  x}+\lambda^2\, \ell_s^3\, R\, \mbox{$\frac{\partial^2}{\partial
    x_4^2}$}+\lambda\,\hbar^2\, \mbf\triangle_{\mbf p} \big) \ .
\eea
We have also introduced the (formal) differential operator
\bea\nonumber
\chi\big(\tilde{\mbf\triangle}_{\vec x}\big) := \tilde{\mbf\triangle}_{\vec x}^{-1}\,\big( \tilde{\mbf\triangle}_{\vec x}^{1/2} \, \coth\tilde{\mbf\triangle}_{\vec x}^{1/2} -1\big) \ .
\label{i4}\eea
We thus find for the algebra of star commutators and Jacobiators
\begin{equation}\label{oct41}
[x^A,x^B]_{\star_\lambda}=2\ii\hbar\, \lambda^{ABC}\, x^C 
 \qquad \mbox{and} \qquad [x^A,x^B,x^C]_{\star_\lambda} = 12\, \hbar^2\, \lambda^{ABCD} \, x^D \ .
\end{equation}
Written in components, these quantum brackets coincide with those of~\cite[eq.~(3.30)]{GLM}.\footnote{Our definition of the Jacobiator
  differs from that of~\cite{GLM} by a factor of $-3$. We have also
  corrected the expression for the 3-bracket
  $[x^i,x^j,x^4]_{\star_\lambda}$ which is missing a factor $\lambda^2$
in~\cite[eq.~(3.30)]{GLM}.}

We will now show that this quantization is the correct M-theory lift of the
quantization of the string theory $R$-flux background from
Section~\ref{sec:Rstar}, in the sense that the star product
\eqref{oct11} reduces to \eqref{starrb1} in the contraction limit
$\lambda=0$; this calculation will also unpackage the formula
\eqref{oct11} somewhat. For this, we need to show that 
the quantity $\vec{\mathcal{ B}}_\eta({\mit\Lambda}\,\vec
k,{\mit\Lambda}\, \vec k'\, )\,\mbf\cdot\, {\mit\Lambda}^{-1}\,\vec x$ reduces
to \eqref{BR} in the $\lambda\to0$ limit. We do this by carefully
computing the contractions of the various vector products comprising
\eqref{Bk} and \eqref{vstar}.

First, let us introduce
\bea\label{l2}
\vec p_{\mit\Lambda}&:=& \frac{\sin\big(\hbar\, |{\mit\Lambda}\,\vec
  k|\big)}{|{\mit\Lambda}\,\vec k|} \ {\mit\Lambda}\,\vec k \\[4pt] 
&=& \frac{\sin\big(\,\frac12\, \sqrt{\lambda\, (\lambda\, \hbar^2\, \mbf
    l^2+\ell_s^3\, R\, \mbf k^2+\lambda^2 \,\ell_s^3\,R\, k_4^2)}\,
  \big)}{\sqrt{\lambda\, (\lambda\, \hbar^2\, \mbf
    l^2+\ell_s^3\, R\, \mbf k^2+\lambda^2 \,\ell_s^3\,R\, k_4^2)}} \
\big(\sqrt{\lambda\,\ell_s^3\,R}\ \mbf
k\,,\, \sqrt{\lambda^3\,\ell_s^3\, R}\ k_4\,,\, -\lambda\,\hbar\ \mbf l\big) \nonumber
\eea
in the conventions of Section~\ref{sec:quaternion}. Evidently
\bea \nonumber
\lim_{\lambda\to0} \, \frac{\sin\big(\hbar\, |{\mit\Lambda}\,\vec
  k|\big)}{|{\mit\Lambda}\,\vec k|} = \hbar
  \ ,
\label{l3}\eea
so that from \eqref{l2} we find the limit
\bea \label{eq:pLambda0}
\lim_{\lambda\to0} \, \vec p_{\mit\Lambda}=\vec 0 \ .
\label{l4}\eea
From the identity \eqref{eq:vstarid} we thus find
\bea\label{eq:vecstar0}
\lim_{\lambda\to0}\, \big|\vec p_{\mit\Lambda}\circledast_\eta\vec
p_{\mit\Lambda}^{ \, \prime} \big|=0
\label{l6}\eea
and
\bea\nonumber
\lim_{\lambda\to0}\, \frac{\sin^{-1}\big|\vec p_{\mit\Lambda}\circledast_\eta\vec
  p_{\mit\Lambda}^{ \, \prime} \big|}{\hbar\, \big|\vec p_{\mit\Lambda}\circledast_\eta\vec
  p_{\mit\Lambda}^{ \, \prime} \big|}= \frac1\hbar \ .
\label{l7}\eea
These limits imply that
\bea\nonumber
\lim_{\lambda\to0} \, \frac{\sin^{-1}\big|\vec p_{\mit\Lambda}\circledast_\eta\vec
  p_{\mit\Lambda}^{ \, \prime} \big|}{\hbar\, \big|\vec p_{\mit\Lambda}\circledast_\eta\vec
  p_{\mit\Lambda}^{ \, \prime} \big|}\, \Big(\sqrt{1-\big|\vec p_{\mit\Lambda}\big|{}^2}\ \vec
p_{\mit\Lambda}^{ \, \prime} +\sqrt{1-\big|\vec p_{\mit\Lambda}^{ \, \prime} \big|{}^2}\ \vec
p_{\mit\Lambda}\Big)\,\mbf\cdot\,{\mit\Lambda}^{-1}\,\vec x
= \big(\vec k+\vec k'\, \big)
\,\mbf\cdot\,\vec x \ .
\label{l8}\eea
Next, using (\ref{oct12}) one easily
finds
\bea\nonumber
2\, \big({\mit\Lambda}\,\vec
k\,\mbf\times_\eta\, {\mit\Lambda}\,\vec k'\,\big)\,\mbf\cdot\,
{\mit\Lambda}^{-1}\,\vec x &=&\lambda\,\mbf
x\,\mbf\cdot\,(\mbf k\,\mbf\times_\varepsilon\,\mbf l'-\mbf l\,\mbf\times_\varepsilon\,\mbf k'\,)+\lambda\, \mbf
x\,\mbf\cdot\,(k_4'\,\mbf l-k_4\, \mbf l'\,)+x^4\,(\mbf k\,\mbf\cdot\, \mbf l'-\mbf
l\,\mbf\cdot\, \mbf k'\,) \\ &&+\,\mbox{$\frac{\ell_s^3\, R}{\hbar^2}$}\, \mbf
p\,\mbf\cdot\,(\mbf k\,\mbf\times_\varepsilon\, \mbf k'\,)-\lambda\,\mbf p\,\mbf\cdot\,(\mbf l\,\mbf\times_\varepsilon\,\mbf
l'\,)+\mbox{$\frac{\lambda\, \ell_s^3\, R}{\hbar^2}$}\, \mbf
p\,\mbf\cdot\,(k_4\, \mbf k'-k_4'\, \mbf k)
\label{oct15}\eea
and from \eqref{oct15} we compute
\bea\nonumber
\lim_{\lambda\to0} \, \frac{\sin^{-1}\big|\vec p_{\mit\Lambda}\circledast_\eta\vec
  p_{\mit\Lambda}^{ \, \prime}\big|}{\hbar\, \big|\vec p_{\mit\Lambda}\circledast_\eta\vec
  p_{\mit\Lambda}^{ \, \prime}\big|}\, \big(\vec p_{\mit\Lambda}\,\mbf\times_\eta\,\vec
p_{\mit\Lambda}^{ \, \prime}\big)\,\mbf\cdot\, {\mit\Lambda}^{-1}\,\vec x &=&
\lim_{\lambda\to0} \, \Big(\, \frac{\sin^{-1}\big|\vec p_{\mit\Lambda}\circledast_\eta\vec
  p_{\mit\Lambda}^{ \, \prime}\big|}{\hbar\, \big|\vec p_{\mit\Lambda}\circledast_\eta\vec
  p_{\mit\Lambda}^{ \, \prime}\big|} \, \frac{\sin\big(\hbar\, |{\mit\Lambda}\,\vec
  k|\big)}{|{\mit\Lambda}\,\vec k|} \, \frac{\sin\big(\hbar\, |{\mit\Lambda}\,\vec
  k'\, |\big)}{|{\mit\Lambda}\,\vec k'\, |} \nonumber \\ && \qquad \qquad \qquad
\times \ \big({\mit\Lambda}\,\vec
k \,\mbf\times_\eta\, {\mit\Lambda}\,\vec k'\, \big)\,\mbf\cdot\,
{\mit\Lambda}^{-1}\,\vec x \, \Big) \nonumber \\[4pt]
&=& \mbox{$\frac1{2\hbar}$}\,\big(\ell_s^3\, R\, \mbf p\,\mbf\cdot\, (\mbf
k\,\mbf\times_\varepsilon\,\mbf k'\,)+\hbar^2\, x^4\, (\mbf k\,\mbf\cdot\,\mbf l'-\mbf l\,\mbf\cdot\,\mbf k'\,)
\big) \ . \nonumber
\label{l9}\eea

Putting everything together we conclude that
\bea\nonumber
\lim_{\lambda\to0} \, \vec{\mathcal{ B}}_\eta({\mit\Lambda}\,\vec
k,{\mit\Lambda}\, \vec k'\, )\,\mbf\cdot\, {\mit\Lambda}^{-1}\,\vec x &=& (\mbf k+\mbf k'\,)
\,\mbf\cdot\,\mbf x + (k_4+k_4'\,)
\, x^4 + (\mbf l+\mbf l'\,)
\,\mbf\cdot\,\mbf p \\
&& -\, \mbox{$\frac1{2\hbar}$}\,\big(\ell_s^3\, R\, \mbf p\,\mbf\cdot\, (\mbf
k\,\mbf\times_\varepsilon\,\mbf k'\,)+\hbar^2\, x^4\, (\mbf k\,\mbf\cdot\,\mbf l'-\mbf l\,\mbf\cdot\,\mbf k'\,)
\big) \ .
\label{l10}\eea
Up to the occurance of the circle fibre coordinate $x^4$, this
expression coincides exactly with (\ref{BR}). In the dimensional reduction of M-theory to IIA string theory
we restrict the algebra of functions to those which are constant along
the $x^4$-direction; they reduce the Fourier space integrations in
\eqref{oct11} to the six-dimensional hyperplanes $k_4=k_4'=0$. From \eqref{poly1} we see that $
x^4\star_\lambda f = x^4\, f + O(\lambda)$,
and hence the coordinate $x^4$ is central in the star product algebra of
functions in the limit $\lambda\to0$; as before we may therefore set
it to any non-zero constant value, which we take to be
$x^4=1$. In this way the $\lambda\to0$ limit of the star
product (\ref{oct11}) reduces exactly to \eqref{starrb1},
\bea\nonumber
\lim_{\lambda\to0} \, (f\star_\lambda g)(\vec x\,) = (f\star g)(x) \ .
\label{l11}\eea
With similar techniques, one shows that the dimensional reduction of the M-theory associator from \eqref{eq:assBk} coincides precisely with the string theory associator \eqref{eq:addassRfluxstring}:
\bea\label{eq:asslambda0}
\lim_{\lambda\to0} \, \vec{\mathcal{A}}_\eta({\mit\Lambda}\,\vec k,{\mit\Lambda}\,\vec k',{\mit\Lambda}\,\vec k^{\prime\prime}\,)\,\mbf\cdot\, {\mit\Lambda}^{-1}\,\vec x = \alg(k,k',k^{\prime\prime}\,) \ .
\eea

\subsection{Closure and cyclicity\label{sec:closure}}

We are not quite done with our phase space quantization of the
non-geometric M-theory background because the star product
\eqref{oct11}, in contrast to its string theory dual counterpart
\eqref{starrb1} at $\lambda=0$, has the undesirable feature that it is
neither 2-cyclic nor 3-cyclic in the sense of~\cite{Mylonas2013};
these properties are essential for a sensible
nonassociative phase space formulation of the quantisation of non-geometric strings~\cite{Mylonas2013}, for matching with the expectations from
worldsheet conformal field theory in non-geometric string backgrounds~\cite{MSS2,Aschieri2015}, and in the
construction of physically viable actions for a nonassociative theory
of gravity underlying the low-energy limit of non-geometric string
theory~\cite{Barnes2016}. It is natural to ask for analogous
features involving M2-branes and a putative nonassociative theory of gravity
underlying the low-energy limit of non-geometric M-theory.

Although for the classical brackets \eqref{oct4a} of Schwartz functions one has $\int\,\dd^7\vec x\ \{f,g\}_\lambda=\int\, \dd^7\vec x\ \partial_A(2\,\lambda^{ABC}\, x^C\, f\, \partial_Bg)=0$, this is no longer true for the quantum brackets $[f,g]_{\star_\lambda}$. The issue is that the star product \eqref{oct11} is not closed with respect to Lebesgue measure on $\real^7$, i.e., $\int\, \dd^7\vec x\ f\star_\lambda g\neq\int\, \dd^7\vec x\ f\,g$, and no modification of the measure can restore closure. In particular, using (\ref{poly1}) a simple integration by parts shows that
\begin{equation}\label{i6}
 \int \, \dd^{7}\vec x \ \big(x^A\star_\lambda f- x^A\, f\big) =6\, \hbar^2\, \int\, \dd^7\vec x \ \chi\big(\hbar^2\, \tilde{\mbf\triangle}_{\vec x}\, \big)\, \tilde\partial^A\triangleright f \ .
\end{equation}

To overcome this problem we seek a gauge equivalent star product
\begin{equation}\label{i7}
f\bullet_\lambda g= \mathcal{D}^{-1}\big( \mathcal{D}f\star_\lambda \mathcal{D}g \big) \qquad \mbox{with} \quad \mathcal{D}=1+{O}(\lambda) \ .
\end{equation}
The construction of the invertible differential operator $\mathcal{D}$ implementing this gauge transformation is analogous to the procedure used in \cite{KV15}. Order by order calculations, see e.g.~\cite{Kup24}, show that $\mathcal{D}$ contains only even order derivatives, and in fact it is a functional $\mathcal{D}=\mathcal{D}(\hbar^2\, \tilde{\mbf\triangle}_{\vec x})$. Since $\mathcal{D}\triangleright x^A=x^A$, we have
\begin{equation}\label{i8}
x^A\bullet_\lambda f=\mathcal{D}^{-1}\big(\mathcal{D}x^A\star_\lambda \mathcal{D}f\big) = \mathcal{D}^{-1} \, \hat x^A\, \mathcal{D} \triangleright f \ . 
\end{equation}
Using now $[\mathcal{D}^{-1},x^C]=-2\,\hbar^2\, \mathcal{D}^{-2}\, \mathcal{D}^\prime\, \tilde\partial^C$, where $\mathcal{D}^\prime$ stands for the (formal) derivative of $\mathcal{D}$ with respect to its argument, together with the explicit form \eqref{poly1} we find
\begin{equation}
\label{i10}
x^A\bullet_\lambda f=x^A\star_\lambda f-2\,\hbar^2\, \mathcal{D}^{-1}\, \mathcal{D}^\prime\, \tilde\partial^A\triangleright f \ .
\end{equation}
The requirement $\int\, \dd^{7}\vec x\ x^A\bullet_\lambda f= \int\,\dd^7\vec x\ x^A\, f$, with the help of (\ref{i6}) and (\ref{i10}), gives the elementary Cauchy problem
 \begin{equation}\label{i12}
\mathcal{D}^{-1}\, {\frac{\dd\mathcal{D}}{\dd t}=3\, \frac{\sqrt{t}\coth\sqrt{t}-1}{t}} \qquad \mbox{with} \quad \mathcal{D}(0)=1 \ ,
\nonumber\end{equation}
whose solution finally yields
\begin{equation}\label{oct13}
\mathcal{D}=\Big( \big(\hbar\,\tilde{\mbf\triangle}_{\vec x}^{1/2}\big)^{-1} \sinh\big(\hbar\, \tilde{\mbf\triangle}_{\vec x}^{1/2}\big) \Big)^{6} \ .
\nonumber\end{equation}

We may therefore write the star product (\ref{i7}) as
\begin{eqnarray}\label{i14}
(f\bullet_\lambda g)(\vec x\, ) &=& \int\, \frac{\dd^{7}\vec k}{( 2\pi
) ^{7}} \ \frac{\dd^{7}\vec k'}{( 2\pi
) ^{7}} \ \tilde{f}( \vec k\, )\, \tilde{g}( \vec k'\, )\, \e^{\ii\vec{\mathcal{ B}}_\eta({\mit\Lambda}\,\vec k,{\mit\Lambda}\, \vec k'\, )\,\mbf\cdot\, {\mit\Lambda}^{-1}\,\vec x} \\ && \qquad\qquad\qquad\qquad \times \ \Big(\, \frac{\sin\big(\hbar\, |{\mit\Lambda}\,\vec k|\big)\sin\big(\hbar\, |{\mit\Lambda}\,\vec k'\, |\big)}{\hbar\, |{\mit\Lambda}\,\vec k| \, |{\mit\Lambda}\,\vec k'\, |} \, \frac{|\vec{\mathcal{ B}}_\eta({\mit\Lambda}\,\vec k,{\mit\Lambda}\, \vec k'\, )|}{\sin\big(\hbar\, |\vec{\mathcal{ B}}_\eta({\mit\Lambda}\,\vec k,{\mit\Lambda}\, \vec k'\, )|\big)}\, \Big)^6 \ .\notag
\end{eqnarray}
This star product still provides a quantization of the brackets (\ref{oct9}),
\begin{equation}\label{i15}
\lim_{\stackrel{\scriptstyle\hbar,\ell_s\to0}{\scriptstyle \ell_s^3/\hbar^2={\rm constant}}} \, \frac{[f,g]_{\bullet_\lambda}}{\ii\hbar} =\{f,g\}_{\lambda} \ .
\nonumber\end{equation}
Using the limits computed in Section~\ref{sec:Mtheorystar}, the extra factors in \eqref{i14} are simply unity in the contraction limit $\lambda\to0$, and so the star product \eqref{i14} still dimensionally reduces to \eqref{starrb1},
\bea\nonumber
\lim_{\lambda\to0} \, (f\bullet_\lambda g)(\vec x\, ) = (f\star g)(x) \ .
\eea
It is Hermitean, $(f\bullet_\lambda g)^\ast=g^\ast \bullet_\lambda
f^\ast,$ and unital, $f\bullet_\lambda 1=f=1\bullet_\lambda f$, but it
is no longer a Weyl star product, i.e., it does not satisfy
(\ref{weyl}); in particular, the star products of plane waves
$\e^{\ii\vec k\,\mbf\cdot\, \vec x}\bullet_\lambda\e^{\ii\vec k'\,\mbf\cdot\,\vec x}$ 
are no longer given simply by the Baker-Campbell-Hausdorff formula. However, it is now closed,
 \begin{equation}\label{i1a}
 \int \, \dd^{7}\vec x\ f\bullet_\lambda g = \int\, \dd^7\vec x\ f\, g \ , \nonumber\end{equation}
which identifies it as the Kontsevich star product; in particular, the
desired 2-cyclicity property follows: $\int\, \dd^7\vec x\
[f,g]_{\bullet_\lambda} = 0$. The closure condition can be regarded
as the absence of noncommutativity (and nonassociativity) among free fields.

Under the gauge transformation (\ref{i7}) the star associator and Jacobiator transform covariantly:
\begin{equation}
\label{i16}
A_{\bullet_\lambda}(f,g,h)=\mathcal{D}^{-1} A_{\star_\lambda}(\mathcal{D}f, \mathcal{D}g,\mathcal{D}h) \qquad \mbox{and} \qquad
[f,g,h]_{\bullet_\lambda}=\mathcal{D}^{-1} [\mathcal{D}f,\mathcal{D}g,\mathcal{D}h]_{\star_\lambda} \ .
\nonumber\end{equation}
The star product (\ref{oct11}) is alternative on the space of Schwartz functions, and since the differential operator $\mathcal{D}$ preserves this subspace it follows that the star product (\ref{i14}) is also alternative on Schwartz functions, i.e., it satisfies
\begin{eqnarray}
\label{i20}
A_{\bullet_\lambda}(f,g,h) = \mbox{$\frac16$}\, [f,g,h]_{\bullet_\lambda} \ .
\end{eqnarray}
By 2-cyclicity the integrated star Jacobiator of Schwartz functions vanishes, and together with (\ref{i20}) we arrive at the desired 3-cyclicity property
\begin{equation}\nonumber
    \int\, \dd^7\vec x\ (f\bullet_\lambda g)\bullet_\lambda h= \int \, \dd^7\vec x\ f\bullet_\lambda (g\bullet_\lambda h) \ .
\label{i21}\end{equation}
This property can be regarded as the absence on-shell of
nonassociativity (but not noncommutativity) among cubic interactions
of fields.

\subsection{Uncertainty relations}

The closure and cyclicity properties of the gauge equivalent
star product $\bullet_\lambda$ enable a consistent formulation of
nonassociative phase space quantum mechanics, along the lines
given in~\cite{Mylonas2013} (see also~\cite[Section~4.5]{Mylonas2014} for a
review). In particular, this framework provides a concrete and rigorous derivation
of the novel uncertainty principles which are heuristically expected
to arise from the commutation relations \eqref{oct41} that quantize
the brackets \eqref{oct9} and \eqref{oct9a} capturing the
nonassociative geometry of the M-theory $R$-flux background. It 
avoids the problems arising from the fact that our nonassociative
algebras are not alternative (which has been the property usually required in previous
treatments of nonassociativity in quantum mechanics).

In this approach, observables $f$ are real-valued functions on
the seven-dimensional phase space that are multiplied together with
the star product \eqref{i14}; dynamics in the quantum theory with
classical Hamiltonian $\sf H$ is then implemented via the time
evolution equations
$$
\frac{\partial f}{\partial t} = \frac\ii\hbar\, [{\sf H},f]_{\bullet_\lambda} \ .
$$
States are characterized by normalized phase space wave functions
$\psi_a$ and statistical probabilities $\mu_a$. Expectation
values are computed via the phase space integral
$$
\langle f\rangle= \sum_a \, \mu_a \ \int\,\dd^7\vec x \
\psi_a^* \bullet_\lambda (f\bullet_\lambda\psi_a) \ ,
$$
which using closure and cyclicity can be expressed in terms of a
normalized real-valued state function $S=\sum_a \,
\mu_a \,\psi_a \bullet_\lambda \psi_a^*$ as $\langle
f\rangle = \int\, \dd^7\vec x \ f\, S$.

From the non-vanishing Jacobiators \eqref{oct9a} in the present case we
expect in fact to obtain a coarse graining of the M-theory phase
space, rather than just the configuration space as it happens in the
reduction to the string theory $R$-flux background~\cite{Mylonas2013}. This can be
quantified by computing the expectation values of oriented area and volume
uncertainty operators following the formalism of~\cite{Mylonas2013}. In this
prescription, we can define the area operator corresponding to
directions $x^A$, $x^B$ as
\begin{equation}
\label{area}
\A^{AB}=\mathfrak{Im}\big( [\tilde x^A,\tilde
x^B]_{\bullet_\lambda}\big)=-\ii\big(\tilde x^A\bullet_\lambda\tilde
x^B-\tilde x^B\bullet_\lambda\tilde x^A\big) \ ,
\end{equation}
while the volume operator in directions $x^A$, $x^B$, $x^C$ is
\begin{equation}
\label{volume}
\V^{ABC}=\mbox{$\frac{1}{3}$}\, \mathfrak{Re}\big(\tilde
x^A\bullet_\lambda [\tilde x^B,\tilde x^C]_{\bullet_\lambda}+\tilde
x^C\bullet_\lambda [\tilde x^A,\tilde x^B]_{\bullet_\lambda}+\tilde
x^B\bullet_\lambda [\tilde x^C,\tilde x^A]_{\bullet_\lambda}\big) \ ,
\end{equation}
where $\tilde x^A:=x^A-\langle x^A\rangle$ are the operator
displacements appropriate to the description of quantum
uncertainties. Explicit computations using the fact that the star
product $\bullet_\lambda$ is alternative on monomials give the
operators \eqref{area} and \eqref{volume} as
\begin{equation}\label{av}
\A^{AB}=2\,\hbar\, \lambda^{ABC}\, \tilde x^C 
 \qquad \mbox{and} \qquad \V^{ABC}=\mbox{$\frac{1}{6}$}\, [\tilde x^A,\tilde x^B,\tilde x^C]_{\bullet_\lambda} = 2\, \hbar^2\, \lambda^{ABCD} \,\tilde x^D \ .
\end{equation}

Let us now write the expectation values of the
operators \eqref{av} in components. For the fundamental area measurement uncertainties (or
minimal areas) we obtain
\begin{eqnarray}\label{au}
\langle \A^{ij}\rangle &=&\mbox{$\frac{\ell_s^3}{\hbar}$}\,\big|
                            R^{4,ijk4}\, \langle p_k\rangle\big|
                            \qquad \mbox{and} \qquad \langle
                            \A^{4i}\rangle \ = \ \mbox{$\frac{\lambda\, \ell_s^3}{\hbar}$}\, \big|R^{4,1234}\,\langle p^i \rangle\big| \ , \\[4pt]
\langle \A^{x^i,p_j}\rangle &=&\hbar\, \big|\delta^i_j\,\langle
                                 x^4\rangle+\lambda\,
                                 \varepsilon^i{}_{jk}\,\langle
                                 x^k\rangle\big| \qquad \mbox{and}
                                 \qquad \langle
                                 \A^{x^4,p_i}\rangle \ = \ \lambda^2\,\hbar\,
                                 \langle x_i \rangle\ , \nonumber\\[4pt]
\langle \A^{p_i,p_j}\rangle &=&\lambda\, \hbar\,\big|
                                 \varepsilon_{ijk}\,\langle
                                 p^k\rangle\big| \ . \nonumber
\end{eqnarray}
The first expression $\langle \A^{ij}\rangle$ demonstrates, as in the
string theory case~\cite{Mylonas2013}, an area uncertainty on $M$ proportional to the magnitude of
the transverse momentum, while the second expression
$\langle\A^{4i}\rangle$ gives area uncertainties along the M-theory
circle proportional to the momentum transverse to the fibre
direction. The third expression $\langle \A^{x^i,p_j}\rangle$
describes phase space cells of position and momentum in the same
direction with area $\hbar\,|\langle x^4\rangle|$, together with new cells
proportional to the transverse directions; in the contraction limit
$\lambda=0$ it reduces to the standard minimal area (for $x^4=1$)
governed by the Heisenberg uncertainty principle. For $\lambda\neq0$
the area uncertainties $\langle \A^{x^\mu,p_i}\rangle$ suggest that there are new limitations to
the simultaneous measurements of transverse position and momentum in
the seven-dimensional M-theory phase space, induced by a non-zero string
coupling constant $g_s$, although we will see below that this
interpretation is somewhat subtle. The final
expression $\langle \A^{p_i,p_j}\rangle$ is also a new area
uncertainty particular to M-theory, yielding cells in
momentum space with area proportional to $\lambda\,\hbar$ (and to the magnitude of the transverse momentum), a
point to which
we return in Section~\ref{sec:polarizations}.

For the fundamental volume measurement uncertainties (or minimal
volumes) we obtain
\begin{eqnarray}\label{vu}
\langle \V^{ijk}\rangle &=& \mbox{$\frac{\ell_s^3}{2}$}\, \big| R^{4,ijk4}
\,\langle x^4 \rangle \big| \qquad \mbox{and} \qquad \langle \V^{ij4}\rangle
                           \ = \ \mbox{$\frac{\lambda^2\,
                           \ell_s^3}{2}$}\, \big| R^{4,ijk4} \,
                           \langle x_k \rangle \big| \ , \\[4pt]
\langle \V^{p_i,x^j,x^k}\rangle &=&
                                   \mbox{$\frac{\lambda\,\ell_s^3}{2}$}
                                   \, \big| R^{4,1234}\,
                                   \big(\delta^j_i\, \langle
                                   p^k\rangle -\delta^k_i\, \langle
                                   p^j\rangle \big) \big| \qquad
                                   \mbox{and} \qquad \langle
                                   \V^{p_i,x^j,x^4}\rangle \ = \
                                   \mbox{$\frac{\lambda^2\,
                                   \ell_s^3}{2}$}\, R^{4,ijk4}\,
                                   \langle p_k \rangle \ , \nonumber \\[4pt]
\langle \V^{p_i,p_j,x^k}\rangle &=& \mbox{$\frac{\lambda\,\hbar^2}2$}
                                     \,\big|
                                   \lambda\, \varepsilon_{ij}{}^{k}\, \langle
                                   x^4\rangle + \delta_j^k\,
                                   \langle x_i\rangle -\delta_i^k\,
                                   \langle x_j\rangle \big| \qquad
                                   \mbox{and} \qquad \langle
                                   \V^{p_i,p_j,x^4}\rangle \ = \ 
                                                              \mbox{$\frac{\lambda^3\,\hbar^2}2$}
                                     \,
                                                              \big|\varepsilon_{ijk}\,
                                                              \langle
                                                              x^k\rangle
                                                              \big| \ . \nonumber 
\end{eqnarray}
They demonstrate volume uncertainties in position coordinates $x^\mu$,
$x^\nu$, $x^\alpha$ proportional to the magnitude of the transverse coordinate direction; in
particular, there is a volume uncertainty on $M$ proportional to
the magnitude of the circle fibre coordinate $x^4$, which reduces for
$x^4=1$ to the expected minimal volume in non-geometric string theory~\cite{Mylonas2013}. A geometric interpretation of these position volume uncertainties will be provided in Section~\ref{sec:TriNAG}. There are also
phase space cubes for position and momentum in the same direction as
well as in transverse directions, reflecting the fact that the
corresponding nonassociating triples of M-theory phase space
coordinates cannot be measured simultaneously to arbitrary precision; these new volume uncertainties
vanish in the contraction limit $\lambda=0$. In the string theory limit
the volume uncertainties can be interpreted as the non-existence of
D-particles in the $R$-flux background due to the Freed-Witten anomaly
in the T-dual $H$-flux frame~\cite{Blumenhagen2013,GLM}; it would interesting to
understand the corresponding meaning in the presence of non-geometric
M-theory $R$-fluxes, which involves the full seven-dimensional
M-theory phase space. However, there are no minimal
volumes in momentum space, as we discuss further in
Section~\ref{sec:polarizations}.

The present situation is much more complicated in the case of the actual
quantum uncertainty principles imposing limitations to position and
momentum measurements; they encode positivity of operators in
nonassociative phase space quantum mechanics~\cite{Mylonas2013}. To
calculate the uncertainty relations amongst phase space coordinates,
we use the Cauchy-Schwarz inequality derived in~\cite{Mylonas2013} to obtain the uncertainty relations
\bea\label{eq:BJHuncertainty}
\Delta x^A\, \Delta x^B\geq \mbox{$\frac12$}\, \big| \big\langle
[x^A,x^B]_{\circ_\lambda} \big\rangle\big| \ ,
\eea
where
\bea\label{eq:bulletcomp}
[x^A,x^B]_{\circ_\lambda}\bullet_\lambda\psi:= x^A\bullet_\lambda(x^B\bullet_\lambda \psi)-x^B\bullet_\lambda(x^A\bullet_\lambda \psi)
\eea
for any phase space wave function $\psi$.

We first observe that from (\ref{poly1}) one obtains the commutator
\begin{equation}\label{u1}
[\hat x^A,\hat x^B]=2\ii\hbar\, \lambda^{ABC}\, \hat x^C-4\, \hbar^2\,
\lambda^{ABDE}\, x^E\,\partial_D \ .
\end{equation}
It is easy check that in the limit $\lambda\to0$ the relations
\eqref{u1} reproduce the algebra of
differential operators \eqref{eq:stringhatxcomm} for the string theory
$R$-flux background. From the contraction identity (\ref{epsilon7}) we
can rewrite \eqref{u1} as
\begin{equation*}
[\hat x^A,\hat x^B]=2\ii\hbar\, \lambda^{ABC}\, \hat x^C+ 4 \, \hbar^2\,
\lambda^{ABC}\, \lambda^{CDE}\, x^E\,\partial_D +x^B\,
\bar\partial^A-x^A\, \bar\partial^B \ ,
\end{equation*}
where
\begin{equation*}
\big(\bar\partial^A\big):=4\, \hbar^2 \, \big(\, \Lambda^{BA}\,
\tilde\partial_B\,\big) =
 \big({\lambda\,
\ell_s^3\, R}\ \nabla_{\mbf x}\,,\, {\lambda^3\,\ell_s^3\, R}\
\mbox{$\frac\partial{\partial x^4}$}\,,\, \lambda^2\,\hbar^2\
\nabla_{\mbf p}\big) \ .
\end{equation*}

Next we calculate
\begin{eqnarray}
x^A\star_\lambda(x^B\star_\lambda
  \psi)-x^B\star_\lambda(x^A\star_\lambda \psi)&=&[\hat x^A,\hat
                                                   x^B]\triangleright
                                                   \psi \label{u3}\\[4pt]
                                               &=&2\ii\hbar\,
                                                   \lambda^{ABC}\, (
                                                   x^C\star_\lambda
                                                   \psi-2\ii\hbar\,
                                                   \lambda^{CDE}\,
                                                   x^E\,\partial_D \psi)
 \nonumber \\ && +\, (x^B\, \bar\partial^A-x^A\, \bar\partial^ B)\triangleright
        \psi \nonumber\\[4pt] &=&2\ii\hbar\, \lambda^{ABC}\,
                              \psi \star_\lambda
                              x^C+(x^B\, \bar\partial^A-x^A\, \bar\partial^
                              B)\triangleright \psi \nonumber \ .
\end{eqnarray}
To translate the expression \eqref{u3} into the definition \eqref{eq:bulletcomp} via the
closed star product $\bullet_\lambda$, we use the gauge transformation
(\ref{i8}) to obtain
\begin{eqnarray}
[x^A,x^B]_{\circ_\lambda}\bullet_\lambda\psi &=& \mathcal{D}^{-1}\big(x^A\star_\lambda(x^B\star_\lambda \mathcal{D}\psi)-x^B\star_\lambda(x^A\star_\lambda \mathcal{D}\psi)\big)\label{u4}\\[4pt]
&=&\mathcal{D}^{-1}\big(2\ii\hbar\, \lambda^{ABC} \, \mathcal{D}\psi
    \star_\lambda x^C+(x^B\, \bar\partial^A-x^A\,
    \bar\partial^B)\triangleright \mathcal{D}\psi \big) \nonumber\\[4pt]
&=&2\ii\hbar\, \lambda^{ABC}\, \psi \bullet_\lambda x^C+(x^B\,
    \bar\partial^A-x^A\, \bar\partial^B)\triangleright \psi \ , \nonumber
\end{eqnarray}
where in the last equality we used $\mathcal{D}=
\mathcal{D}(\hbar^2\, \tilde{\mbf\triangle}_{\vec x})$. 

The explicit computation of the uncertainty relations
\eqref{eq:BJHuncertainty} is complicated by the second term in the
last line of \eqref{u4}. The differential operator $x^A\,
\bar\partial^B-x^B\, \bar\partial^A$ is of order $O(\lambda)$, and it
can be regarded as a generator of ``twisted'' rotations in the
phase space plane spanned by the vectors $x^A$ and $x^B$; in the
limit $\lambda\to0$, the result (\ref{u4}) reproduces exactly the
corresponding calculation from \cite[eq.~(5.33)]{Mylonas2013} where
this problem does not arise. If we restrict to states which are
rotationally invariant in this sense, so that the corresponding wave
functions $\psi$ obey $(x^B\, \bar\partial^A-x^A\,
\bar\partial^B)\triangleright \psi=0$, and which obey the ``symmetry'' condition of~\cite{Mylonas2013}, then the corresponding uncertainty
relations \eqref{eq:BJHuncertainty} for phase space coordinate measurements reads as
\begin{equation*}
\Delta x^A\, \Delta x^B\geq\hbar\, \big| \lambda^{ABC}\, \langle
x^C\rangle \big| \ ,
\end{equation*}
with similar interpretations as those of the area measurement uncertainties
derived in \eqref{au}. However, the uncertainty relations
\eqref{eq:BJHuncertainty} seem too complicated to suggest a universal
lower bound which does not depend on the choice of state.

\subsection{Configuration space triproducts and nonassociative geometry\label{sec:TriNAG}}

Thus far all of our considerations have applied to phase space, and it is now natural to look at polarizations which suitably reduce the physical degrees of freedom as is necessary in quantization. 
From the perspective of left-right asymmetric worldsheet conformal field theory, closed strings probe the nonassociative deformation of the $R$-flux background through phase factors that turn up in off-shell correlation functions of tachyon vertex operators, which can be encoded in a triproduct of functions on configuration space $M$~\cite{Blumenhagen2011}; this
triproduct originally appeared
in~\cite{Takhtajan1993,Dito1996} (see also~\cite{Blumenhagen2010}) as a candidate 
deformation quantization of the canonical Nambu-Poisson bracket on
$\real^3$. This geometric structure was generalised to curved spaces
with non-constant fluxes within the framework of double field theory
in~\cite{Blumenhagen2013}, and in~\cite{Aschieri2015} it was shown
that these triproducts descend precisely from polarisation of the
phase space star product along the leaf of zero momentum $\mbf p=\mbf
0$ in phase space $T^*M$. Although at present we do not have available a quantum
theory that would provide an M2-brane analog of the computation of conformal field
theory correlation functions for closed strings propagating in
constant non-geometric $R$-flux
compactifications, we can imitate this latter reduction of the phase space star product in our case and derive triproducts which geometrically describe the quantization of the four-dimensional M-theory configuration space.

For this, we consider functions which depend only on configuration space coordinates, that we denote by $\vec x_0=(\mbf x,x^4,\mbf 0)\in\real^4$, and define the product
\bea\nonumber
\big(f\vartriangle_\lambda^{\!{}_{(2)}}g\big)(\vec x_0):= (f\star_\lambda
g)(\mbf x,x^4,\mbf p)\big|_{\mbf p= \mbf 0} = \int\, \frac{\dd^{4}\vec k}{( 2\pi
) ^{4}} \ \frac{\dd^{4}\vec k'}{( 2\pi
) ^{4}} \ \tilde{f}( \vec k\, )\, \tilde{g}( \vec k'\, )\,
\e^{\ii\vec{\mathcal{ B}}_\eta({\mit\Lambda}\,\vec k,{\mit\Lambda}\,
  \vec k'\, )\,\mbf\cdot\, {\mit\Lambda}^{-1}\,\vec x_0} \ .
\eea
From \eqref{oct12} we see that the cross product of four-dimensional vectors $\vec k=(\mbf 0,\mbf k,k_4)\in\real^4$ gives
\bea\nonumber
{\mit\Lambda}\,\vec k\,\mbf\times_\eta\,{\mit\Lambda}\,\vec k'= \mbox{$\frac{\lambda\,\ell_s^3\,R}{4\hbar^2}$} \, \big(-\mbf k\,\mbf\times_\varepsilon\,\mbf k' +
  \lambda\, k_4'\, \mbf k-\lambda\, k_4\,\mbf k' \,,\,\mbf 0\,,\,0 \big) \ ,
\eea
and it is therefore orthogonal to ${\mit\Lambda}^{-1}\,\vec x_0$,
i.e., $\big({\mit\Lambda}\,\vec
k\,\mbf\times_\eta\,{\mit\Lambda}\,\vec k'\,\big)\,\mbf\cdot\,
{\mit\Lambda}^{-1}\,\vec x_0=0$. Hence the source of noncommutativity
and nonassociativity vanishes in this polarization, and using the
variables \eqref{l2} we can write the product succinctly as
\bea\label{eq:triprod2}
\big(f\vartriangle_\lambda^{\!_{(2)}}g\big)(\vec x_0) = \int\, \frac{\dd^{4}\vec k}{( 2\pi
) ^{4}} \ \frac{\dd^{4}\vec k'}{( 2\pi
) ^{4}} \ \tilde{f}( \vec k\, )\, \tilde{g}( \vec k'\, )\, \e^{\ii\vec{\mathcal{ T}}_{\mit\Lambda}^{{}_{(2)}}(\vec k, \vec k'\, )\,\mbf\cdot\, {\mit\Lambda}^{-1}\,\vec x_0} \ ,
\eea
where we introduced the deformed vector sum
\bea\nonumber
\vec{\mathcal{ T}}_{\mit\Lambda}^{{}_{(2)}}(\vec k, \vec k'\, ) = \frac{\sin^{-1}\big|\vec p_{\mit\Lambda}\circledast_\eta\vec
  p_{\mit\Lambda}^{ \, \prime} \big|}{\hbar\, \big|\vec p_{\mit\Lambda}\circledast_\eta\vec
  p_{\mit\Lambda}^{ \, \prime} \big|}\, \Big(\sqrt{1-\big|\vec p_{\mit\Lambda}^{ \, \prime} \big|{}^2}\ \vec
p_{\mit\Lambda} +\sqrt{1-\big|\vec p_{\mit\Lambda} \big|{}^2}\ \vec
p_{\mit\Lambda}^{ \, \prime}\Big) \ .
\eea
It has a perturbative expansion given by
\bea\nonumber
{\mit\Lambda}^{-1}\,\vec{\mathcal{ T}}_{\mit\Lambda}^{{}_{(2)}}(\vec k, \vec k'\, ) = \vec k + \vec k' + O\big(\sqrt{\lambda}\,\big) \ .
\eea
The product \eqref{eq:triprod2} inherits properties of the phase space
star product $\star_\lambda$; in particular, since $\vec{\mathcal{
    T}}_{\mit\Lambda}^{{}_{(2)}}(\vec k, \vec 0 \, ) =\vec k=\vec{\mathcal{
    T}}_{\mit\Lambda}^{{}_{(2)}}(\vec 0, \vec k\, )$, it is unital:
\bea\nonumber
f\vartriangle_\lambda^{\!_{(2)}}1=f=1\vartriangle_\lambda^{\!_{(2)}}f \ .
\eea
It is commutative and associative, as expected from the area
uncertainties $\langle \A^{\mu\nu}\rangle$ of \eqref{au} in this
polarisation; from the limits \eqref{eq:pLambda0} and
\eqref{eq:vecstar0} it follows that it reduces at $\lambda=0$ to the
ordinary pointwise product of fields on $\real^3$, as anticipated from
the corresponding string theory
result~\cite{Blumenhagen2011,Aschieri2015}. However, the product
\eqref{eq:triprod2} is \emph{not} generally the pointwise product of
functions on $\real^4$, $f\vartriangle_\lambda^{\!_{(2)}}g\neq f\,g$;
in particular
\bea\nonumber
x^{\mu}\vartriangle_\lambda^{\!_{(2)}}f =x^\mu \, f
+\hbar^2\, \big( x^\mu \, {\tilde{\mbf\triangle}_{\vec x_0}}- (\vec
x_0 \,\mbf\cdot\, \tilde\nabla_{\vec x_0} ) \, \tilde\partial^\mu \big)\, \chi\big(\hbar^2\,\tilde{\mbf\triangle}_{\vec x_0}\big)\triangleright f \ ,
\eea
so that off-shell membrane amplitudes in this case experience a commutative and associative deformation. Moreover, $\int\,\dd^4\vec x_0 \ f\vartriangle_\lambda^{\!_{(2)}}g\neq \int\,\dd^4\vec x_0 \ f\, g$, but this can be rectified by defining instead a product $\blacktriangle_\lambda^{\!_{(2)}}$ based on the closed star product \eqref{i14}; then $\int\,\dd^4\vec x_0 \ f\blacktriangle_\lambda^{\!_{(2)}}g= \int\,\dd^4\vec x_0 \ f\, g$.

Next we define a triproduct for three functions $f$, $g$ and $h$ of $\vec x_0=(\mbf x,x^4)\in\real^4$ by a similar rule:
\bea\nonumber
\big(f\vartriangle_\lambda^{\!{}_{(3)}}g\vartriangle_\lambda^{\!{}_{(3)}}h\big)(\vec
x_0):= \big((f\star_\lambda g)\star_\lambda h
\big)(\mbf x,x^4,\mbf p)\big|_{\mbf p= \mbf 0} \ .
\eea
As before, the Fourier integrations truncate to four-dimensional subspaces and we can write
\bea\nonumber
\big(f\vartriangle_\lambda^{\!{}_{(3)}}g\vartriangle_\lambda^{\!{}_{(3)}}h\big)(\vec x_0)= \int\, \frac{\dd^{4}\vec k}{( 2\pi
) ^{4}} \ \frac{\dd^{4}\vec k'}{( 2\pi
) ^{4}} \ \frac{\dd^{4}\vec k''}{( 2\pi
) ^{4}} \ \tilde{f}( \vec k\, )\, \tilde{g}( \vec k'\, )\,\tilde{h}(
\vec k''\, )\, \e^{\ii\vec{\mathcal{ B}}_\eta(\vec{\mathcal{
      B}}_\eta({\mit\Lambda}\,\vec k,{\mit\Lambda}\, \vec k'\, )
  ,{\mit\Lambda}\, \vec k''\, )\,\mbf\cdot\, {\mit\Lambda}^{-1}\,\vec
  x_0} \ .
\eea
As in the calculation which led to \eqref{eq:assBk}, we can compute the deformed vector addition from products of octonion exponentials
\bea\nonumber
\e^{X_{\vec{\mathcal{ B}}_\eta(\vec{\mathcal{
        B}}_\eta({\mit\Lambda}\,\vec k,{\mit\Lambda}\, \vec k'\, )
    ,{\mit\Lambda}\, \vec k''\, )}}= \big(\e^{X_{{\mit\Lambda}\, \vec
    k}}\, \e^{X_{{\mit\Lambda}\, \vec k'}}\,\big)\,
\e^{X_{{\mit\Lambda}\, \vec k^{\prime\prime}}}
\eea
using \eqref{eq:octexp} and \eqref{eq:octexpprod}, together with the
identities \eqref{eq:vstarid} and \eqref{eq:trigid}. The final result
is a bit complicated in general, but is again most concisely expressed in
terms of the variables \eqref{l2}. Exploiting again the property that
the vector cross product $\vec p_{\mit\Lambda}^{ \, \prime}\,\mbf\times_\eta\,\vec p_{\mit\Lambda}$ lives in the orthogonal complement $\real^3$ to the four-dimensional subspace $\real^4$ in $\real^7$ containing $\vec p_{\mit\Lambda}$, so that $\big(\vec p_{\mit\Lambda}^{ \, \prime}\,\mbf\times_\eta\,\vec p_{\mit\Lambda}\big)\,\mbf\cdot\, {\mit\Lambda}^{-1}\,\vec x_0=0$, after a bit of calculation one finds that the triproduct can be written as
\bea\label{eq:triprod3}
\big(f\vartriangle_\lambda^{\!{}_{(3)}}g\vartriangle_\lambda^{\!{}_{(3)}}\,h\big)(\vec x_0)= \int\, \frac{\dd^{4}\vec k}{( 2\pi
) ^{4}} \ \frac{\dd^{4}\vec k'}{( 2\pi
) ^{4}} \ \frac{\dd^{4}\vec k''}{( 2\pi
) ^{4}} \ \tilde{f}( \vec k\, )\, \tilde{g}( \vec k'\, )\,\tilde{h}( \vec k''\, )\, \e^{\ii\vec{\mathcal{T}}_{\mit\Lambda}^{{}_{(3)}}( \vec k, \vec k', \vec k''\, )\,\mbf\cdot\, {\mit\Lambda}^{-1}\,\vec x_0} \ ,
\eea
where we defined the deformed vector sum
\bea\nonumber
\vec{\mathcal{T}}_{\mit\Lambda}^{{}_{(3)}}(\vec k,\vec k',\vec k^{\prime\prime}\,) &=&
\frac{\sin^{-1}\big|(\vec p_{\mit\Lambda}\circledast_\eta \vec p_{\mit\Lambda}^{\,\prime})
    \circledast_\eta\vec p_{\mit\Lambda}^{\,\prime\prime} \big|}{\hbar \, \big|(
    \vec p_{\mit\Lambda}\circledast_\eta \vec p_{\mit\Lambda}^{\,\prime} ) \circledast_\eta\vec p_{\mit\Lambda}^{\,\prime\prime}\big|} \, \Big( \vec A_\eta\big(\vec p_{\mit\Lambda},\vec p_{\mit\Lambda}^{\,\prime},\vec p_{\mit\Lambda}^{\,\prime\prime}\, \big) + \epsilon_{\vec p_{\mit\Lambda}^{ \, \prime},\vec p_{\mit\Lambda}^{ \, \prime\prime}}\, \sqrt{1-\big|\vec p_{\mit\Lambda}^{ \, \prime} \circledast_\eta\vec p_{\mit\Lambda}^{\,\prime\prime}\,\big|{}^2}\ \vec
p_{\mit\Lambda} \\ && +\,\epsilon_{\vec p_{\mit\Lambda},\vec p_{\mit\Lambda}^{ \, \prime\prime}}\, \sqrt{1-\big|\vec p_{\mit\Lambda}\circledast_\eta\vec p_{\mit\Lambda}^{ \, \prime\prime} \,\big|{}^2}\ \vec
p_{\mit\Lambda}^{ \, \prime} + \epsilon_{\vec p_{\mit\Lambda},\vec p_{\mit\Lambda}^{ \, \prime}}\, \sqrt{1-\big|\vec p_{\mit\Lambda}\circledast_\eta\vec p_{\mit\Lambda}^{ \, \prime} \big|{}^2}\ \vec
p_{\mit\Lambda}^{ \, \prime\prime}\Big) 
\label{eq:TLambda3}\eea
which contains the associator \eqref{eq:assBk}. It has a perturbative expansion given by
\bea\nonumber
{\mit\Lambda}^{-1}\,\vec{\mathcal{T}}_{\mit\Lambda}^{{}_{(3)}}(\vec k,\vec k',\vec k^{\prime\prime}\,) &=& \vec k+\vec k'+\vec k'' + \mbox{$ \frac{\hbar^2}{2}$}\, \big(2{\mit\Lambda}^{-1}\, \vec A_\eta({\mit\Lambda}\,\vec k, {\mit\Lambda}\,\vec k^{\prime}, {\mit\Lambda}\,\vec k^{\prime\prime}\,)\\ && \nonumber +\, |{\mit\Lambda}\,\vec k'+{\mit\Lambda}\,\vec k''\,|^2\ \vec k+ |{\mit\Lambda}\,\vec k+ {\mit\Lambda}\,\vec k''\,|^2\ \vec k'+ |{\mit\Lambda}\,\vec k+{\mit\Lambda}\,\vec k'\,|^2\ \vec k\,''\,\big) + O(\lambda) \ .
\eea

Using the triproduct \eqref{eq:triprod3}, we then define a completely
antisymmetric quantum 3-bracket in the usual way by
\bea\nonumber
[f_1,f_2,f_3]_{\vartriangle_\lambda^{\!{}_{(3)}}}:= \sum_{\sigma\in S_3}\, (-1)^{|\sigma|}\,
 f_{\sigma(1)}\vartriangle_\lambda^{\!{}_{(3)}}f_{\sigma(2)}\vartriangle_\lambda^{\!{}_{(3)}}f_{\sigma(3)} \ .
\eea
It reproduces the 3-brackets from \eqref{oct41} amongst linear functions that encodes the nonassociative geometry of configuration space,
\bea\nonumber
[x^\mu,x^\nu,x^\alpha]_{\vartriangle_\lambda^{\!{}_{(3)}}} = -12\,\hbar^2\,\lambda^{\mu\nu\alpha\beta}\, x^\beta
\eea
for $\mu,\nu,\alpha,\beta=1,2,3,4$. For $\lambda=1$, these are just the brackets (up to rescaling) of the 3-Lie algebra $A_4$,
\bea\label{eq:3LieA4}
[x^\mu,x^\nu,x^\alpha]_{\vartriangle_1^{\!{}_{(3)}}} = 3\,\ell_s^3\,R \,\varepsilon^{\mu\nu\alpha\beta}\, x^\beta \ ,
\eea
familiar from studies of multiple M2-branes in M-theory where it
describes the polarisation of open membranes ending on an M5-brane
into fuzzy three-spheres~\cite{Bagger2012}; indeed, the brackets
\eqref{eq:3LieA4} quantize the standard Nambu-Poisson structure on the
three-sphere $S^3\subset\real^4$ of radius
$\sqrt{3\,\ell_s^3\,R/\hbar^2}$. In the present case we are in a
sector that involves only membranes of M-theory and excludes
M5-branes, but we can nevertheless interpret the nonassociative
geometry modelled on \eqref{eq:3LieA4}: It represents a (discrete) foliation of
the M-theory configuration space $\real^4$ by fuzzy membrane
worldvolume three-spheres,\footnote{See e.g.~\cite{DeBellis2010} for
  an analogous description of a noncommutative deformation of
  $\real^3$ in terms of discrete foliations by fuzzy two-spheres.} and in this sense our triproduct
\eqref{eq:triprod3} gives a candidate deformation quantization of the standard
Nambu-Poisson structure on $S^3$. We will say more about this perspective in Section~\ref{sec:Mtheory3alg}.

From the limits \eqref{eq:pLambda0} and \eqref{eq:vecstar0}, together
with \eqref{eq:asslambda0}, we see that the triproduct
\eqref{eq:triprod3} reproduces that of the string theory configuration
space $\real^3$ in the contraction limit
$\lambda\to0$~\cite{Blumenhagen2011,Aschieri2015}; in particular
\bea\label{eq:stringtribracket}
\lim_{\lambda\to0} \,
[x^i,x^j,x^k]_{\vartriangle_\lambda^{\!{}_{(3)}}} = -3\,\ell_s^3\, R\,
\varepsilon^{ijk} \ .
\eea
Thus while the string theory triproduct represents a deformation
quantization of the Nambu-Heisenberg 3-Lie algebra, its lift to M-theory represents a deformation quantization of the 3-Lie algebra $A_4$. Since the associator $\vec A_\eta\big(\vec p_{\mit\Lambda},\vec p_{\mit\Lambda}^{\,\prime},\vec p_{\mit\Lambda}^{\,\prime\prime} \big)$ is $\vec0$ whenever any of its arguments is the zero vector, using \eqref{eq:vecstarsumid} we have $
\vec{\mathcal{T}}_{\mit\Lambda}^{{}_{(3)}}(\vec k,\vec k',\vec 0\, )=\vec{\mathcal{T}}_{\mit\Lambda}^{{}_{(3)}}(\vec k,\vec 0,\vec k^{\prime}\,)=\vec{\mathcal{T}}_{\mit\Lambda}^{{}_{(3)}}(\vec 0,\vec k,\vec k^{\prime}\,)=\vec{\mathcal{T}}_{\mit\Lambda}^{{}_{(2)}}(\vec k,\vec k' \,)$ 
and so we obtain the unital property
\bea\nonumber
f\vartriangle_\lambda^{\!{}_{(3)}}g\vartriangle_\lambda^{\!{}_{(3)}}1= f\vartriangle_\lambda^{\!{}_{(3)}}1\vartriangle_\lambda^{\!{}_{(3)}}\,g= 1\vartriangle_\lambda^{\!{}_{(3)}}f\vartriangle_\lambda^{\!{}_{(3)}}g=f\vartriangle_\lambda^{\!_{(2)}}g \ ,
\eea
as expected from the $\lambda\to0$
limit~\cite{Blumenhagen2011,Aschieri2015}. However, in contrast to the
string theory triproduct, here the M-theory triproduct does \emph{not}
trivialise on-shell, i.e., $\int\, \dd^4\vec x_0\
f\vartriangle_\lambda^{\!{}_{(3)}}g\vartriangle_\lambda^{\!{}_{(3)}}h\neq \int\,
\dd^4\vec x_0\ f\,g\,h$. This is again related to the fact that the
precursor phase space star product $\star_\lambda$ is not closed, and
presumably one can find a suitable gauge equivalent triproduct $\blacktriangle_\lambda^{\!{}_{(3)}}$
analogous to the closed star product $\bullet_\lambda$ that we derived
in Section~\ref{sec:closure}. In Section~\ref{sec:Mtheory3alg} we will
give a more intrinsic definition of this triproduct in terms of an
underlying $Spin(7)$-symmetric 3-algebra on the membrane phase space.

We close the present discussion by sketching two generalisations of these constructions in light of the results of~\cite{Blumenhagen2011,Blumenhagen2013,Aschieri2015}. Firstly, one can generalize these derivations to work out explicit $n$-triproducts for any $n\geq4$, which in the string theory case would represent the off-shell contributions to $n$-point correlation functions of tachyon vertex operators in the $R$-flux background~\cite{Blumenhagen2011}. For functions $f_1,\dots,f_n$ of $\vec x_0=(\mbf x,x^4)\in\real^4$ we set
\bea\nonumber
\big(f_1\vartriangle_\lambda^{\!{}_{(n)}}\cdots\vartriangle_\lambda^{\!{}_{(n)}}f_n\big)(\vec
x_0)&:=& \big(\big(\cdots\big((f_1\star_\lambda f_2)\star_\lambda
f_3\big)\star_\lambda\cdots\big)\star_\lambda f_n\big)(\mbf x,x^4,\mbf
p) \big|_{\mbf p=\mbf 0} \\[4pt] &=& \nonumber \int \ \prod_{a=1}^n\, \frac{\dd^{4}\vec k_a}{( 2\pi
) ^{4}} \ \tilde f_a(\vec k_a) \ \e^{\ii\vec\CB_\eta(\vec\CB_\eta(\dots,(\vec\CB_\eta({\mit\Lambda}\,\vec k_1,{\mit\Lambda}\,\vec k_2),{\mit\Lambda}\,\vec k_3),\dots),{\mit\Lambda}\,\vec k_n)\,\mbf\cdot\, {\mit\Lambda}^{-1}\,\vec x_0}
\eea
and as before the nested compositions of vector additions can be computed from products of corresponding octonion exponentials $\big(\cdots\big((\e^{X_{{\mit\Lambda}\, \vec
    k_1}}\, \e^{X_{{\mit\Lambda}\, \vec k_2}})\,\e^{X_{{\mit\Lambda}\, \vec
    k_3}} \big)\cdots\big)\,
\e^{X_{{\mit\Lambda}\, \vec k_n}}$. The calculation simplifies again by dropping all vector cross products ${\mit\Lambda}\,\vec
k_a\,\mbf\times_\eta\,{\mit\Lambda}\,\vec k_b$ (which do not contribute to the inner product with ${\mit\Lambda}^{-1}\,\vec x_0$), by correspondingly dropping many higher iterations of associator terms using the contraction identity \eqref{eq:eta34}, and by making repeated use of the identities from Section~\ref{sec:crossvector}. Here we only quote the final result:
\bea\nonumber
\big(f_1\vartriangle_\lambda^{\!{}_{(n)}}\cdots\vartriangle_\lambda^{\!{}_{(n)}}f_n\big)(\vec x_0) = 
\int \ \prod_{a=1}^n\, \frac{\dd^{4}\vec k_a}{( 2\pi
) ^{4}} \ \tilde f_a(\vec k_a) \ \e^{\ii\vec{\mathcal{T}}_{\mit\Lambda}^{{}_{(n)}}( \vec k_1, \dots, \vec k_n)\,\mbf\cdot\, {\mit\Lambda}^{-1}\,\vec x_0} \ ,
\eea
where
\bea\nonumber
&& 
\vec{\mathcal{T}}_{\mit\Lambda}^{{}_{(n)}}( \vec k_1, \dots, \vec k_n) \ = \  
\frac{\sin^{-1}\big|\big(\cdots\big((\vec p_{1\mit\Lambda}\circledast_\eta \vec p_{2\mit\Lambda})
    \circledast_\eta\vec p_{3\mit\Lambda}\big)\circledast_\eta\cdots\big)\circledast_\eta \vec p_{n\mit\Lambda}\big|}{\hbar \, \big| \big(\cdots\big((\vec p_{1\mit\Lambda}\circledast_\eta \vec p_{2\mit\Lambda})
    \circledast_\eta\vec p_{3\mit\Lambda}\big)\circledast_\eta\cdots\big)\circledast_\eta \vec p_{n\mit\Lambda}\big|}\\ && \nonumber \times \ \Big(\, \sum_{a=1}^n\, \sqrt{1-\big|\big(\cdots\big((\vec p_{1\mit\Lambda}\circledast_\eta \vec p_{2\mit\Lambda})
    \circledast_\eta\cdots\big)\circledast_\eta \widehat{\vec
      p_{a\mit\Lambda}}\, \big)\circledast_\eta\cdots\big)\circledast_\eta \vec p_{n\mit\Lambda}\big|^2} \ \epsilon_a \ \vec
p_{a\mit\Lambda} \\ && \nonumber \times \ \sum_{a<b<c}\,
\sqrt{1-\big|\big(\cdots\big(\vec
  p_{1\mit\Lambda}\circledast_\eta\cdots\big)\circledast_\eta
  \widehat{\vec p_{a\mit\Lambda}}\, \big)\circledast_\eta\cdots\big)\circledast_\eta \widehat{\vec p_{b\mit\Lambda}}\big)\circledast_\eta\cdots\big)\circledast_\eta \widehat{\vec p_{c\mit\Lambda}}\big)\circledast_\eta\cdots\big)\circledast_\eta \vec p_{n\mit\Lambda}\big|^2} \\ && \nonumber \hspace{4cm} \ \times \ \epsilon_{abc} \ \vec A_\eta\big(\vec p_{a\mit\Lambda},\vec p_{b\mit\Lambda} ,\vec p_{c\mit\Lambda} \big)\, \Big)
\eea
and $\widehat{\vec p_{a\mit\Lambda}}$ denotes omission of $\vec p_{a\mit\Lambda}$ for $a=1,\dots,n$; here we abbreviated signs of square roots analogous to those in \eqref{eq:TLambda3} by $\epsilon_a$ and $\epsilon_{abc}$. Again we see that
\bea\nonumber
\vec{\mathcal{T}}_{\mit\Lambda}^{{}_{(n)}}\big( \vec k_1, \dots,(\vec k_a{=}\vec 0\,),\dots ,\vec k_n\big) = \vec{\mathcal{T}}_{\mit\Lambda}^{{}_{(n-1)}}\big( \vec k_1, \dots,\widehat{\vec k_a},\dots ,\vec k_n\big) \ ,
\eea
which implies that the $n$-triproducts obey the expected unital property
\bea\nonumber
f_1\vartriangle_\lambda^{\!{}_{(n)}}\cdots\vartriangle_\lambda^{\!{}_{(n)}}(f_a{=}1)\vartriangle_\lambda^{\!{}_{(n)}} \cdots\vartriangle_\lambda^{\!{}_{(n)}}f_n
= f_1\vartriangle_\lambda^{\!{}_{(n-1)}}\cdots\vartriangle_\lambda^{\!{}_{(n-1)}}\widehat{f_a}\vartriangle_\lambda^{\!{}_{(n-1)}}\cdots\vartriangle_\lambda^{\!{}_{(n-1)}} f_n
\eea
for $a=1,\dots,n$. As previously, these $n$-triproducts reduce to those of the string theory $R$-flux background in the limit $\lambda\to0$~\cite{Blumenhagen2011,Aschieri2015}.

Secondly, one can consider more general foliations of the M-theory phase space by leaves of constant momentum $\mbf p=\bar{\mbf p}$. This would modify the product \eqref{eq:triprod2} by introducing phase factors
\bea\nonumber
&& \exp\Big(\, \frac{\sin^{-1}\big|\vec p_{\mit\Lambda}\circledast_\eta\vec
  p_{\mit\Lambda}^{ \, \prime} \big|}{\hbar\, \big|\vec p_{\mit\Lambda}\circledast_\eta\vec
  p_{\mit\Lambda}^{ \, \prime} \big|}\, \frac{\ii\ell_s^3}{2\hbar^2}
\, R \, \bar{\mbf p} \,\mbf\cdot\, \big(\mbf p_{\mit\Lambda}\,\mbf\times_\varepsilon\,\mbf p_{\mit\Lambda}' +
 \lambda\, p_{{\mit\Lambda}4} \, \mbf p_{\mit\Lambda}'-\lambda\,
 p_{{\mit\Lambda}4}'\,\mbf p_{\mit\Lambda} \big) \, \Big) \\ && \nonumber \hspace{4cm} \ = \ \exp\Big(\, \frac{\ii\ell_s^3}{2\hbar}
\, R \, \bar{\mbf p} \,\mbf\cdot\, \big(\mbf k\,\mbf\times_\varepsilon\,\mbf k' +
 \lambda\, k_{4} \, \mbf k'-\lambda\,
 k_{4}'\,\mbf k \big) +O\big(\sqrt{\lambda}\,\big) \, \Big)
\eea
into the integrand, exactly as for the Moyal-Weyl type deformation of
the string theory $R$-flux background which is obtained at
$\lambda=0$~\cite{Aschieri2015} (see \eqref{BR}). This turns the
product $\vartriangle_\lambda^{\!_{(2)}}$ into a noncommutative (but
still associative) star product. One can likewise include such phase
factors into the calculations of higher $n$-triproducts
$\vartriangle_\lambda^{\!{}_{(n)}}$ to obtain suitable noncommutative
deformations. In the string theory setting, the physical meaning of
these non-zero constant momentum deformations is explained in~\cite{Aschieri2015}, and it would be interesting to understand their interpretation in the M-theory lift.

\subsection{Noncommutative geometry of momentum space\label{sec:polarizations}}

Polarisation along leaves of constant momentum is of course not the only possibility; see~\cite{Aschieri2015} for a general discussion of polarised phase space geometry in our context. A natural alternative polarisation is to set $x^\mu=0$ and restrict to functions on momentum space $\mbf p\in\real^3$. This is particularly interesting in the M-theory $R$-flux background: In the string theory case momentum space itself undergoes no deformation, whereas here we see from the brackets \eqref{oct9} and \eqref{oct9a} that momentum space experiences a noncommutative, but associative, deformation by the M-theory radius $\lambda$ alone. As pointed out already in Section~\ref{sec:quaternion}, in this polarisation the nonassociative star product (\ref{oct11})
reproduces the associative star product on $\real^3$ for quantisation
of the dual of the Lie algebra $\mathfrak{su}(2)$. Unlike the star
product of configuration space functions, the star product
$\star_\lambda$ restricts to the three-dimensional momentum space, so the projections employed in Section~\ref{sec:TriNAG} are not necessary and one can work directly with the star product restricted to $\real^3$; for functions $f$ and $g$ of $\mbf p\in\real^3$ it reads as
\bea\label{eq:QGstar}
(f\star_{\lambda}g)(\mbf p) = \int\, \frac{\dd^3\mbf
  l}{(2\pi)^3} \ \frac{\dd^3\mbf l'}{(2\pi)^3} \ \tilde f(\mbf l)\,
\tilde g(\mbf l'\,) \, \e^{-\frac{2\ii}\lambda\, {\mbf{\mathcal{
        B}}}_\varepsilon(-\frac\lambda2\, {\mbf l},-\frac\lambda2\,
  {\mbf l}'\, ) \,\mbf\cdot\, \mbf p} \ .
\eea
Thus the M-theory momentum space itself experiences an associative,
noncommutative deformation of its geometry, independently of the $R$-flux. We will say more about this purely membrane deformation in Section~\ref{sec:Mtheory3alg}.

We can understand this noncommutative deformation by again restricting
configuration space coordinates to the three-sphere $(\mbf x,x^4)\in
S^3\subset\real^4$ of radius $\frac1\lambda$, similarly to the fuzzy
membrane foliations we described in Section~\ref{sec:TriNAG}, although
the present discussion also formally applies to a vanishing $R$-flux. Then on
the upper hemisphere $x^4\geq0$ the quantization of the brackets
\eqref{oct9} yields the star commutators
\bea\nonumber
[p_i,p_j]_{\star_\lambda} = -\ii\hbar\, \lambda\, \varepsilon_{ijk}\,
p^k \qquad \mbox{and} \qquad [x^i,p_j]_{\star_\lambda}=
\ii\hbar\,\lambda\, 
\sqrt{\mbox{$\frac1{\lambda^2}$} -|\mbf x|^2}\ \delta^i_j+\ii\hbar\,
\lambda\, \varepsilon^i{}_{jk}\, x^k \ .
\eea
These commutation relations reflect the fact that the configuration
space is curved: They show that the momentum coordinates $\mbf p$ are
realised as right-invariant derivations on configuration space
$S^3\subset\real^4$, and as a result generate the brackets of the Lie
algebra $\mathfrak{su}(2)$. This noncommutativity has bounded position
coordinates, which is consistent with the minimal momentum space areas
$\langle \A^{p_i,p_j}\rangle$ computed in \eqref{au}. At weak string coupling $\lambda=0$, the
three-sphere decompactifies and one recovers the canonical quantum phase
space algebra of flat space $\real^3$. The star product
\eqref{eq:QGstar} enables order by order computations of
M-theory corrections to closed string amplitudes in this sense. The noncommutative geometry here parallels
that of three-dimensional quantum gravity~\cite{Freidel2005}, where
however the roles of position and momentum coordinates are
interchanged. 

\newsection{$Spin(7)$-structures and M-theory 3-algebra\label{sec:Mtheory3alg}}

In this final section we shall describe some preliminary steps towards
extending the quantum geometry of the $R$-flux compactification
described in Section~\ref{sec:Mtheory} to the full eight-dimensional
M-theory phase space. It involves replacing the notion of
$G_2$-structure with that of a $Spin(7)$-structure and the
quasi-Poisson algebra with a suitable 3-algebra, as anticipated on
general grounds in lifts of structures from string theory to M-theory. We show, in
particular, that quantisation of this 3-algebra naturally encompasses
the triproducts from Section~\ref{sec:TriNAG} and the deformed
geometry of the membrane momentum space from Section~\ref{sec:polarizations}.

\subsection{Triple cross products\label{sec:triplecross}}

The constructions of this section will revolve around the linear
algebraic notion of a \emph{triple cross product} on an
eight-dimensional real inner product space
$W$~\cite{HL82,Joyce,Salamon2010}. For three vectors $K,K',K''\in
W=\real^8$, their triple cross product $K\,\mbf\times_\phi\, K'\,\mbf\times_\phi\,
K''\in W$ is defined by
\bea\label{eq:triplecross}
(K\,\mbf\times_\phi\, K'\,\mbf\times_\phi\, K''\,)^{\hat A}:= \phi^{\hat A\hat B\hat C\hat D}\, K^{\hat B}\, K^{\prime\,\hat C}\, K^{\prime\prime\,\hat D} \ ,
\eea
where $\hat A,\hat B,\dots=0,1,\dots,7$ and $\phi_{\hat A\hat B\hat C\hat D}$ is a completely antisymmetric tensor of rank four with the nonvanishing values
\bea\nonumber
\phi_{\hat A\hat B\hat C\hat D}\ =\ +1 \qquad \mbox{for} \quad \hat
A\hat B\hat C\hat D &=& 0123, \ 0147, \ 0165, \ 0246, \ 5027, \ 3045, \ 3076, \\ \nonumber
&& 4567, \ 2365, \ 2374, \ 1537, \ 2145, \ 2176, \ 3164 \ .
\eea
It can be written more succinctly in terms of the structure constants of the octonion algebra $\oct$ as
\bea\label{eq:phieta}
\phi_{0ABC} = \eta_{ABC} \qquad \mbox{and} \qquad \phi_{ABCD} = \eta_{ABCD} \ .
\eea
Following~\cite{Kar10,Kar08}, the tensor $\phi_{\hat A\hat B\hat C\hat D}$ satisfies the self-duality relation
\bea\nonumber
\varepsilon_{\hat A\hat B\hat C\hat D\hat E\hat F\hat G\hat H}\, \phi_{\hat E\hat F\hat G\hat H} = \phi_{\hat A\hat B\hat C\hat D} \ ,
\eea
where $\varepsilon_{\hat A\hat B\hat C\hat D\hat E\hat F\hat G\hat H}$
is the alternating symbol in eight dimensions normalized as
$\varepsilon_{01234567}=+1$. It also obeys the contraction identity
\bea\label{eq:phicontraction}
\phi_{\hat A\hat B\hat C\hat D}\, \phi_{\hat A'\hat B'\hat C'\hat D} &=& \delta_{\hat A\hat A'}\, \delta_{\hat B\hat B'}\, \delta_{\hat C\hat C'} +\delta_{\hat A\hat B'}\, \delta_{\hat B\hat C'}\, \delta_{\hat C\hat A'}
+\delta_{\hat A\hat C'}\, \delta_{\hat B\hat A'}\, \delta_{\hat C\hat B'} \\
&& \nonumber -\, \delta_{\hat A\hat A'}\, \delta_{\hat B\hat C'}\, \delta_{\hat C\hat B'} -\delta_{\hat A\hat B'}\, \delta_{\hat B\hat A'}\, \delta_{\hat C\hat C'}
-\delta_{\hat A\hat C'}\, \delta_{\hat B\hat B'}\, \delta_{\hat C\hat A'} \\
&& \nonumber -\,\delta_{\hat A\hat A'}\,\phi_{\hat B\hat C\hat B'\hat C'}
-\delta_{\hat B\hat A'}\,\phi_{\hat C\hat A\hat B'\hat C'}
-\delta_{\hat C\hat A'}\,\phi_{\hat A\hat B\hat B'\hat C'} \\
&& \nonumber -\, \delta_{\hat A\hat B'}\,\phi_{\hat B\hat C\hat C'\hat A'}
-\delta_{\hat B\hat B'}\,\phi_{\hat C\hat A\hat C'\hat A'}
-\delta_{\hat C\hat B'}\,\phi_{\hat A\hat B\hat C'\hat A'} \\
&& \nonumber -\, \delta_{\hat A\hat C'}\,\phi_{\hat B\hat C\hat A'\hat B'}
-\delta_{\hat B\hat C'}\,\phi_{\hat C\hat A\hat A'\hat B'}
-\delta_{\hat C\hat C'}\,\phi_{\hat A\hat B\hat A'\hat B'} \ .
\eea

Despite its appearance, the triple cross product is
\emph{not} a simple iteration of the cross product $\,\mbf\times_\eta\,$, but
it can also be expressed in terms of the algebra of octonions. For this, we choose a split
$W=\real\oplus V$ with $K=(k_0,\vec k\, )\in W$, and consider the
octonion $X_K:=k_0\, \unit+ k^A\,e_A$ together with its conjugate $\bar
X_K:=k_0\, \unit- k^A\,e_A$. As the commutator of two octonions is
purely imaginary, the cross product from
\eqref{eq:Xveckcomm}  can in fact be expressed in terms of eight-dimensional
vectors as
\bea\label{eq:XKcomm}
X_{\vec k \,\mbf\times_\eta\, \vec k'}=\mbox{$\frac12$}\, \big[X_{K},X_{K'} \big] \ .
\eea
The natural extension of \eqref{eq:XKcomm} to three vectors $K,K',K''\in W$ is the antisymmetrization of $\big(X_K\,\bar X_{K'}\big)\, X_{K''}$, which can be simplified by repeated use of properties of the octonion algebra to give
\bea\label{eq:XKtriple}
X_{K\,\mbf\times_\phi\, K'\,\mbf\times_\phi\, K''}= \mbox{$\frac12$}\, \big((X_K\,\bar X_{K'})\, X_{K''}-(X_{K''}\, \bar X_{K'})\, X_K\big) \ .
\eea

This trilinear product satisfies the defining properties of triple cross products~\cite{Salamon2010}:
\begin{description}
\item[(TC1)]  $K_1\,\mbf\times_\phi\, K_2\,\mbf\times_\phi\, K_3=(-1)^{|\sigma|}\, 
  K_{\sigma(1)}\,\mbf\times_\phi\, K_{\sigma(2)}\,\mbf\times_\phi\, K_{\sigma(3)}$ for
  all permutations $\sigma\in S_3$\,;
\item[(TC2)] $K_1\,\mbf\cdot\,(K_2\,\mbf\times_\phi\, K_3 \,\mbf\times_\phi\, K_4)=-
  K_2\,\mbf\cdot\,(K_1\,\mbf\times_\phi\, K_3\,\mbf\times_\phi\, K_4)$\, ;
\item[(TC3)]  $|K_1\,\mbf\times_\phi\, K_2\,\mbf\times_\phi\, K_3| = |K_1\wedge
  K_2\wedge K_3|$\,.
 \end{description}
From the definitions above it follows that
\bea\label{eq:triplecrossexpl}
K\,\mbf\times_\phi\, K'\,\mbf\times_\phi\, K''&=& \big(\vec k\,\mbf\cdot\,(\vec k'\,\mbf\times_\eta\, \vec k''\,) \ , \\ && \nonumber \qquad \mbox{$\frac1{12}$}\, \vec{J}_\eta(\vec k,\vec k ',\vec k^{\prime\prime}\,) -k_0\, (\vec k' \,\mbf\times_\eta\, \vec k''\,)-k_0'\, (\vec k'' \,\mbf\times_\eta\, \vec k \,) -k_0''\, (\vec k \,\mbf\times_\eta\, \vec k'\,) \big) \ .
\eea
Only rotations in the $21$-dimensional spin group $Spin(7)\subset SO(8)$ preserve the
triple cross product, where the action of $Spin(7)$ can be described
as the transitive action on the unit sphere $S^7\subset W$
identified with the homogeneous space $S^7\simeq Spin(7)/G_2$; the Lie group $Spin(7)$ is isomorphic to the double cover of $SO(7)$, with the two copies of $SO(7)$ corresponding to the upper and lower hemispheres of $S^7$ and $G_2$ the unique lift to a subgroup of $Spin(7)$. A
\emph{$Spin(7)$-structure} on an oriented eight-dimensional vector space
$W$ is the choice of a triple cross product that can be written as
\eqref{eq:triplecross} in a suitable oriented frame.

An important feature of a $Spin(7)$-structure is that it can be used
to generate $G_2$-structures. For this, let $\hat k\in W$ be a fixed unit vector and let $V_{\hat k}$ be the orthogonal complement to the real line spanned by $\hat k$ in $W$. Then $W=\real\oplus V_{\hat k}$, and using \eqref{eq:phieta} the seven-dimensional subspace $V_{\hat k}$ carries a cross product~\cite[Theorem~6.15]{Salamon2010}
\bea\label{eq:crosshatk}
\vec k\,\mbf\times_{\hat k}\,\vec k':= \hat k\,\mbf\times_\phi\, \vec k\,\mbf\times_\phi\,\vec
k' \ ,
\eea
for $\vec k,\vec k'\in V_{\hat k}$.  

\subsection{Phase space 3-algebra\label{sec:M3algebra}}

Let us set $\lambda=1$, and consider the symmetries
underlying the quasi-Poisson bracket relations \eqref{oct9} and
\eqref{oct9a}. For this, we rewrite the bivector
\eqref{eq:Thetaeta} in component form as
\bea\nonumber
\Theta_\eta= \mbox{$\frac12\, \varepsilon_{ijk}\,
  \xi_k\,\frac\partial{\partial\xi_i}\wedge\frac\partial{\partial
    \xi_j}+\xi_i\,\big(\frac\partial{\partial \sigma^4}\wedge\frac\partial{\partial\sigma^i}+\varepsilon_{ijk}\, \frac\partial{\partial\sigma^j}\wedge\frac\partial{\partial\sigma^k}\big) -\frac\partial{\partial\xi_i}\wedge \big(\sigma^4\,\frac{\partial}{\partial\sigma^i}+
  \varepsilon_{ijk}\, \sigma^j\, \frac\partial{\partial \sigma^k}\big)$} \ .
\eea
From our discussion of $G_2$-structures from
Section~\ref{sec:crossproducts}, it follows that this bivector is
invariant under the subgroup $G_2\subset SO(7)$ which is generated by
antisymmetric $7\times7$ matrices $S=(s_{AB})$ satisfying $\eta_{ABC}\, s_{BC}=0$ for $A=1,\dots,7$. Applying the affine transformation \eqref{oct6} (with $\lambda=1$) generically breaks this symmetry to an $SO(4)\times SO(3)$ subgroup of $SO(7)$. As $G_2$ contains no nine-dimensional subgroups (the maximal compact subgroup $SU(3)\subset G_2$ is eight-dimensional), the residual symmetry group is $G_2\cap\big( SO(4)\times SO(3) \big)$ (see e.g.~\cite{Mayanskiy2016} for a description of the corresponding regular subalgebra $G_2[\alpha]$ of the Lie algebra of $G_2$). This exhibits the non-invariance of the quasi-Poisson algebra under $SL(4)$ and $SL(3)$ observed in~\cite{GLM}; however, as also noted by~\cite{GLM}, the $SL(3)$ symmetry is restored in the contraction limit by the discussion of Section~\ref{sec:quaternion}, and indeed the string theory quasi-Poisson algebra and its quantization from Section~\ref{sec:NAdef} are $SL(3)$-invariant, being based on the three-dimensional cross product.

This symmetry breaking may be attributed to the specific choice of
frame wherein the $R$-flux has non-vanishing components
$R^{4,\mu\nu\alpha\beta}$ and the momentum constraint
\eqref{eq:Rpconstraint} is solved by $p_4=0$. In~\cite{GLM} it is
suggested that this constraint could be implemented in a covariant
fashion on the eight-dimensional phase space by the introduction of
some sort of ``Nambu-Dirac bracket''. Here we will offer a slightly different, but related, explicit
proposal for such a construction based on the $Spin(7)$-structures
introduced in Section~\ref{sec:triplecross}. The impetus behind this
proposal is that, in the lift from string theory to M-theory,
2-brackets should be replaced by suitable 3-brackets, as has been
observed previously on many occasions (see e.g.~\cite{Bagger2012,Ho2016} for
reviews); it is naturally implied by the Lie 2-algebra structure discussed at the beginning of Section~\ref{sec:G2star}. This is prominent in the lift via T-duality of the $SO(3)$-invariant D1--D3-brane system in IIB string theory to the $SO(4)$-symmetric M2--M5-brane system wherein the underlying Lie algebra $\mathfrak{su}(2)$, representing the polarisation of D1-branes into fuzzy two-spheres $S_F^2$, is replaced by the 3-Lie algebra $A_4$, representing the polarization of M2-branes into fuzzy three-spheres $S_F^3$ (see e.g.~\cite{DeBellis2010,Saemann2011,Saemann2012,Saemann2016} for reviews in the present context); we have already adapted a similar point of view in our considerations of Sections~\ref{sec:TriNAG} and~\ref{sec:polarizations}. Based on this, the BLG model uses a
3-algebra for the underlying gauge symmetry to construct the
$\mathcal{N}=2$ worldvolume theory on the M-theory membrane, which is
related to $\mathcal{N}=6$ Chern-Simons theories, while the Moyal-Weyl type deformation of the coordinate algebra of 
D3-branes in a flat two-form $B$-field background of $10$-dimensional
supergravity lifts to Nambu-Heisenberg 3-Lie algebra type deformations of the
coordinate algebra of M5-branes in a flat 
three-form $C$-field background of $11$-dimensional supergravity~\cite{Chu2009}.
If we moreover adopt the point of view of~\cite{MSS2} that closed strings in $R$-flux compactifications should be regarded as boundary degrees of freedom of open membranes whose topological sector is described by an action that induces a phase space quasi-Poisson structure on the boundary, then this too has a corresponding lift to M-theory: In that case the action for an open topological 4-brane induces a 3-bracket structure on the boundary~\cite{Park2000}, regarded as the worldvolume of M2-branes in the M-theory $R$-flux background.

Taking this perspective further, we will adapt the point of view
of~\cite{Chu2010}: In some systems with gauge symmetry, 3-brackets
$\{\!\{f,g,h\}\!\}$ of fields can be defined without gauge-fixing, in
contrast to quasi-Poisson brackets which depend on a gauge choice,
such that for any gauge-fixing condition $G=0$ the quasi-Poisson
bracket $\{f,g\}_G$ in that gauge is simply given by
$\{f,g\}_G = \{\!\{f,g,G\}\!\}$. This is analogous to the
procedure of reducing a 3-Lie algebra to an ordinary Lie algebra by
fixing one slot of the 3-bracket (see e.g.~\cite{Takhtajan1993,DeBellis2010}), and
it can be used to dimensionally reduce the BLG theory of
M2-branes to the
maximally supersymmetric Yang-Mills theory of D2-branes~\cite{Mukhi2008}.
It is also reminescent of the relation \eqref{eq:crosshatk} between cross
products and triple cross products, which motivates an application of
this construction to the full eight-dimensional M-theory phase space
subjected to the constraint \eqref{eq:Rpconstraint}. Writing
${\mit\Xi}=(\xi_0,\vec\xi\ )\in\real^8$, we extend the $G_2$-symmetric bivector $\Theta_\eta$ to the $Spin(7)$-symmetric trivector
\bea\label{eq:Phi}
\Phi:=\mbox{$\frac13\,\phi_{\hat A\hat B\hat C\hat D}\, {\mit\Xi}^{\hat D}\, \frac\partial{\partial {\mit\Xi}^{\hat A}}\wedge \frac\partial{\partial {\mit\Xi}^{\hat B}}\wedge\frac\partial{\partial {\mit\Xi}^{\hat C}}= \frac\partial{\partial\xi_0}\wedge\Theta_\eta+\eta_{ABCD}\, \xi_D\, \frac\partial{\partial\xi_A}\wedge\frac\partial{\partial\xi_B}\wedge \frac\partial{\partial\xi_C}$}
\eea
which generates a 3-algebra structure on coordinate functions $\complex[{\mit\Xi}]$ with 3-brackets
\bea\nonumber
\big\{\!\big\{{\mit\Xi}_{\hat A},{\mit\Xi}_{\hat B},{\mit\Xi}_{\hat C}\big\}\!\big\}_\phi = 2\, \phi_{\hat A\hat B\hat C\hat D}\, {\mit\Xi}_{\hat D} \ .
\eea
The trivector \eqref{eq:Phi} is a Nambu-Poisson tensor if these 3-brackets satisfy the fundamental identity~\cite{Takhtajan1993}, thus defining a 3-Lie algebra structure on $\complex[{\mit\Xi}]$, so the Jacobiator \eqref{cjac} is now replaced by the 5-bracket 
\bea\nonumber
\{f_1,f_2,g,h,k\}_\phi&:=& \{\!\{f_1,f_2,\{\!\{g,h,k\}\!\}_\phi\}\!\}_\phi- \{\!\{ \{\!\{f_1,f_2,g\}\!\}_\phi,h,k\}\!\}_\phi \\ && \nonumber -\, \{\!\{ g ,\{\!\{f_1,f_2,h\}\!\}_\phi,k\}\!\}_\phi- \{\!\{ g,h,\{\!\{f_1,f_2,k\}\!\}_\phi\}\!\}_\phi \ ,
\eea
which is natural from the index structure of the M-theory
$R$-flux. One can compute it explicitly on linear functions
${\mit\Xi}_{\hat A}$ by using the contraction
identity~\eqref{eq:phicontraction} to get
\bea\nonumber
\{{\mit\Xi}_{\hat A},{\mit\Xi}_{\hat B},{\mit\Xi}_{\hat
  C},{\mit\Xi}_{\hat D},{\mit\Xi}_{\hat E}\}_\phi&=&\,
12\,\big(\,\delta_{\hat A\hat C}\,\phi_{\hat D\hat E\hat B\hat
  F}+\delta_{\hat A\hat D}\,\phi_{\hat E\hat C\hat B\hat
  F}+\delta_{\hat A\hat E}\,\phi_{\hat C\hat D\hat B\hat F} \\ && \nonumber
\qquad \qquad -\, \delta_{\hat B\hat C}\,\phi_{\hat D\hat E\hat A\hat F} 
-\delta_{\hat B\hat D}\,\phi_{\hat E\hat C\hat A\hat F}
-\delta_{\hat B\hat E}\,\phi_{\hat C\hat D\hat A\hat
  F}\,\big)\,{\mit\Xi}^{\hat F}\\ &&\nonumber
-\, 12\,\big( \, {\mit\Xi}_{\hat A}\,\phi_{\hat B\hat C\hat D\hat
  E}-{\mit\Xi}_{\hat B}\,\phi_{\hat A\hat C\hat D\hat E}\,\big) \ .
\eea

To write the 3-brackets of phase space coordinates, we use \eqref{oct8} to define an affine transformation of the vector space $\real^8$ given by
$X=(\vec x,p_4)=(x^\mu,p_\mu)= ({\mit\Lambda}\,
\vec\xi,-\frac\lambda2\, \xi_0)$, which we have chosen to preserve
$SO(4)$-symmetry of momentum space; this breaks (at $\lambda=1$) the
symmetry of the trivector \eqref{eq:Phi} to a subgroup $Spin(7)\cap\big(SO(4)\times SO(4)\big)$ of $SO(8)$. Then the 3-brackets are given by
\bea\nonumber
\big\{\!\big\{ x^A,x^B,x^C\big\}\!\big\}_\phi &=& -\lambda^{ABCD}\, x^D-\mbox{$\frac2\lambda$}\, {\mit\Lambda}^{AA^\prime}\,{\mit\Lambda}^{BB^\prime}\,{\mit\Lambda}^{CC^\prime}\,\eta_{A^\prime B^\prime C^\prime}\, p_4 \ , \\[4pt] \nonumber \big\{\!\big\{ p_4, x^A,x^B\big\}\!\big\}_\phi &=& -\mbox{$\frac\lambda2$}\, \lambda^{ABC}\, x^C \ .
\eea
Altogether, the phase space 3-algebra of the non-geometric M-theory
$R$-flux background is summarised by the 3-brackets
\bea\label{eq:M3algebra}
\{\!\{x^i,x^j,x^k\}\!\}_\phi &=& \mbox{$\frac{\ell_s^3}{2\hbar^2}$} \,
R^{4,ijk4}\, x^4 \qquad \mbox{and} \qquad \{\!\{x^i,x^j,x^4\}\!\}_\phi \ =
\ -\mbox{$\frac{\lambda^2\,\ell_s^3}{2\hbar^2}$}\, R^{4,ijk4}\, x_k \
, \nonumber \\[4pt]
\{\!\{p^i,x^j,x^k\}\!\}_\phi &=&
\mbox{$\frac{\lambda^2\,\ell_s^3}{2\hbar^2}$}\, R^{4,ijk4}\,
p_4+\mbox{$\frac{\lambda\,\ell_s^3}{2\hbar^2}$}\, R^{4,ijk4}\, p_k \ ,
\nonumber\\[4pt]
\{\!\{p_i,x^j,x^4\}\!\}_\phi &=&
 \mbox{$\frac{\lambda^2\,\ell_s^3}{2\hbar^2}$}\, R^{4,1234}\,
\delta_i^j\, p_4+\mbox{$\frac{\lambda^2\,\ell_s^3}{2\hbar^2}$}\,
R^{4,ijk4}\, p_k \ , \nonumber \\[4pt]
\{\!\{p_i,p_j,x^k\}\!\}_\phi &=& -\mbox{$\frac{\lambda^2}2$}\,
\varepsilon_{ij}{}^k\, x^4-\mbox{$\frac\lambda2$}\,\big(\delta_j^k\,
x_i-\delta_i^k\, x_j\big) \ , \nonumber \\[4pt]
\{\!\{p_i,p_j,x^4\}\!\}_\phi &=& \mbox{$\frac{\lambda^3}2$}\,
\varepsilon_{ijk}\, x^k \qquad \mbox{and} \qquad
\{\!\{p_i,p_j,p_k\}\!\}_\phi \ = \ 2\,\lambda \,
\varepsilon_{ijk}\, p_4 \ , \nonumber\\[4pt]
\{\!\{p_4,x^i,x^j\}\!\}_\phi &=& -\mbox{$\frac{\lambda\, \ell_s^3}{2\hbar^2}$}\,
R^{4,ijk4}\, p_k \qquad \mbox{and} \qquad \{\!\{p_4,x^i,x^4\}\!\}_\phi \ =
\ \mbox{$\frac{\lambda^2\, \ell_s^3}{2\hbar^2}$}\, R^{4,1234}\, p^i \ ,
\nonumber \\[4pt]
\{\!\{p_4,p_i,x^j\}\!\}_\phi &=& \mbox{$\frac\lambda2$}\, \delta_i^j\,
x^4+ \mbox{$\frac{\lambda^2}2$} \, \varepsilon_i{}^{jk}\, x_k \ , \nonumber
\\[4pt]
\{\!\{p_4,p_i,x^4\}\!\}_\phi &=& \mbox{$\frac{\lambda^3}2$} \, x_i \qquad
\mbox{and} \qquad \{\!\{p_4,p_i,p_j\}\!\}_\phi \ = \
\mbox{$\frac{\lambda^2}2$} \, \varepsilon_{ijk}\, p^k \ .
\eea

For any constraint
\bea\nonumber
G(X)=0
\eea
on the eight-dimensional phase space, one can now define quasi-Poisson brackets through
\bea\nonumber
\{f,g\}_G := \{\!\{f,g,G\}\!\}_\phi \ .
\eea
In particular, for the constraint function $G(X)=
-\frac2\lambda\, p_4$ these brackets reproduce the quasi-Poisson
brackets $\{f,g\}_\lambda$ of the seven-dimensional phase space from
Section~\ref{sec:QPAMtheory}; the more general choice $G(X) =
R^{\mu,\nu\rho\alpha\beta} \, p_\mu$ constrains the 3-algebra of the
eight-dimensional phase space to the codimension one hyperplane
defined by
\eqref{eq:Rpconstraint} (which in the gauge $p_4=0$ is orthogonal to
the $p_4$-direction). Moreover, setting $p_4=0$ in the remaining
3-brackets from \eqref{eq:M3algebra} reduces them to the
seven-dimensional phase space Jacobiators from
Section~\ref{sec:QPAMtheory}; we shall see this feature at the quantum
level later on. 
Finally, we observe that in the limit $\lambda\to0$ reducing M-theory
to IIA string theory, the only non-vanishing 3-brackets from
\eqref{eq:M3algebra} are $\{\!\{x^i,x^j,x^k\}\!\}_\phi $, which at
$x^4=1$ reproduce the string theory Jacobiator \eqref{cjac}, and
$\{\!\{p_4,x^I,x^J\}\!\}_\phi$, which at $x^4=1$ reproduce the
bivector \eqref{rb1}. This framework thereby naturally explains the
$SO(4)$-invariance (at $\lambda=1$) of the M-theory 
brackets \eqref{oct9} and \eqref{oct9a} described in \cite{GLM}: The trivector \eqref{eq:Phi} can be
alternatively modelled on the space $S_-(\real^4)$ of
negative chirality spinors over $\real^4$ (which is a quaternionic
line bundle over $\real^4$) with respect to the
splitting $W=\real^4\oplus\real^4$ (see e.g.~\cite{Kar10} for details).

These consistency checks support our proposal for the M-theory phase
space 3-algebra. It suggests that in the absence of $R$-flux, $R^{\mu,\nu\rho\alpha\beta}=0$, the brackets \eqref{eq:M3algebra} describe the free phase space 3-algebra structure of M2-branes with the non-vanishing 3-brackets
\bea\nonumber
\{\!\{p_i,p_j,x^k\}\!\} &=& -\mbox{$\frac{\lambda^2}2$}\,
\varepsilon_{ij}{}^k\, x^4-\mbox{$\frac\lambda2$}\,\big(\delta_j^k\,
x_i-\delta_i^k\, x_j\big) \qquad \mbox{and} \qquad
\{\!\{p_i,p_j,x^4\}\!\}  \ = \ \mbox{$\frac{\lambda^3}2$}\,
\varepsilon_{ijk}\, x^k \ , \\[4pt] \label{eq:Mphase3alg}
\{\!\{p_i,p_j,p_k\}\!\} &=& \ 2\,\lambda \,
\varepsilon_{ijk}\, p_4 \qquad \mbox{and} \qquad
\{\!\{p_4,p_i,x^j\}\!\} \ = \ \mbox{$\frac\lambda2$}\, \delta_i^j\,
x^4+ \mbox{$\frac{\lambda^2}2$} \, \varepsilon_i{}^{jk}\, x_k \ , 
\\[4pt] \nonumber
\{\!\{p_4,p_i,x^4\}\!\} &=& \mbox{$\frac{\lambda^3}2$} \, x_i \qquad
\mbox{and} \qquad \{\!\{p_4,p_i,p_j\}\!\} \ = \
\mbox{$\frac{\lambda^2}2$} \, \varepsilon_{ijk}\, p^k \ ,
\eea
which as expected all vanish in the weak string coupling limit $\lambda\to0$. Quantization of this 3-algebra then provides a higher version of the noncommutative geometry of the M-theory momentum space discussed in Section~\ref{sec:polarizations}.
In the following we will provide further evidence for these assertions.

\subsection{Vector trisums\label{sec:trisums}}

At this stage the next natural step is to quantise the 3-algebra
\eqref{eq:M3algebra}. Although at present we do not know how to do this
in generality, we can show how the star product of
Section~\ref{sec:Mtheorystar} and the
configuration space triproducts of Section~\ref{sec:TriNAG} naturally
arise from the 3-algebraic structure of the
full membrane phase space; this is based on a natural
ternary extension of the vector star sum described in
Section~\ref{sec:crossvector}. To motivate its construction, let us first provide an
alternative eight-dimensional characterisation of the binary operation
\eqref{vstar} on the unit ball $B^7\subset V\subset W$. For this, we
note that for arbitrary vectors $P,P'\in W$ one has~\cite{Salamon2010}
\bea\label{eq:XP8D}
X_P\, X_{P'} = \big(p_0\,p_0'-\vec p \,\mbf\cdot\, \vec p\,'\,\big)\, \unit
+p_0\, X_{\vec p\,'}+p_0'\, X_{\vec p} +X_{\vec p\, \mbf\times_\eta\,
  \vec p\,'} \ .
\eea
By restricting to vectors $P$ from the unit sphere $S^7\subset W$,
i.e., $|P|=1$, we
can easily translate this identity to the vector star sum
\eqref{vstar}: We fix a hemisphere $p_0=\pm\, \sqrt{1-|\vec p\,|^2}$,
such that for vectors $\vec p,\vec p\,'$ in the ball $B^7 = S^7/\zed_2
\simeq SO(7)/G_2$ we
reproduce the seven-dimensional vector star sum through
\bea\label{1}
X_{\vec p\circledast_\eta\vec
  p\,^{\prime}}=\mathfrak{Im}\big( X_{P\,^{\prime}}\, X_P\big) =
\mbox{$\frac12$}\, \big(X_{P\,^{\prime}}\, X_{P}-\bar X_{P}\, \bar X_{P\,^{\prime}}\big) \ ,
\eea
where the sign factor $\epsilon_{\vec p,\vec p\,'}$ from
\eqref{vstar}, which is the sign of the real part of \eqref{eq:XP8D}, ensures that the result of the vector star sum remains in
the same hemisphere. This interpretation of the vector star sum is
useful for deriving various properties. For example, since the algebra of
octonions $\oct$ is a normed algebra, for $|P|=1$ we have $|X_P|=1$
and
\bea\nonumber
\big|X_{P\,^{\prime}}\,
X_{P}\big|^2=\big|\mathfrak{Re}(X_{P\,^{\prime}}\, X_{P})
\big|^2+\big|\mathfrak{Im}(X_{P\,^{\prime}}\, X_{P}) \big|^2=1 \ .
\eea
From \eqref{eq:XP8D} and \eqref{1} we then immediately infer the
identity \eqref{eq:vstarid}.

By comparing the representation \eqref{eq:XKcomm} of the vector cross product
with the definition \eqref{1} of the vector star sum, we can apply
the same reasoning to the
triple cross product represented through \eqref{eq:XKtriple}. We 
define a ternary operation on $B^7=S^7/\zed_2\subset W$ by
\bea\label{5}
X_{\vec p\circledast_\phi\vec
    p\,^{\prime}\circledast_\phi \vec
    p\,^{\prime\prime}}:=\mathfrak{Im}\big( (X_{P} \,
  X_{P\,^{\prime}})\, X_{P\,^{\prime\prime}}\big) \ ,
\eea
and call it a \emph{vector trisum}. To obtain an explicit
expression for it we observe that
\begin{eqnarray*}
(X_{P} \, X_{P\,^{\prime}})\, X_{P\,^{\prime\prime}}=
  \mbox{$\frac12$}\, \big((X_{P}\, X_{P\,^{\prime}})\,
  X_{P\,^{\prime\prime}}+(X_{P''}\, X_{P'})\, X_P\big) +X_{P\,
  \mbf\times_\phi\, \bar P' \,\mbf\times_\phi\, P''} \ ,
\end{eqnarray*}
which leads to
\begin{eqnarray}\label{6}
\vec p\circledast_\phi\vec p\,^{\prime}\circledast_\phi \vec
  p\,^{\prime\prime}&=&\epsilon_{\vec p,\vec p\,',\vec p\,''}\,\Big(\epsilon_{\vec p\,',\vec p\,^{\prime\prime}}\,
                        \sqrt{1-\big|\vec  p\,^{\prime}
                        \circledast_\eta \vec p\,^{\prime\prime}\, \big|^2}
                        \ \vec p+ \epsilon_{\vec p,-\vec p\,^{\prime\prime}}\, \sqrt{1-\big|\vec
   p\circledast_\eta (-\vec
   p\,^{\prime\prime}\,) \big|^2} \ \vec p\,^{\prime} \\
&&+\, \epsilon_{\vec p,\vec p\,'}\, \sqrt{1-\big|\vec
   p\circledast_\eta \vec p\,^{\prime}\,
   \big|^2} \ \vec
   p\,^{\prime\prime}+\vec A_\eta(\vec p, \vec
   p\,^{\prime}, \vec p\,^{\prime\prime}\, )\nonumber\\
&&+\, \sqrt{1-|\vec  p\, |^2}\, \big(\vec p\,^{\prime}
   \,\mbf\times_\eta\, \vec p\,^{\prime\prime}\, \big)+ \sqrt{1-|\vec
   p\,^{\prime} |^2}\, \big(\vec p\,^{\prime\prime}
   \,\mbf\times_\eta\, \vec p\,\big)+\sqrt{1-|\vec  p\,^{\prime\prime}
   |^2}\, \big(\vec p\, \mbf\times_\eta\, \vec p\,^{\prime}\, \big)
   \Big) \ . \nonumber
\end{eqnarray}
Here $\epsilon_{\vec p,\vec p\,',\vec p\,''}=\pm\,1$ is the sign of $\mathfrak{Re}\big((X_{P} \,
 X_{P\,^{\prime}})\, X_{P\,^{\prime\prime}}\big)$ satisfying
\bea\nonumber
\epsilon_{\vec p,\vec p\,'}=\epsilon_{\vec p,\vec p\,',\vec 0}=\epsilon_{\vec p,\vec 0,\vec p\,'}=\epsilon_{\vec 0,\vec p,\vec p\,'} \ ,
\eea
which as previously ensures that the result of the vector trisum remains in
the same hemisphere of $S^7$.
As before, since $|X_P|=1$ we have
\begin{equation*}
\big| (X_{P} \, X_{P\,^{\prime}})\,
X_{P\,^{\prime\prime}}\big|^2=\big| \mathfrak{Re}\big((X_{P} \,
X_{P\,^{\prime}})\, X_{P\,^{\prime\prime}}\big) \big|^2+\big|
\mathfrak{Im}\big((X_{P}\, X_{P\,^{\prime}})\,
X_{P\,^{\prime\prime}}\big)\big|^2=1 \ .
\end{equation*}
This implies that
\begin{eqnarray}\nonumber
1-\big|\vec p\circledast_\phi\vec p\,^{\prime}\circledast_\phi\vec
  p\,^{\prime\prime}\big|^2&=&\big| \mathfrak{Re}(X_{P}\,
                               X_{P\,^{\prime}})\, X_{P\,^{\prime\prime}}\big|^2\\[4pt]
&=& \big(\, \sqrt{1-|\vec  p\,|^2}\, \sqrt{1-|\vec  p\,^{\prime}
    |^2}\, \sqrt{1-|\vec p\,^{\prime\prime}\, |^2}-\sqrt{1-|\vec
    p\,^{\prime\prime}|^2}\ \vec p\,\mbf\cdot\, \vec p \nonumber\\
&&-\, \sqrt{1-|\vec  p\,|^2}\ \vec p\,^{\prime} \,\mbf\cdot\, \vec
   p\,^{\prime\prime} -\sqrt{1-|\vec  p\,^{\prime}|^2}\ \vec p
   \,\mbf\cdot\, \vec p\,^{\prime\prime} -\vec p\,^{\prime\prime}
   \,\mbf\cdot\, (\vec p \,\mbf\times_\eta\, \vec
   p\,^{\prime}\,)\big)^2 \ \geq \ 0 \ , \nonumber
\end{eqnarray}
which shows that the vector $\vec p\circledast_\phi\vec
p\,^{\prime}\circledast_\phi\vec p\,^{\prime\prime}$ indeed also
belongs to the unit ball $B^7\subset V$.

The relation between the vector trisum $\vec
p\circledast_\phi\vec p\,^{\prime}\circledast_\phi\vec
p\,^{\prime\prime}$ and the vector star sum $\vec
p\circledast_\eta\vec p\,^{\prime}$ can be described as follows. From
the definitions \eqref{1} and \eqref{5} it easily follows that
\begin{equation}
\label{7}
\vec p\circledast_\eta\vec p\,^{\prime}=\vec p\circledast_\phi\vec
0\circledast_\phi \vec p\,^{\prime}=\vec 0\circledast_\phi\vec
p\,^{\prime}\circledast_\phi\vec p\,=\vec
p\,^{\prime}\circledast_\phi\vec p\circledast_\phi \vec 0 \ .
\end{equation}
By \eqref{eq:crosshatk}, for $\hat p=(1,\vec 0\, )\in W$ there is an analogous relation
\begin{equation*}
\vec p \,\mbf\times_\eta\, \vec p\,^{\prime}=\vec p
\,\mbf\times_\phi\, \hat p \,\mbf\times_\phi\, \vec p\,^{\prime}=\hat
p \,\mbf\times_\phi\, \vec p\,^{\prime} \mbf\times_\phi\, \vec
p\,=\vec p\,^{\prime} \,\mbf\times_\phi\, \vec p \,\mbf\times_\phi\,
\hat p \ .
\end{equation*}
From \eqref{eq:triplecrossexpl} it follows that the antisymmetrization of the vector trisum $\vec
p\circledast_\phi \vec p\,^{\prime}\circledast_\phi \vec
p\,^{\prime\prime}$ reproduces the imaginary part of the triple cross
product $\mathfrak{Im}(X_{P \,\mbf\times_\phi\, P' \,\mbf\times_\phi\,
  P''})$; this is analogous to the relation \eqref{eq:vstarcomm} between the vector star
sum and the cross product. However, in
contrast to \eqref{eq:XKcomm}, for the antisymmetrization of the
product of three octonions one sees
\begin{equation*}
\mathfrak{Im}\big(X_{P \,\mbf\times_\phi\, P' \,\mbf\times_\phi\,
  P''}\big) \neq \mathfrak{Im}\big(X_{\vec p \,\mbf\times_\phi\, \vec
  p\,' \,\mbf\times_\phi\, \vec p\,''} \big) \ ,
\end{equation*}
because the antisymmetrization of the vector trisum $\vec
p\circledast_\phi\vec p\,^{\prime}\circledast_\phi\vec
p\,^{\prime\prime}$ contains terms involving $p_0=\pm\, \sqrt{1-| \vec
  p\,|^2}$.

To extend the vector trisum (\ref{6}) over the entire
vector space $V\subset W$, we again apply the map
\eqref{eq:pkmap}. Then for $\vec k,\vec k',\vec k''\in V$, the
corresponding mapping of the vector trisum is given by
\begin{equation}\label{Bkphi}
\vec{\mathcal{ B}}_\phi(\vec k,\vec k',\vec k''\,):=
\left.\frac{\sin^{-1}|\vec p\circledast_\phi\vec
    p\,^{\prime}\circledast_\phi\vec
    p\,^{\prime\prime}\,|}{\hbar\,|\vec p\circledast_\phi\vec
    p\,^{\prime}\circledast_\phi \vec p\,^{\prime\prime}\,|}\ \vec
  p\circledast_\phi\vec p\,^{\prime}\circledast_\phi \vec
  p\,^{\prime\prime}\, \right|_{ \vec p=\vec k\sin(\hbar\, |\vec
  k|)/|\vec k|} \ .
\end{equation}
From (\ref{Bkphi}) one immediately infers the following properties:
\begin{description}
\item[(TB1)] $\vec{\mathcal{B}}_\phi(\vec k_1,\vec k_2,\vec
  k_3)=(-1)^{|\sigma|}\, \vec{\mathcal{B}}_\phi(-\vec k_{\sigma(1)},-\vec k_{\sigma(2)},-\vec
  k_{\sigma(3)}) $ for all permutations $\sigma\in S_3$\ ;
\item[(TB2)]  $\vec{\mathcal{ B}}_\eta(\vec k,\vec
  k'\,)=\vec{\mathcal{ B}}_\phi(\vec k,\vec 0, \vec
  k'\,)=\vec{\mathcal{ B}}_\phi(\vec 0, \vec k',\vec
  k\,)=\vec{\mathcal{ B}}_\phi( \vec k',\vec k,\vec 0\, )$\, ;
\item[(TB3)] Perturbative expansion:
\begin{eqnarray*}
 \vec{\mathcal{ B}}_\phi(\vec k,\vec k',\vec k''\,)&=&\vec k+\vec
                                                       k'+\vec k''+{
                                                       \hbar}\,\big(
                                                       \vec
                                                       k\times_\eta\vec
                                                       k'+ \vec
                                                       k''\times_\eta\vec
                                                       k+\vec
                                                       k'\times_\eta\vec
                                                       k''\, \big)\\
 &&+\, \mbox{$ \frac{\hbar^2}{2}$}\,\big(2\,\vec A_\eta(\vec k, \vec
    k\,^{\prime}, \vec k\,^{\prime\prime}\, )-\big|\vec k\,'+\vec k\,
    ''\,\big|^2\,\vec k-\big|\vec k+\vec k\, ''\, \big|^2\,\vec k\,'-
    \big|\vec k\,'+\vec k\,\big|^2\,\vec k\,''\, \big) \\ &&
                                                             +\, O\big(\hbar^3
                                                             \big) \ ;
\end{eqnarray*}
\item[(TB4)] The higher associator
\bea\nonumber
\vec\CA_\phi(\vec k_1,\vec k_2,\vec k_3,\vec k_4,\vec k_5)&:=&
\vec\CB_\phi\big(\vec k_1\,,\,\vec k_2\,,\,\vec\CB_\phi(\vec k_3,\vec k_4,\vec
k_5)\big)- \vec\CB_\phi\big(\vec\CB_\phi(\vec k_1,\vec k_2,\vec k_3)\,,\,\vec
k_4\,,\,\vec k_5\big) \\ && \nonumber
-\, \vec\CB_\phi\big(\vec k_3\,,\,\vec\CB_\phi(\vec k_1,\vec k_2,\vec k_4)\,,\,\vec
k_5\big) - \vec\CB_\phi\big(\vec k_3\,,\,\vec k_4\,,\,\vec\CB_\phi(\vec k_1,\vec
k_2,\vec k_5)\big)
\eea
is antisymmetric in all arguments.
 \end{description}

\subsection{Phase space triproducts\label{sec:triproducts}}

We now define a triproduct on the M-theory phase
space, analogously to the star product of
Section~\ref{sec:Mtheorystar}, by setting
\begin{equation}\label{tsp}
(f\diamond_\lambda g\diamond_\lambda h)( \vec x\, ) =\int \, \frac{\dd^{7}\vec k}{( 2\pi
) ^{7}} \ \frac{\dd^{7}\vec k'}{( 2\pi
) ^{7}} \ \frac{\dd^{7}\vec k''}{( 2\pi
) ^{7}} \ \tilde{f}( \vec k\,)\, \tilde{g}(\vec k'\,)\,\,
\tilde{h}(\vec k''\,)\, \e^{\ii\vec{\mathcal{ B}}_\phi({\mit\Lambda}\,
  \vec k,{\mit\Lambda}\, \vec k',{\mit\Lambda}\, \vec k''\,)
  \,\mbf\cdot\, {\mit\Lambda}^{-1}\,\vec x} \ ,
\end{equation}
where as before $\tilde{f}$ stands for the Fourier transform of the
function $f$ and $\vec{\mathcal{ B}}_\phi(\vec k,\vec k',\vec k''\,)$
is the deformed vector sum \eqref{Bkphi}. Property {\bf{(TB2)}} from Section~\ref{sec:trisums}
implies that the triproduct $f\diamond_\lambda g\diamond_\lambda h$ is
related to the star product $f\star_\lambda g$, in precisely the same
way that the triple cross product is related to the cross product,
through the unital property
\begin{equation}\label{10}
f\star_\lambda  g=f\diamond_\lambda 1\diamond_\lambda
g=1\diamond_\lambda g\diamond_\lambda f=g\diamond_\lambda
f\diamond_\lambda 1 \ .
\end{equation}
If we define a quantum 3-bracket in the usual way by 
\begin{equation}\label{12}
[\![f_1,f_2,f_3]\!]_{\diamond_\lambda} :=\sum_{\sigma\in S_3}\, (-1)^{|\sigma|} \
f_{\sigma(1)}\diamond_\lambda f_{\sigma(2)} \diamond_\lambda
f_{\sigma(3)} \ ,
\end{equation}
then by property {\bf{(TB3)}} it has a perturbative expansion given to lowest
orders by
\begin{equation}\label{13}
     [\![f,g,h]\!]_{\diamond_\lambda}=-\ii\hbar \, \big(f\, \{
     g,h\}_\lambda+g\, \{h,f\}_\lambda + h\, \{f,g\}_\lambda\big)-3\, \hbar^2\,\{f,g,h\}_\lambda+O(\hbar,\lambda) \ .
\end{equation}
The fact that the perturbative expansion \eqref{13} contains terms without derivatives can be easily understood from the property (\ref{10}), which together with (\ref{12}) implies
\begin{equation}
\label{14}
 [\![f,g,1]\!]_{\diamond_\lambda}=-3\,[f,g]_{\star_\lambda} \ .
\end{equation}
Although this feature may seem unusual from the conventional perspective of
deformation quantization, it is exactly the quantum version of the 
gauge fixing of 3-brackets to 2-brackets on the reduced M-theory phase
space that we discussed in Section~\ref{sec:M3algebra}.

All this generalises the properties of the triproducts described in
Section~\ref{sec:TriNAG}; indeed it is easy to see that the phase space
triproduct $\diamond_\lambda$ reduces at $\mbf p=\mbf 0$ to the
configuration space triproduct $\vartriangle_\lambda^{\!{}_{(3)}}$ by
comparing \eqref{6} and \eqref{Bkphi} with \eqref{eq:TLambda3}. In
particular, it is straightforward to show that the 3-bracket
\eqref{12} reproduces the string theory 3-bracket on configuration
space in the collapsing limit of the M-theory circle, as in
\eqref{eq:stringtribracket}: From the calculations of
Section~\ref{sec:Mtheorystar} we find
\begin{eqnarray*}
\lim_{\lambda\to0} \, \vec{\mathcal{ B}}_\phi({\mit\Lambda}\,\vec
k,{\mit\Lambda}\, \vec k'\,,{\mit\Lambda}\, \vec k''\, )\,\mbf\cdot\, {\mit\Lambda}^{-1}\,\vec x &=& (\mbf k+\mbf k'\,+\mbf k''\,)
\,\mbf\cdot\,\mbf x + (k_4+k_4'\,+k_4''\,)
\, x^4 + (\mbf l+\mbf l'\,+\mbf l''\,)
\,\mbf\cdot\,\mbf p \\
&& -\, \mbox{$\frac{\ell_s^3\, R}{2\hbar}$}\, \mbf p\,\mbf\cdot\, (\mbf
k'\,\mbf\times_\varepsilon\,\mbf k''\,+\mbf
k\,\mbf\times_\varepsilon\,\mbf k''\,+\mbf
k\,\mbf\times_\varepsilon\,\mbf k'\,)\\ &&
+\, \mbox{$\frac{\hbar}{2}$} \, x^4\, (\mbf k'\,\mbf\cdot\,\mbf l''-\mbf l'\,\mbf\cdot\,\mbf k''\,-\mbf k''\,\mbf\cdot\,\mbf l+\mbf l''\,\mbf\cdot\,\mbf k\,+\mbf k\,\mbf\cdot\,\mbf l'-\mbf l\,\mbf\cdot\,\mbf k'\,)\\&&
-\, \mbox{$\frac{\ell_s^3\, R}{2}$}\,x^4\,\mbf k\,\mbf\cdot\, (\mbf
k'\,\mbf\times_\varepsilon\,\mbf k''\,)
\ ,
\end{eqnarray*}
so that in this limit the 3-bracket (\ref{12}) for the configuration
space coordinates ${\mbf x}$ upon setting ${\mbf p}=\mbf 0$ yields
\begin{equation*}
\lim_{\lambda\to0} \, [\![x^i,x^j,x^k]\!]_{\diamond_\lambda}\, \Big|_{\mbf
  p=\mbf 0}=-3\, \ell_s^3\, R \, \varepsilon^{ijk} \ .
\end{equation*}

The phase space triproduct \eqref{tsp} also naturally
quantises the 3-algebraic structure \eqref{eq:Mphase3alg} of the
membrane momentum space.
Restricting \eqref{6} to vectors $\vec p=(\mbf q,\mbf 0,0)$ yields a
non-vanishing vector trisum of vectors $\mbf q\in\real^3$ with
vanishing associator $\mbf J_\varepsilon$ and cross products
$\,\mbf\times_\eta\,$ replaced with $\,\mbf\times_\varepsilon\,$. The
restriction of the triproduct \eqref{tsp} to functions of $\mbf p$
alone thereby produces a non-trivial momentum space triproduct which
quantises the 3-algebra \eqref{eq:Mphase3alg}.

\subsection*{Acknowledgments}

We thank Peter Schupp for helpful discussions. The authors acknowledge support by the Action MP1405 QSPACE from 
the European Cooperation in Science and Technology (COST), the Consolidated Grant ST/L000334/1 
from the UK Science and Technology Facilities Council (STFC), 
the Visiting Researcher Program
Grant 2016/04341-5 from the Funda\c{c}\~{a}o de Amparo \'a Pesquisa do
Estado de S\~ao Paulo (FAPESP, Brazil), the Grant 443436/2014-2 from the Conselho Nacional de Pesquisa (CNPq, Brazil), and the Capes-Humboldt Fellowship 0079/16-2.

\end{document}